\documentclass[twocolumns,10pt,final]{IEEEtran}

\usepackage{times}
\usepackage{amsmath,dsfont}
\usepackage{amssymb,amsthm}
\usepackage{epsfig,verbatim}
\usepackage{mathrsfs}
\usepackage{ifthen}
\usepackage{syntonly}
\usepackage{graphicx}
\usepackage{epstopdf}
\usepackage{subfig}
\usepackage[font=footnotesize]{caption}
\usepackage{multirow}
\usepackage{stfloats}
\usepackage{float}
\usepackage{latexsym}
\usepackage{cite}
\usepackage{psfrag}
\usepackage{subfloat}
\usepackage{subfig}
\usepackage{mnsymbol}
\usepackage{setspace}
 \allowdisplaybreaks

\newtheorem{remark}{Remark}

\newtheorem{definition}{Definition}

\newtheorem{theorem}{Theorem}
\newtheorem{lemma}[theorem]{Lemma}

\newtheorem{corollary}[theorem]{Corollary}

\floatstyle{ruled}%
\newfloat{floatequation}{thbp}{leq}
\floatname{floatequation}{}

\begin{document}

\newcommand\numberthis{\addtocounter{equation}{1}\tag{\theequation}}
\newcommand*{\medcup}{\mathbin{\scalebox{1.3}{\ensuremath{\bigcup}}}}%
\newcommand{\be}{\begin{equation}}
\newcommand{\ee}{\end{equation}}
\newcommand{\bea}{\begin{eqnarray}}
\newcommand{\eea}{\end{eqnarray}}
\newcommand{\beaa}{\begin{eqnarray*}}
\newcommand{\eeaa}{\end{eqnarray*}}
\newcommand{\p}[1]{\left(#1\right)}
\newcommand{\pp}[1]{\left[#1\right]}
\newcommand{\ppp}[1]{\left\{#1\right\}}
\def\ci{\perp\!\!\!\perp} 
\newcommand\independent{\protect\mathpalette{\protect\independenT}{\perp}} 
\def\independenT#1#2{\mathrel{\rlap{$#1#2$}\mkern2mu{#1#2}}}

\title{The Finite State MAC with Cooperative Encoders and Delayed CSI}
\author{Ziv Goldfeld, Haim H. Permuter and Benjamin M. Zaidel
\thanks{
Manuscript received March 31, 2013; revised December 15, 2013; accepted July 18, 2014. The work was supported by the European Research Council (ERC) starting grant, ISF grant no. 684/11 and the IMOD. This paper was presented in part at the IEEE International Symposium on Information Theory 2012, Cambridge, MA, USA, July, 2012, and in part at the 2012 IEEE 27-th Convention of Electrical and Electronics Engineers in Israel, Eilat, Israel, November, 2012.
\par Ziv Goldfeld and Haim Permuter are with the department of Electrical and Computer Engineering,  Ben-Gurion University of the Negev, Beer-Sheva, Israel (e-mails: zgzg1984@gmail.com,  haimp@bgu.ac.il).
\par Benjamin M.\ Zaidel is an independent researcher (e-mail: benjamin.zaidel@gmail.com).
}}
\maketitle

\begin{abstract}
In this paper, we consider the finite-state multiple access channel (MAC) with partially cooperative encoders and delayed channel state information (CSI). Here partial cooperation refers to the communication between the encoders via finite-capacity links. The channel states are assumed to be governed by a Markov process. Full CSI is assumed at the receiver, while at the transmitters, only delayed CSI is available. The capacity region of this channel model is derived by first solving the case of the finite-state MAC with a common message. Achievability for the latter case is established using the notion of strategies, however, we show that optimal codes can be constructed directly over the input alphabet. This results in a single codebook construction that is then leveraged to apply simultaneous joint decoding. Simultaneous decoding is crucial here because it circumvents the need to rely on the capacity region's corner points, a task that becomes increasingly cumbersome with the growth in the number of messages to be sent. The common message result is then used to derive the capacity region for the case with partially cooperating encoders. Next, we apply this general result to the special case of the Gaussian vector MAC with diagonal channel transfer matrices, which is suitable for modeling, e.g., orthogonal frequency division multiplexing (OFDM)-based communication systems. The capacity region of the Gaussian channel is presented in terms of a convex optimization problem that can be solved efficiently using numerical tools. The region is derived by first presenting an outer bound on the general capacity region and then suggesting a specific input distribution that achieves this bound. Finally, numerical results are provided that give valuable insight into the practical implications of optimally using conferencing to maximize the transmission rates.
\end{abstract}




\section{\textsc{Introduction}}
\par

\begin{figure*}[ht]
\begin{center}
\begin{psfrags}
    \psfragscanon
    \psfrag{A}{Encoder 1}
    \psfrag{B}{Encoder 2}
    \psfrag{C}{$\ P_{Y|X_1,X_2,S}$}
    \psfrag{D}{\hspace{.7mm}Decoder}
    \psfrag{E}{$M_1$}
    \psfrag{F}{$M_2$}
    \psfrag{G}{$\ \ \ X_{1}^{n}$}
    \psfrag{H}{$\ \ \ X_{2}^{n}$}
    \psfrag{I}{$Y^{n}$}
    \psfrag{J}{$(\hat{M}_1,\hat{M}_2)$}
    \psfrag{K}{$S_{i-d_1}$}
    \psfrag{L}{$S_{i-d_2}$}
    \psfrag{M}{\ \ $S_{i}$}
    \psfrag{N}{ \ \ Channel}
    \psfrag{O}{\hspace{-1.2mm}$V_1^\ell$}
    \psfrag{P}{$V_2^\ell$}
    \psfrag{Q}{\hspace{-1.2mm}$C_{12}$}
    \psfrag{R}{$C_{21}$}
\includegraphics[scale=0.65]{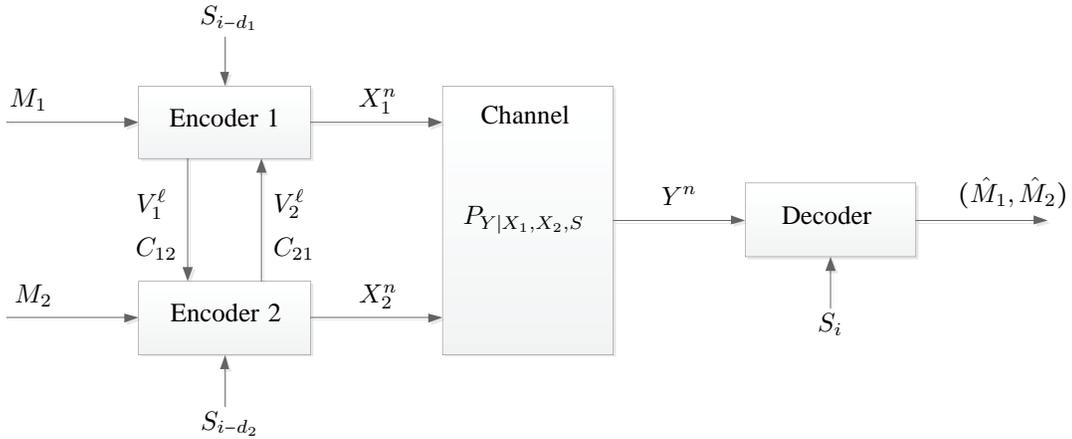}
\caption{FSM-MAC with partially cooperative encoders, CSI at the decoder and delayed CSI at the encoders with delays $d_1$ and $d_2$.} \label{MAC Conference Delay_int}
\psfragscanoff
\end{psfrags}
\end{center}
\end{figure*}

Temporal variations, a characteristic typical of wireless channels, may occur due to atmospheric changes, changes in the environment, the mobility of transmitters and/or receivers or time-varying intentional or unintentional interference. Since accurate channel state information (CSI) at both the transmitting and the receiving ends is crucial for efficient communications, measures are commonly incorporated in the communication protocol to enable channel state estimation. For example, the long term evolution (LTE) cellular communication standard relies on pilot signals transmitted at pre-scheduled time intervals and frequency slots to estimate the channel's state \cite{LTE}. Performed at the receiver, these estimations are then typically fed back to the transmitter, but obtaining perfect CSI at both ends of the channel in practical systems is a formidable challenge. More often than not, CSI is subject to channel estimation errors and feedback is not instantaneous due to some inevitable processing delay, and as a result, receivers and transmitters typically have access to only partial CSI. The impact of such partial CSI on the achievable performance, therefore, has attracted much attention in recent years. In the case of multiuser communication, performance is affected not only by channel characteristics, but also by interactions between the users. In particular, different forms of cooperation between the transmitting and receiving ends, a subject of growing interest in recent years (e.g., \cite{Maric_Yates_Kramer2007,Benny_Coop2011}), may significantly enhance performance. This paper aims to investigate the combined impact of both partial CSI and cooperation. More specifically, we focus on a two-user finite state Markov multiple access channel (FSM-MAC), with \emph{partially} cooperative encoders and \emph{delayed} CSI, as illustrated in Fig. \ref{MAC Conference Delay_int} and explained in the following text.

\par In the communication scenario under discussion, each of the two encoders wishes to send an independent private message through a time-varying MAC to the decoder. Delayed CSI is assumed to be available at the encoders, while full delayless CSI is assumed at the decoder. Different users may be subject to different CSI delays. It is further assumed that prior to each transmission block, the two encoders are allowed to hold a conference. More specifically, it is assumed that the encoders can communicate with each other over noise-free communication links of given capacities. We restrict the discussion to the case in which the conference held between the encoders is independent of the CSI.

\par The non-state-dependent MAC with partially cooperative encoders was first introduced by Willems \cite{Willems83_cooperating}, who also derived the capacity region for the discrete memoryless setting.
Special cases of this channel model include that in which the encoders are ignorant of each other's messages (i.e., the capacities of the communication links between them are both zero) and that in which the encoders fully cooperate (i.e., the capacities of the communication links are infinite).
The first setting, where no conference is held, corresponds to the classical MAC, for which the capacity region was determined by Ahlswede \cite{Ahlswede73MAC} and Liao \cite{Liao}. In contrast, in the second setting, where total cooperation is available, the encoders can act as one by fully sharing their private messages via the conference. The capacity region for this case is the part of the first quadrant below the so-called total cooperation line. This triangle-shaped region always contains the capacity region for the classical MAC.

\par In his proof of achievability for the conferencing MAC, Willems \cite{Willems83_cooperating} introduced a coding scheme based on the capacity region for the MAC with a common message, derived by Slepian and Wolf in \cite{Slepian_Wolf_MAC73}. Willems showed that in order to achieve the capacity region, the encoders should use the cooperation link to share parts of their private messages and then use a coding scheme for the ordinary MAC with a common message. Although Willems's model allows interactive communication between the encoders, it was shown both in \cite{Willems83_cooperating} and later in \cite{Willems85_cribbing_encoders} that a single round of communication between the encoders (referred to as a ``pair of simultaneous monologues'' in \cite{Willems83_cooperating}) suffices to achieve optimality.

\par Additional multiuser settings that involve cooperation between users through communication links of finite capacities have been extensively treated in the literature. See, for example, \cite{Willems1982} and \cite{Simeone:2009} for studies of the MAC, \cite{Maric_Yates_Kramer2007} and \cite{Prabhakaran2009,Prabhakaran2011,Wang_Tse,Bagheri2009,Ng_Jindal2007,Maric_Kramer_Shamai2007} for studies of the interference channel with cooperating nodes, \cite{DaboraServetto2006} for the broadcast channel, \cite{GunduzErkip2007} and \cite{SankarKramer2008} for cooperative relaying and \cite{Benny_Coop2011} and \cite{SimeoneSomekhPoorShamai} and references therein for cooperation in cellular architectures. A comprehensive survey of cooperation and its role in communication can be found in \cite{Kramer_Maric}. It is important to note, however, that in all of the above settings the channel was not assumed to be time-varying.

\par Multiuser settings that combine both time-varying channels and user cooperation are obviously of major interest as well. A Gaussian fading MAC with cooperating encoders that have access to delayless CSI was considered in \cite{HaghiMarvasti2011Corr} and in \cite{HaghiAref10_coopertaive_mac_state_fading_isit}. As in our case, these works assume that cooperation is allowed only before the CSI becomes available at the encoders. The case in which the CSI becomes available to the encoders prior to transmission is treated in \cite{PermuterShamaiSomekh_MessageStateCooperation}, where a MAC with perfect noncausal CSI is considered. The coding scheme introduced in \cite{PermuterShamaiSomekh_MessageStateCooperation} uses conferencing  to share parts of the messages as well as CSI.

\par The notion of modeling time-varying channels as state-dependent channels dates back to Shannon \cite{Shannon58}, who characterized the capacity of the state-dependent, memoryless point-to-point channel with independent and identically distributed (i.i.d.) states available causally at the encoder. To establish achievability, Shannon presented a code construction that relied on ``strategies'' (or ``strategy letters'') \cite{Shamai99}, a notion we also exploit in this paper. Gelfand and Pinsker \cite{GePi80}, and later Heegard and El Gamal \cite{HeegardElGamal_state_encoded83}, studied the case in which the encoder observes the channel states noncausally. In both \cite{GePi80} and \cite{HeegardElGamal_state_encoded83} a single letter expression for the capacity is derived using random binning. In \cite{Goldsmith_Varaiya}, Goldsmith and Varaiya considered a fading channel with perfect CSI at both the transmitter and the receiver. It was shown that in such a case, the optimal strategy is to employ waterfilling over time.

\par As was already stated, because perfect CSI is difficult to obtain in practical systems, models that involve partial or imperfect CSI have attracted a lot of attention in recent years. At first, different settings involving an i.i.d. state sequence with imperfect CSI were treated. Initially, various point-to-point channel scenarios with partial CSI were studied. Among others, the causal, noncausal, rate-limited and noisy cases were addressed \cite{Rosenzweig2005,Jafar2006,Salehi1992}. Extension of the result to the MAC with rate-limited CSI can be found in \cite{Steinberg_Cemal_MAC05}. In \cite{Como_Yuksel_MAC2011}, the authors derive the capacity region for the MAC with asymmetric quantized CSI at the encoders, where the quantization models the imperfection in the channel state estimation (full CSI at the decoder is assumed). Later, in \cite{Lapidoth2013MAC} Lapidoth and Steinberg provided an inner bound for the capacity region of the MAC with strictly causal CSI at the encoders. In contrast to the point-to-point setting, where strictly causal CSI regarding an i.i.d. state sequence does not increase capacity, the capacity region of the MAC with causal CSI is strictly larger than the corresponding region without CSI.
Li \emph{et al}. presented an improved inner bound for the same setting in \cite{Simeone_StrictlyCausalCSI}. A comprehensive monograph on channel coding in the presence of side information can be found in \cite{Keshet_Steinberg_Merhav2007}, where an i.i.d. state sequence is assumed. An information theoretic model for a single user channel involving delayed CSI and a state process that is no longer restricted to be memoryless and i.i.d. was first introduced by Viaswanathan \cite{Viswanathan99}, who derived the capacity while assuming a FSM channel. This result was later generalized by Caire and Shamai in \cite{Shamai99}, where they addressed a point-to-point channel in which the CSIs at both encoder and decoder admit some general joint probability law. A general capacity formula, which relies on the notion of \emph{inf-information rate} \cite{Verdu94}, is then provided for the case of state processes with memory. The result is then shown to boil down to a single-letter characterization in the case in which perfect CSI is available to the receiver, the CSI at the transmitter is given by a deterministic function of the channel state, and the two processes are jointly stationary and ergodic. By an appropriate choice of the above deterministic function, the result for Viswanathan's delayed CSI model \cite{Viswanathan99} is obtained as a special case of the result in \cite{Shamai99}. A generalization of the point-to-point results of \cite{Shamai99} to the MAC was presented by Das and Narayan in \cite{Das_Narayan_MAC2002}. The generality of the channel model therein leads to multiletter characterization of the capacity region in various settings, which unfortunately provides limited insight into practical encoding schemes for channel models in this framework.

\par Taking a practically oriented approach, we focus in this paper on a specific channel definition that leads to single-letter results. Following \cite{Viswanathan99}, we model temporal variations by means of a FSM channel \cite{Gallager1968,H.S._Wang_and_N.Moayeri}. The channel state is determined on a per symbol basis and governed by the underlying FSM process. An important extension of this idea to the multiuser case was introduced by Basher \emph{et al}. in \cite{Basher_Permuter}, presenting the FSM-MAC with delayed CSI and non-cooperating encoders, i.e., where no conference is held (see also \cite{Permuter_Simeone2011} for a related source coding analysis). In the proof of the capacity region for this model, achievability was established by employing a coding scheme based on rate-splitting and multiplexing-coding combined with successive decoding at the receiver. Successive decoding was used in \cite{Basher_Permuter} to demonstrate that the two corner points of the capacity region are achievable. The whole capacity region is then achievable via time-sharing. Although the setting in \cite{Basher_Permuter} constitutes a special case of the general model in \cite{Das_Narayan_MAC2002}, the main contribution of \cite{Basher_Permuter} is the single-letter characterization of the capacity region and the detailed construction of the coding scheme.

\par In the current paper, accounting for the availability of a conferencing link between the encoders, we take a different approach than that taken in \cite{Basher_Permuter}. We base the proof of achievability on the coding scheme for the MAC with a common message as presented in \cite{Willems83_cooperating}, and therefore, we start by deriving the capacity region for the FSM-MAC with a common message and the same CSI properties as in \cite{Basher_Permuter}. We thus provide a solution to what has been, until now, an unsolved problem. Next, using the achievable scheme for the common message setting, the achievability of the conferencing region is established. We note that the large number of corner points induced by the presence of an additional transmission rate (namely, the rate of the common message) render the provision of an achievable coding scheme for the common message setting based on achieving the region's corner points an awkward task. Moreover, the use of rate-splitting and multiplexing-coding when a common message is involved yields a rather complex coding scheme which we sought to avoid.

\par Therefore, we present an alternative coding scheme that employs strategy letters in the code construction (cf., e.g., \cite{Shannon58,Shamai99} and \cite{Das_Narayan_MAC2002}) and simultaneous decoding. However, unlike the case of Shannon's classical result for the point-to-point channel with causal encoder CSI, here we show that optimal codes can be constructed directly over the input alphabet (as also shown for certain special cases in \cite{Das_Narayan_MAC2002}). Namely, a single codebook is generated for each of the three messages over a super-alphabet that corresponds to the different realizations of the delayed CSI available at the encoders. At each time instance, a symbol that is correlated with the \emph{current} available delayed CSI is selected by the encoders and transmitted to the channel.
Thus, in contrast to previous works involving delayed CSI (cf., \cite{Viswanathan99} and \cite{Basher_Permuter}), here rate-splitting is no longer required. The decoder then uses its access to full CSI (which deterministically defines the delayed state sequences as well) to reduce each codeword (originally constructed over a super-alphabet) to a sequence over the input alphabet and executes a simultaneous decoding scheme based on joint typicality. Indeed, one of the most signiﬁcant contributions of our paper is this coding scheme for the MAC with a common message and delayed CSI.
Not only does it successfully avoid the unnecessary complexity of its rate-splitting and multiplexing counterpart and relies on a simpler codebook construction, it also achieves every possible point in the region rather than only the corner points. Furthermore, this two-user coding scheme is easily extendable to the case of multiple users with a \emph{single} common message.

\begin{figure*}[ht]
\begin{center}
\begin{psfrags}
    \psfragscanon
    \psfrag{A}{Encoder 1}
    \psfrag{B}{Encoder 2}
    \psfrag{C}{$\ P_{Y|X_1,X_2,S}$}
    \psfrag{D}{\hspace{.7mm}Decoder}
    \psfrag{E}{$(M_0,M_1)$}
    \psfrag{F}{$(M_0,M_2)$}
    \psfrag{G}{$\ \ \ X_{1}^{n}$}
    \psfrag{H}{$\ \ \ X_{2}^{n}$}
    \psfrag{I}{$Y^{n}$}
    \psfrag{J}{$(\hat{M}_0,\hat{M}_1,\hat{M}_2)$}
    \psfrag{K}{$S_{i-d_1}$}
    \psfrag{L}{$S_{i-d_2}$}
    \psfrag{M}{\ \ $S_{i}$}
    \psfrag{N}{\ \ \ Channel}
\includegraphics[scale=0.65]{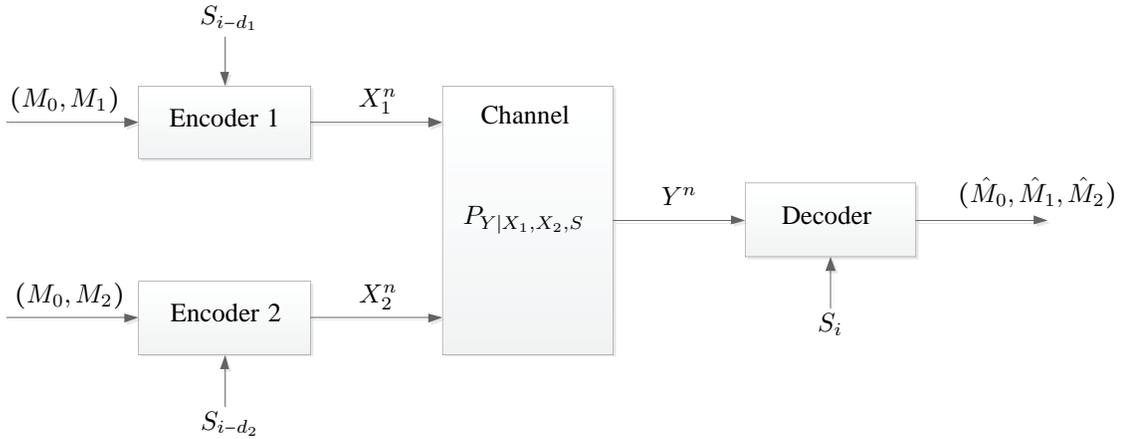}
\caption{FSM-MAC with a common message, full CSI at the decoder and delayed CSI at the encoders with delays $d_1$ and $d_2$.} \label{MAC Common Delay}
\psfragscanoff
\end{psfrags}
\end{center}
\end{figure*}

\par Based on the general results for the FSM-MAC with conferencing, we continue with the derivation of the capacity region for the special case of a vector Gaussian FSM-MAC with diagonal channel transfer matrices. This channel model can be used to represent an orthogonal frequency-division multiplexing (OFDM)-based communication system, employing single receive and transmit antennas, where the diagonal entries of the channel matrices represent the orthogonal sub-channels used by the OFDM scheme.

\par To derive the capacity region for the latter channel, we use a multivariate extension of a novel tool first derived in \cite{Venkatesan} (namely, a necessary and sufficient condition for a Gaussian triplet of random variables to satisfy a certain Markov relation), and demonstrate that Gaussian multivariate distributions maximize certain mutual information expressions under a Markovity constraint. The scalar version of this tool was employed by Lapidoth \emph{et al}. \cite{Wigger_gaussian_cop} to provide an outer bound for the capacity region of the scalar Gaussian non-state-dependent MAC with conferencing encoders. Wigger and Kramer also used this tool in their solution for the capacity region of the three-user, non-state-dependent MIMO MAC with conferencing \cite{ThreeUserMAC_Conf}. The need to use the tool from \cite{Venkatesan} stems from the fact that the input distribution of the conferencing channel must admit a certain Markovity constraint. For cases in which no Markov relation needs to be satisfied, the traditional approach to proving the optimality of Gaussian multivariate distributions involves employing either the Vector Max-Entropy Theorem (a direct extension of \cite[Theorem 12.1.1]{CovThom06}) or a conditional version of it. Here, however, this approach fails since replacing a non-Gaussian vector satisfying the Markovity condition by a Gaussian vector of the same covariance matrix may result in a Gaussian vector that violates the Markovity condition. To overcome this issue we use a sufficient and necessary condition on the (auto- and cross-) covariance matrices of the involved Gaussian random vectors for them to admit a Markov relation~\cite[Section 2, Theorem~1]{Markov_Gauss}.

\par We note that although Gaussian input vectors are shown to be optimal in this setting, the original form of the capacity region involves a non-convex optimization problem. To circumvent this difficulty, new variables are introduced to convert the optimization problem into a convex problem that can then be solved using numerical tools such as CVX \cite{cvx}. The capacity region for the corresponding scalar Gaussian channel can be immediately derived from the result for the vector channel setting and serves as an extension of the result in \cite{Wigger_gaussian_cop} to the state-dependent case. The capacity region of the vector Gaussian FSM-MAC with a common message and the same CSI properties can also be easily derived from the result for the conferencing channel by exploiting the strong correspondence between the two models and using a simple analogy.

\par To gain some insight into the practical implications of the results we conclude this paper with a specific example, namely, a scalar AWGN channel with two possible states (`Good' and `Bad'). Numerical results are included to demonstrate the impact of different channel parameters on the capacity region and the optimal input distribution. Our interpretation of interactions between the different parameters produces valuable insights.

\par The remainder of the paper is organized as follows. In Section \ref{Channel Model} we describe the two communication models of interest -- the FSM-MAC with a common message and delayed CSI and the FSM-MAC with partially cooperative encoders and delayed CSI. In Sections \ref{Common_Result} and \ref{Conf_Result}, we state the capacity results for the common message and conferencing models, respectively. Each result is followed by its proof. Section \ref{Gauss_Diagonal_Model} follows with the definition of the vector Gaussian FSM-MAC with diagonal channel transfer matrices and the derivation of the maximization problem defining its capacity region. The regions for the corresponding common message model and the scalar setting are given as special cases. The two-state Gaussian example is discussed in this section as well. Finally, Section \ref{summary} summarizes the main achievements and insights presented in this paper along with some possible future research directions and extensions.


\section{\textsc{Channel Models and Notation}}\label{Channel Model}

In this paper, we investigate the capacity region of the FSM-MAC with partially cooperative encoders, full CSI at the decoder (receiver) and delayed CSI at the encoders (transmitters), as illustrated in Fig. \ref{MAC Conference Delay_int}. To this end, we first consider a different setting, which is the FSM-MAC with a common message and the same CSI properties, as depicted in Fig. \ref{MAC Common Delay}. The derivation of the capacity region for the latter common message setting forms the basis for the achievability proof for the former setting where a conferencing link exists between the encoders. Since most definitions for both channels follow similar lines, we start by defining the common message setting and then extend the description for the setting of partially cooperative encoders.

\newcounter{newcounter}
\begin{figure*}[!b]
\normalsize
\setcounter{newcounter}{\value{equation}}
\setcounter{equation}{7}

\vspace*{4pt}
\hrulefill
\vspace{-0.3cm}

\begin{equation}
P_{e}^{(n)}= \frac{1}{2^{n(R_0+R_1+R_2)}}\sum_{(m_0,m_1,m_2)}\sum_{s^{n}}P_{S^{n}}(s^{n})\mathbb{P}\Big[ \psi(Y^{n},s^{n})\neq(m_0,m_1,m_2)\ \Big|\ (m_0,m_1,m_2) \mbox{ was sent}\ ,\ S^n=s^n\Big]\label{Common_Pe}
\end{equation}

\vspace*{4pt}
\vspace{-0.3cm}
\setcounter{equation}{0}
\end{figure*}
\par We use the following notations. Matrices are denoted by nonitalicized capital letters, e.g., $\mathrm{X}$. Calligraphic letters denote sets, e.g., $\mathcal{X}$, while the cardinality of a set $\mathcal{X}$ is denoted by $|\mathcal{X}|$. $\mathcal{X}^n$ stands for the $n$-fold Cartesian product of $\mathcal{X}$. An element of $\mathcal{X}^n$ is denoted by $x^n=(x_1,x_2,\ldots,x_n)$, and its substrings as $x_i^j=(x_i,x_{i+1},\ldots,x_j)$; when $i=1$, the subscript is omitted. We use the notation $x^{n\backslash i}=(x_1,\ldots,x_{i-1},x_{i+1},\ldots,x_n)$. Whenever the dimension $n$ is clear from the context, vectors (or sequences) are denoted by boldface letters, e.g., $\mathbf{x}$. Random variables are denoted by uppercase letter, e.g., $X$, with similar conventions for random vectors. $X_i^j$ stands for the sequence of random variables $(X_i,X_{i+1},\ldots,X_j)$, while $\mathbf{X}$ stands for $X^n$. The probability of an event $\mathcal{A}$ is denoted by $\mathbb{P}[\mathcal{A}]$, while $\mathbb{P}[\mathcal{A}\big|\mathcal{B}\mspace{2mu}]$ denotes conditional probability of $\mathcal{A}$ given $\mathcal{B}$. Probability mass functions (PMFs) are denoted by the capital letter $P$ with a subscript that identifies the random variable and its possible conditioning. For example, for two jointly distributed random variables $X$ and $Y$, let $P_X$, $P_{X,Y}$, and $P_{X|Y}$ denote, respectively, the PMF of $X$, the joint PMF of $(X,Y)$, and the conditional PMF of $X$ given $Y$. In particular, when $X$ and $Y$ are discrete, $P_{X|Y}$ represents the stochastic matrix whose elements are given by $P_{X|Y}(x|y)=\mathbb{P}\big[X = x|Y = y\big]$. We omit the subscripts if the arguments of the distribution are lower case versions of the random variables.


\subsection{FSM-MAC with a Common Message and Delayed CSI}\label{CommonDef}

\par The FSM-MAC with a common message considered in this paper is illustrated in Fig. \ref{MAC Common Delay}. The MAC setting consists of two senders and one receiver. Each sender $j\in\{1,2\}$ chooses a pair of indices, $(m_0,m_j)$, uniformly from
the set $\left\{1, ..., 2^{nR_0}\right\}\times\left\{1, ..., 2^{nR_j}\right\}$, where $m_0$ denotes the common message and $m_j$, $j\in\{1,2\}$, denotes the private message of the corresponding sender. The choices of $m_0$, $m_1$ and $m_2$ are independent. The input to the channel from encoder $j\in\{1,2\}$ is denoted by $X_j^n=(X_{j,1},X_{j,2}, \ldots,X_{j,n})$, and the output of the channel is denoted
by $Y^n=(Y_1, Y_2,\ldots,Y_n)$.

\par At each instance of time, the FSM channel is assumed to be in one of a finite number of states $\mathcal{S}=\{s_1,s_2,...,s_k\}$. In each state, the channel
is a discrete memoryless channel (DMC),
with input alphabets $\mathcal{X}_1,\mathcal{X}_2$ and output alphabet $\mathcal{Y}$.
Let the random variable $S_i$ denote the channel state at time $i$.
Similarly, we denote by $X_{1,i},X_{2,i}$ and $Y_i$ the inputs and the output of the channel at time $i$.
The channel transition probability distribution at time $i$ depends on the
state $S_i$ and the inputs $X_{1,i},X_{2,i}$ at time $i$, and it is given by $P(y_i|x_{1,i},x_{2,i},s_i)$.
The channel output at any time $i$ is assumed to depend only on the channel inputs and
state at time $i$. Hence,
\begin{eqnarray}
P(y_i|x_{1}^{i},x_{2}^{i},s^{i})=P(y_i|x_{1,i},x_{2,i},s_i).
\end{eqnarray}
The state process, $\{S_i\}_{i=1}^n$, is assumed to be an irreducible, aperiodic,
finite-state, homogeneous and stationary Markov chain and is therefore ergodic.
The state process is independent of the channel inputs and output when
conditioned on the previous states, i.e.,
\begin{eqnarray}
P(s_i|s^{i-1},x_{1}^{i-1},x_{2}^{i-1},y^{i-1})=P(s_i|s_{i-1}).
\end{eqnarray}
Furthermore, we assume that the state process is independent of the messages $M_0$, $M_1$ and $M_2$, i.e.,
\begin{equation}
P(s^n,m_0,m_1,m_2)=\prod_{i=1}^{n}P(s_i|s_{i-1})P(m_0)P(m_1)P(m_2).
\end{equation}

We assume that full CSI is available at the decoder (i.e., the decoder knows $S_i$ at each time instance $i$). However, the encoders are only assumed to have access to delayed CSI, with delays $d_1$ and $d_2$ for Encoder 1 and Encoder 2, respectively. We let $S_{i-d_j}$, $j\in{1,2}$, denote the channel state at time $i-d_j$, and assume without loss of generality that $d_1\ge d_2$.
Now, let $\mathrm{K}$ be the one-step state-transition probability matrix of the Markov process that governs the channel states, and let $\pi$ be its steady state probability distribution. The joint distribution of $(S_{i},S_{i-d})$ is stationary and is given by
\begin{eqnarray}
\pi_d(S_i=s_{l},S_{i-d}=s_j)=\pi(s_j)\mathrm{K}^{d}(s_{l},s_{j}),
\end{eqnarray}
where $\mathrm{K}^{d}(s_{l},s_j)$ is the $(l,j)$-th element of the d-step transition
probability matrix $\mathrm{K}^{d}$ of the Markov state process.
To simplify the notation, we define the joint distribution of the random variables $(S,\tilde{S}_1,\tilde{S}_2)$
as the joint distribution of $(S_{i},S_{i-d_1},S_{i-d_2})$, i.e.,
\begin{equation}
P_{S,\tilde{S}_1,\tilde{S}_2}(s_l,s_j,s_v)=\pi(s_j)\mathrm{K}^{d_1-d_2}(s_{v},s_{j})\mathrm{K}^{d_2}(s_{l},s_{v}),\label{s1s2sdefinition}
\end{equation}
where $(s_j,s_l,s_v)\in \mathcal{S}^3.$

\begin{definition}[Code Description] A~$(\mspace{-2mu}n,\mspace{-2mu}2^{nR_0}\mspace{-2mu},\mspace{-2mu}2^{nR_1}\mspace{-2mu},\mspace{-2mu}2^{nR_2}\mspace{-2mu},\mspace{-2mu}d_1,d_2\mspace{-2mu})$ code for the FSM-MAC with CSI at the decoder and delayed CSI at the encoders with delays $d_1$ and $d_2$ consists of:
\begin{enumerate}
\item
Three sets of integers $\mathcal{M}_0=\{1,2,...,2^{nR_0}\}$, $\mathcal{M}_1=\{1,2,...,2^{nR_1}\}$ and $\mathcal{M}_2=\{1,2,...,2^{nR_2}\}$, referred to as the {\it{message sets}}.
\item
Two encoding functions $f_j$, $j\in \{1,2\}$. Each function $f_j$ is defined by means of a sequence of functions $f_{j,i}$, $i\in\{1,2,\ldots,n\}$, that depend only on the pair of messages $(M_0,M_j)$, and the channel states up to time $i-d_j$. The output of Encoder $j$ at time $i$, $X_{j,i}\in\mathcal{X}_j$, is given by
\begin{eqnarray}
X_{j,i}=\begin{cases}
f_{j,i}(M_0,M_j),& 1\leq i\leq d_j\\
f_{j,i}(M_0,M_j,S^{i-d_j}),& d_j+1\leq i\leq n
\end{cases}.
\end{eqnarray}
\item
A decoding function:
\begin{eqnarray}
\psi : {\cal Y}^{n}\times {\cal S}^{n} \rightarrow \mathcal{M}_{0}\times \mathcal{M}_{1}\times \mathcal{M}_{2}\ .
\end{eqnarray}
\end{enumerate}
\end{definition}

\par \emph{The average probability of error} for the  $(n,2^{nR_0},2^{nR_1},2^{nR_2},d_1,d_2)$ code is given in (\ref{Common_Pe}) at the bottom of the page. We use standard definitions of achievability and of the capacity region \cite{CovThom06}. Namely, a rate triplet $(R_0,R_1,R_2)$ is \emph{achievable} for the FSM-MAC if there exists a sequence of $(n,2^{nR_0},2^{nR_1},2^{nR_2},d_1,d_2)$ codes with $P_{e}^{(n)}\rightarrow0$ as $n\rightarrow \infty$. \emph{The capacity region} is the closure of the set of achievable rates $(R_0,R_1,R_2)$.

\begin{figure*}[!b]
\hrulefill
\normalsize
\setcounter{newcounter}{\value{equation}}
\setcounter{equation}{13}
\begin{equation}
P_{e}^{(n)}= \frac{1}{2^{n(R_1+R_2)}}\sum_{(m_1,m_2)}\sum_{s^{n}}P_{S^{n}}(s^{n})\mathbb{P}\Big[ \psi(Y^{n},s^{n})\neq(m_1,m_2)\ \Big|\ (m_1,m_2) \mbox{ was sent}\ ,\ S^n=s^n\Big]\label{Conf_Pe}
\end{equation}

\end{figure*}


\subsection{FSM-MAC with Partially Cooperative Encoders and Delayed CSI}\label{ConferenceDef}

The FSM-MAC with partially cooperative encoders and delayed CSI is depicted in Fig. \ref{MAC Conference Delay_int}. The channel definition relies on Subsection \ref{CommonDef}, while taking the common message set to be $\mathcal{M}_0=\emptyset$. Here, however, conferencing between the encoders is introduced under the assumption that conferencing links of fixed and finite capacities $C_{12}$ and $C_{21}$ exist between the encoders. Accordingly, the amount of information exchanged between the encoders during the conference is bounded by $C_{12}$ and $C_{21}$. The conference is assumed to take place prior to the transmission of a codeword through the channel and consists of $\ell$ consecutive pairs of communications, simultaneously transmitted by the encoders. Each communication depends on the message to be transmitted by the sending encoder and  previously \emph{received} communications from the other encoder. We denote the communications transmitted from encoder $j\in\{1,2\}$ to the other encoder by $V_j^{\ell}$. Note that here the state process is also assumed to be independent of the conference communications, i.e.,
\setcounter{equation}{8}

\begin{equation}
P(s^n,v_1^{\ell},v_2^{\ell})=P(s^n)P(v_1^{\ell},v_2^{\ell})=\prod_{i=1}^{n}P(s_i|s_{i-1})P(v_1^{\ell},v_2^{\ell}).
\end{equation}


\begin{definition}[Code Description]
A $(n,\ell,2^{nR_1},2^{nR_2},d_1,d_2)$ code for the FSM-MAC with CSI at the decoder, delayed CSI at the encoders with delays $d_1$ and $d_2$, and conferencing links with capacities $C_{12}$ and $C_{21}$ consists of:
\begin{enumerate}
\item
Two sets of integers $\mathcal{M}_1=\{1,2,...,2^{nR_1}\}$ and $\mathcal{M}_2=\{1,2,...,2^{nR_2}\}$, referred to as the {\it{message sets}}.
\item
Two encoders, where each encoder is completely described by an encoding function, $f_j$, and a set of $\ell$ ($\ell\geq1$) communication functions, $\{h_{j,1},h_{j,2},\ldots,h_{j,\ell}\}$, $j\in \{1,2\}$ (similar definitions were also used in \cite{Willems83_cooperating}).
\item
The encoding function, $f_j$, maps the message $M_j$, $j\in\{1,2\}$, and what was learned from the conference with the other encoder into channel codewords of length $n$. Each function $f_j$ is defined by means of a sequence of functions $f_{j,i}$ that depend only on the message $M_j$, the received communications from the other encoder in the conferencing stage, and the channel states up to time $i-d_j$. We emphasize that since encoding occurs only after the conferencing stage has finished, each $f_{j,i}$ depends on all received communications.
\item
Each of the two communication functions $h_{1,i}$ and $h_{2,i}$, $i\in\{1,2,\ldots,\ell\}$, maps the message $M_1$ (or $M_2$, respectively) and the sequence of previously received communications from the other encoder $V_2^{i-1}$ (or $V_1^{i-1}$, respectively), onto the $i$-th communication $V_{1,i}$ (or $V_{2,i}$, respectively). More specifically, the communications are defined as:
\begin{eqnarray}
V_{1,i}=h_{1,i}(M_1,V_2^{i-1}) \mbox{ } ; \mbox{ } V_{2,i}=h_{2,i}(M_2,V_1^{i-1}).
\end{eqnarray}
\item
The encoding function for Encoder $1$ satisfies
\begin{eqnarray}
X_{1,i}=\begin{cases}
f_{1,i}(M_1,V_2^{\ell}),& 1\leq i\leq d_1\\
f_{1,i}(M_1,V_2^{\ell},S^{i-d_1}),& d_1+1\leq i\leq n\\
\end{cases},
\end{eqnarray}
and the encoding function for Encoder $2$ is defined analogously (using the private message $M_2$, the communications $V_1^{\ell}$ and the delay $d_2$).\\
\item
The random variable $V_{j,i}$, for $j\in\{1,2\}$ and $i\in\{1,2,\ldots,\ell\}$ ranges over the finite alphabet $\mathcal{V}_{j,i}$. A conference is $(C_{12},C_{21})$-permissible if the sets of communication functions are such that \cite{Willems83_cooperating}:
\begin{eqnarray}
\sum_{i=1}^{\ell}{\log|\mathcal{V}_{1,i}|} \leq nC_{12} \ ; \ \sum_{i=1}^{\ell}{\log|\mathcal{V}_{2,i}|} \leq nC_{21}. \label{finite_cap_def}
\end{eqnarray}

\item
A decoding function:
\begin{eqnarray}
\psi : {\cal Y}^{n}\times {\cal S}^{n} \rightarrow \mathcal{M}_{1}\times \mathcal{M}_{2}.
\end{eqnarray}
\end{enumerate}
\end{definition}

\par \emph{The average probability of error} for the $(n,\ell,2^{nR_1},2^{nR_2},d_1,d_2)$ code is given by  (\ref{Conf_Pe}) at the bottom of the page. The \emph{achievable rates} and the \emph{capacity region} for this channel are defined analogously to their definitions in Section \ref{CommonDef}.


\section{\textsc{The Capacity Region of the FSM-MAC with a Common Message and Delayed Transmitter CSI}}\label{Common_Result}
In this section we state the capacity region of the FSM-MAC with a common message and delayed transmitter CSI, after which we present its proof.

\setcounter{equation}{14}
\begin{theorem}\label{MAC Common Capacity}
The capacity region of the FSM-MAC with a common message, CSI at the decoder and asymmetrically delayed  CSI at the encoders with delays $d_1$ and $d_2$, such that $d_1\ge d_2$, is the union of all sets of rate triplets $(R_0,R_1,R_2)\in\mathbb{R}^3_+$ satisfying:
\begin{subequations}
\begin{align}
R_1 &\leq I(X_1;Y|X_2,U,S,\tilde{S}_1,\tilde{S}_2)\\
R_2 &\leq I(X_2;Y|X_1,U,S,\tilde{S}_1,\tilde{S}_2)\\
R_1+R_2 &\leq I(X_1,X_2;Y|U,S,\tilde{S}_1,\tilde{S}_2)\\
R_0+R_1+R_2 &\leq I(X_1,X_2;Y|S,\tilde{S}_1,\tilde{S}_2)\mspace{4mu},
\end{align}\label{T1region}
\end{subequations}

\vspace{-4mm}
\noindent where the union is over all joint distributions $P_{U|\tilde{S}_1}P_{X_1|\tilde{S}_1,U}P_{X_{2}|\tilde{S}_1,\tilde{S}_2,U}$.
The joint distribution of $(S,\tilde{S}_1,\tilde{S}_2)$ is specified in (\ref{s1s2sdefinition}) and $|\mathcal{U}|\leq|\mathcal{X}_1|\cdot|\mathcal{X}_2|\cdot|\mathcal{S}|^3+2$. Furthermore, the capacity region is convex.
\end{theorem}



\begin{IEEEproof}

\subsection{Converse} \label{MAC Common Converse}
We need to show that for every achievable rate triplet~$(R_{0},R_{1},R_{2})$, there exists a joint distribution
$P_{S,\tilde{S}_1,\tilde{S}_2}P_{U|\tilde{S}_1}P_{X_1|\tilde{S}_1,U}P_{X_2|\tilde{S}_1,\tilde{S}_2,U}P_{Y|X_1,X_2,S}$ such~that the inequalities in (\ref{T1region}) are satisfied.
Since $(R_0,R_1,R_2)$ is an achievable rate triplet, there exists
a $(n,2^{nR_0},2^{nR_1},2^{nR_2},d_1,d_2)$ code  with a probability of error  $P_{e}^{(n)}$ that becomes arbitrarily small with the increase of the block length (see (\ref{Common_Pe})). By Fano's inequality,
\begin{align*}
H(M_0,M_1,M_2|Y^n,S^n)&\leq n(R_0+R_1+R_2)P_{e}^{(n)}+H(P_{e}^{(n)})\\&\triangleq n\epsilon_n,\numberthis
\end{align*}
where clearly $\epsilon_n\rightarrow0$ as $P_{e}^{(n)}\rightarrow 0$. It therefore follows that
\begin{align}
H(M_1|Y^n,S^n)&\leq H(M_0,M_1,M_2|Y^n,S^n)\leq n\epsilon_n \label{fano1}\\
H(M_2|Y^n,S^n)&\leq H(M_0,M_1,M_2|Y^n,S^n)\leq n\epsilon_n \\
H(M_1,M_2|Y^n,S^n)&\leq H(M_0,M_1,M_2|Y^n,S^n)\leq n\epsilon_n.
\end{align}

For the sake of brevity we focus here on the upper bound on $R_1$, while noting that all other upper bounds in (\ref{T1region}) can be analogously derived using the same auxiliary random variable definition. It now follows that
\begin{align*}
  nR&_{1}= H(M_1)\\
        &\stackrel{(a)}\leq I(M_1;Y^n,S^{n})+n \epsilon_n\\
        &\stackrel{(b)}= I(M_1;Y^n|S^{n})+n \epsilon_n\\
        &\stackrel{(c)}\leq H(M_1|S^n,M_0,M_2)\mspace{-2mu}-\mspace{-2mu}H(M_1|Y^n,S^n,M_0,M_2)\mspace{-2mu}+\mspace{-2mu}n \epsilon_n\\
        &\stackrel{(d)}= \sum_{i=1}^n{I(M_1;Y_i|S^n,M_0,M_2,Y^{i-1})}+n \epsilon_n\\
        &\stackrel{(e)}= \sum_{i=1}^n\Big[H(Y_i|S^n,X_2^n,M_0,M_2,Y^{i-1})\\&\mspace{65mu}-H(Y_i|S^n,X_1^n,X_2^n,M_0,M_1,M_2,Y^{i-1})\Big]+n \epsilon_n\\
        &\stackrel{(f)}\leq \sum_{i=1}^n\Big[H(Y_i|X_{2,i},S_i,S_{i-d_1},S_{i-d_2},M_0,S^{i-d_1-1})\\&-H(Y_i|X_{1,i},X_{2,i},S_i,S_{i-d_1},S_{i-d_2},M_0,S^{i-d_1-1})\Big]+n \epsilon_n\\
        &\stackrel{(g)}= \sum_{i=1}^n{I(X_{1,i};Y_i|X_{2,i},S_i,S_{i-d_1},S_{i-d_2},U_i)}+n \epsilon_n\numberthis
  \end{align*}
where:\\
(a) follows from (\ref{fano1});\\
(b) follows because $M_1$ and $S^{n}$ are independent;\\
(c) follows because $M_1$ and $(M_0,M_2)$ are independent given $S^{n}$ (first term) and since conditioning reduces entropy (second term);\\
(d) follows by the mutual information chain rule;\\
(e) follows because $X_{1}^{n}$ is a deterministic function of $(M_0,M_1,S^n)$ and $X_{2}^{n}$ is a deterministic function of $(M_0,M_2,S^n)$;\\
(f) follows since conditioning reduces entropy (first term), and because when conditioned on $(X_{1,i},X_{2,i})$ and $S_i$, the channel output at time $i$ is independent of $(M_0,M_1,M_2,S^{i-1},S_{i+1}^n,X_1^{i-1},X_{1,i+1}^n,X_2^{i-1},X_{2,i+1}^n,Y^{i-1})$ (second term);\\
(g) follows by defining $U_i\triangleq(M_0,S^{i-d_1-1})$\label{U_def_common}.

\par Note that the definition of the auxiliary random variable $U_i$ represents the common message and the common knowledge of the state sequence at time $i$ (except for $S_{i-d_1}$), which, in fact, encompasses all common information shared by the two encoders at this instant of time. We can therefore conclude that the rate $R_1$ must satisfy the following upper bound:
\begin{equation}
  R_{1}\leq \frac{1}{n}\sum_{i=1}^{n}{I(X_{1,i};Y_{i}|X_{2,i},S_{i},S_{i-d_1},S_{i-d_2},U_i)}+ \epsilon_n.  \label{r1com}
\end{equation}

\par In a completely analogous manner it can be shown that
\begin{gather}
R_{2}\leq \frac{1}{n}\sum_{i=1}^{n}{I(X_{2,i};Y_{i}|X_{1,i},S_{i},S_{i-d_1},S_{i-d_2},U_i)}+ \epsilon_n  \label{r2com}\\
R_{1}\mspace{-4mu}+\mspace{-3mu}R_{2}\leq\mspace{-3mu} \frac{1}{n}\mspace{-3mu}\sum_{i=1}^{n}I(X_{1,i},X_{2,i};Y_{i}|S_{i},S_{i-d_1},S_{i-d_2},U_i)\mspace{-3mu}+\mspace{-3mu} \epsilon_n \label{r1+r2com}\\
R_{0}+R_{1}+R_{2}\leq \frac{1}{n}\sum_{i=1}^{n}I(X_{1,i},X_{2,i};Y_{i}|S_{i},S_{i-d_1},S_{i-d_2})+ \epsilon_n. \label{r0+r1+r2}
\end{gather}

The upper bounds in (\ref{r1com})-(\ref{r0+r1+r2}) can also be rewritten by introducing a new time sharing random variable $Q$ that is uniformly distributed over the set $\{1,2,...,n\}$. For example, the upper bound in (\ref{r1com}) can be rewritten as
\begin{align}
  R_{1}\mspace{-3mu}&\leq   \mspace{-3mu}\frac{1}{n}\mspace{-2mu}\sum_{i=1}^{n}\mspace{-2mu}I(Y_{Q};\mspace{-3mu}X_{1,Q}|X_{2,Q},\mspace{-3mu}S_{Q},S_{Q-d_1}\mspace{-3mu},\mspace{-2mu}S_{Q-d_2},\mspace{-2mu}U_Q,Q\mspace{-2mu}=\mspace{-2mu}i\mspace{-2mu})\mspace{-3mu}+ \mspace{-3mu} \epsilon_n \nonumber\\
        &= I(Y_{Q};X_{1,Q}|X_{2,Q},S_{Q},S_{Q-d_1},S_{Q-d_2},U_Q,Q)+ \epsilon_n \label{Uconverse_com}.
\end{align}
By rewriting the rate bounds (\ref{r2com})-(\ref{r0+r1+r2}) in the same manner as (\ref{r1com}) is rewritten into (\ref{Uconverse_com}), it is clear that the obtained region is convex. This follows directly by the presence of the time sharing random variable $Q$ in the conditioning of all the mutual information terms.
\par Next, by denoting $ X_{1}\triangleq X_{1,Q},\ X_{2} \triangleq X_{2,Q},\ Y \triangleq Y_{Q},\ S\triangleq S_{Q},\ \tilde{S}_1\triangleq S_{Q-d_1},\ \tilde{S}_2\triangleq S_{Q-d_2}$ and  $U\triangleq (U_Q,Q)$, we get:
\begin{subequations}
\begin{align}
R_{1}&\leq I(X_{1};Y|X_{2},U,S,\tilde{S}_1,\tilde{S}_2)+ \epsilon_n \\
R_{2} &\leq I(X_{2};Y|X_{1},U,S,\tilde{S}_1,\tilde{S}_2) + \epsilon_n\\
R_{1}+R_{2}&\leq I(X_{1},X_{2};Y|U,S,\tilde{S}_1,\tilde{S}_2) + \epsilon_n\\
R_{0}+R_{1}+R_{2}&\leq I(X_{1},X_{2};Y|S,\tilde{S}_1,\tilde{S}_2) + \epsilon_n,\label{r0r1r2_noQ}
\end{align}
\end{subequations}
\noindent where (\ref{r0r1r2_noQ}) holds due to
\begin{align}
R_{0}&\mspace{-2mu}+\mspace{-2mu}R_{1}\mspace{-2mu}+\mspace{-2mu}R_{2}\leq I(X_{1,Q},X_{2,Q};Y_Q|S_{Q},S_{Q-d_1},S_{Q-d_2},Q)+ \epsilon_n\nonumber\\
                 &\stackrel{(a)}\leq I(X_{1,Q},X_{2,Q},U_Q;Y_Q|S_{Q},S_{Q-d_1},S_{Q-d_2},Q)+ \epsilon_n\nonumber\\
                 &\stackrel{(b)}\leq I(X_{1,Q},X_{2,Q},U_Q,Q;Y_Q|S_{Q},S_{Q-d_1},S_{Q-d_2})+ \epsilon_n\nonumber\\
                 &\stackrel{(c)}\leq I(X_1,X_2,U;Y|S,\tilde{S}_1,\tilde{S}_2)+ \epsilon_n\nonumber\\
                 &\stackrel{(d)}= I(X_1,X_2;Y|S,\tilde{S}_1,\tilde{S}_2)+ \epsilon_n\label{r0r1r2_q_justification}.
\end{align}
Here:\\
(a) and (b) follow from the fact that conditioning reduces entropy;\\
(c) follows from the definition of $(X_1,X_2,Y,U,S,\tilde{S}_1,\tilde{S}_2)$;\\
(d) follows from the Markov relation $Y-(X_1,X_2,S)-(U,\tilde{S}_1,\tilde{S}_2)$, which is induced from the channel model.\\Taking the limit as $n\rightarrow\infty$, one obtains the bounds as in (\ref{T1region}).
\ \\
\par To complete the proof of the converse, it is left to show that the following Markov relations hold:
\begin{subequations}
\begin{gather}
U-\tilde{S}_1-(S,\tilde{S}_2)\label{common_markov1}\\
X_1-(\tilde{S}_1,U)-(S,\tilde{S}_2)\label{common_markov2}\\
X_2-(\tilde{S}_1,\tilde{S}_2,U)-(X_1,S).\label{common_markov3}
\end{gather}\label{common_markov}
\end{subequations}

The proof of (\ref{common_markov}) is given in Appendix \ref{common_markov_proof}.


\begin{figure*}[t!]
\begin{center}
\begin{psfrags}
    \psfragscanon
    \psfrag{A}{$\vdots$}
    \psfrag{B}{$\cdots$}
    \psfrag{C}[][][0.75]{\ \ \ \ \ \ $\tilde{S}_1=1$}
    \psfrag{D}[][][0.75]{\ \ \ \ \ \ $\tilde{S}_1=2$}
    \psfrag{E}[][][0.75]{\ \ \ \ \ \ $\tilde{S}_1=3$}
    \psfrag{F}[][][0.75]{\ \ \ \ \ \ $\tilde{S}_1=k$}
    \psfrag{G}[][][0.75]{\ \ \ \ \ \ $t_{0,1}$}
    \psfrag{H}[][][0.75]{\ \ \ \ \ \ $t_{0,2}$}
    \psfrag{I}[][][0.75]{\ \ \ \ \ \ $t_{0,3}$}
    \psfrag{J}[][][0.75]{\ \ \ \ \ \ $t_{0,n}$}
    \psfrag{K}[][][0.75]{\ \ \ \ \ \ $(\tilde{S}_1=1, U=u_1)$}
    \psfrag{L}[][][0.75]{\ \ \ \ \ \ $(\tilde{S}_1=2, U=u_1)$}
    \psfrag{M}[][][0.75]{\ \ \ \ \ \ $(\tilde{S}_1=k, U=u_1)$}
    \psfrag{N}[][][0.75]{\ \ \ \ \ \ $(\tilde{S}_1=1, U=u_2)$}
    \psfrag{O}[][][0.75]{\ \ \ \ \ \ $(\tilde{S}_1=2, U=u_2)$}
    \psfrag{P}[][][0.75]{\ \ \ \ $(\tilde{S}_1=k, U=u_\mathcal{|U|})$}
    \psfrag{Q}[][][0.75]{\ \ \ \ \ \ $t_{1,1}$}
    \psfrag{R}[][][0.75]{\ \ \ \ \ \ $t_{1,2}$}
    \psfrag{S}[][][0.75]{\ \ \ \ \ \ $t_{1,3}$}
    \psfrag{T}[][][0.75]{\ \ \ \ \ \ $t_{1,n}$}
    \psfrag{X}[][][0.75]{$n$}
    \psfrag{Y}[][][0.75]{$|\mathcal{S}|$}
    \psfrag{W}[][][0.75]{\ \ \ \ \ $|\mathcal{U}|\cdot|\mathcal{S}|$}
    \psfrag{U}[][][0.75]{$\mathbf{t}_0(m_0)$}
    \psfrag{V}[][][0.75]{$\mathbf{t}_1(m_1)$}
\includegraphics[scale=0.92]{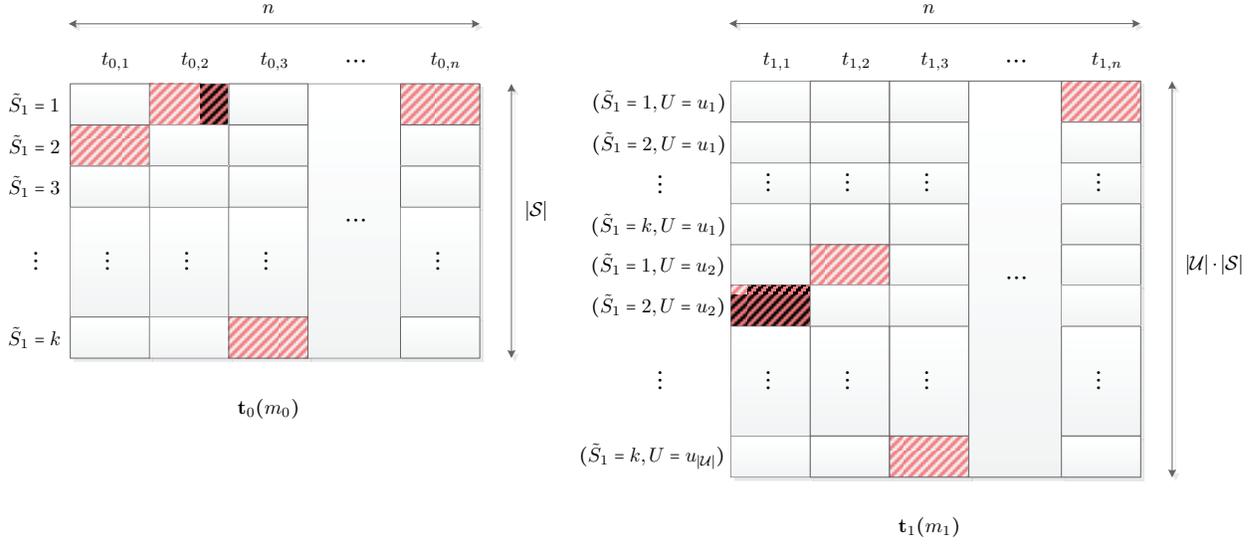}
\caption{The codewords $\mathbf{t}_0(m_0)$ and $\mathbf{t}_1(m_1)$ that are associated with some message pair $(m_0,m_1)\in\mathcal{M}_0\times\mathcal{M}_1$. The shaded regions correspond to the symbols chosen by Encoder 1 when the sequence of delayed CSI realizations is $\tilde{\mathbf{s}}_1=(2,1,k,\ldots,1)$, and when $t_{0,1}(m_0,2)=u_2$, $t_{0,2}(m_0,1)=u_2$, $t_{0,3}(m_0,k)=u_\mathcal{|U|}$ and $t_{0,n}(m_0,1)=u_1$.} \label{Code_Construction}
\psfragscanoff
\end{psfrags}
\end{center}
\end{figure*}

\subsection{Achievability}  \label{MAC Common Achievability}
To establish achievability, we need to show that for a fixed $\epsilon>0$, a fixed distribution
\begin{equation} P_{U|\tilde{S}_1}P_{X_1|U,\tilde{S}_1}P_{X_2|U,\tilde{S}_1,\tilde{S}_2},\label{input_joint_dist}
\end{equation}
and rates $(R_0,R_1,R_2)$ that satisfy the inequalities in (\ref{T1region}), there exists a sequence of $(n,2^{nR_{0}},2^{nR_{1}},2^{nR_{2}},d_1,d_2)$ codes such that $P_{e}^{(n)}\rightarrow0$ as $n\rightarrow\infty$.

\par Without loss of generality, we assume that the finite-state space is the set $\mathcal{S}=\left\{1,2,...,k\right\}$. By the underlying assumptions of the channel model, we take the delays $(d_1,d_2)$ to be fixed and finite integers. Moreover, throughout this proof we use the following notations. For an arbitrary finite set $\mathcal{A}=\{a_1,a_2,\ldots,a_{\mathcal{|A|}}\}$, we denote by  $(x_a)_{a\in\mathcal{A}}$ a column vector of size $\mathcal{|A|}$ with elements $\{x_{a_1},x_{a_2},\ldots,x_{a_\mathcal{|A|}}\}$. As stated in Section \ref{Channel Model}, sequences of length $n$ are denoted by bold lowercase letters, while random sequences are denoted by bold uppercase letters. Consider now the following encoding and decoding scheme.


\subsubsection{Codebook Generation}


Generate a common message codebook that comprises $2^{nR_0}$ codewords $\mathbf{t}_0(m_0)$, $m_0\in\mathcal{M}_0$, assembled from $n$ symbols from the super-alphabet $\mathcal{U}^\mathcal{|S|}$, which are drawn in an i.i.d. manner. Each codeword $\mathbf{t}_0(m_0)$ is distributed according to the product distribution
\begin{equation}
\mathbb{P}\big[\mathbf{T}_0=\mathbf{t}_0\big]=\prod_{i=1}^nP_{T_0}\big(t_{0,i}\big),
\end{equation}
where $t_{0,i}\in\mathcal{U}^{\mathcal{|S|}}$, for $i\in\{1,2,\ldots,n\}$ (each $t_{0,i}$ can thus be treated as a column vector of size $|\mathcal{S}|$ with elements in $\mathcal{U}$ ordered by the natural order of the set $\mathcal{S}$), and
\begin{equation}
P_{T_0}\big(t_0\big)=\mathbb{P}\big[T_0=t_0\big]=\prod_{\tilde{s}_1\in\mathcal{S}}P_{U|\tilde{S}_1=\tilde{s}_1}(u_{\tilde{s}_1}|\tilde{s}_1),
\end{equation}
where $t_0=(u_{\tilde{s}_1})_{\tilde{s}_1\in\mathcal{S}}$. Each codeword can hence be viewed as a matrix of dimension $|\mathcal{S}|\times n$ with elements in $\mathcal{U}$, where each row is associated with a different (delayed) state $\tilde{s}_1\in\mathcal{S}$. Accordingly, we denote by
$t_{0,i}(m_0,\tilde{s}_{1,i})$ the $\tilde{s}_{1,i}$th element of the $i$th symbol of the codeword $\mathbf{t}_0(m_0)$.

\begin{figure*}[!b]
\hrulefill
\normalsize
\setcounter{newcounter}{\value{equation}}
\setcounter{equation}{35}
\begin{equation}
\Big(\mathbf{u}(\hat{m}_0,\tilde{\mathbf{s}}_1),\mathbf{x}_1\big(\hat{m}_1,\mathbf{u}(\hat{m}_0,\tilde{\mathbf{s}}_1),\tilde{\mathbf{s}}_1\big),\mathbf{x}_2\big(\hat{m}_2,\mathbf{u}(\hat{m}_0,\tilde{\mathbf{s}}_1),\tilde{\mathbf{s}}_1,\tilde{\mathbf{s}}_2\big),\mathbf{s},\tilde{\mathbf{s}}_1,\tilde{\mathbf{s}}_2,\mathbf{y}\Big)\in\mathcal{T}_\epsilon^{(n)}(U,X_1,X_2,S,\tilde{S}_1,\tilde{S}_2,Y),\label{Typ_test}
\end{equation}

\end{figure*}
\setcounter{equation}{31}

\par Next, generate the codebook associated with the private message set $\mathcal{M}_1$ that comprises $2^{nR_1}$ codewords $\mathbf{t}_1(m_1)$, $m_1\in\mathcal{M}_1$, assembled from $n$ symbols from the super-alphabet $\mathcal{X}_1^{\mathcal{|U|}\cdot\mathcal{|S|}}$ drawn in an i.i.d. manner. Accordingly, the distribution of each codeword $\mathbf{t}_1(m_1)$ is given by
\begin{equation}
\mathbb{P}\big[\mathbf{T}_1=\mathbf{t}_1\big]=\prod_{i=1}^nP_{T_1}\big(t_{1,i}\big),
\end{equation}
where $t_{1,i}\in\mathcal{X}_1^{\mathcal{|U|}\cdot\mathcal{|S|}}$, for $i\in\{1,2,\ldots,n\}$ (here each $t_{1,i}$ can be treated as a column vector of size $\mathcal{|U|}\cdot\mathcal{|S|}$ with elements in $\mathcal{X}_1$), and
\begin{align*}
P_{T_1}(t_1)&=\mathbb{P}\big[T_1=t_1\big]\\&=\prod_{(u,\tilde{s}_1)\in\mathcal{U}\times\mathcal{S}}P_{X_1|U=u,\tilde{S}_1=\tilde{s}_1}(x_{1,(u,\tilde{s}_1)}|u,\tilde{s}_1),\numberthis
\end{align*}
where $t_1=(x_{1,(u,\tilde{s}_1)})_{(u,\tilde{s}_1)\in\mathcal{U}\times\mathcal{S}}$. Each codeword can therefore be viewed as a matrix of dimension $\big(\mathcal{|U|}\cdot\mathcal{|S|}\big)\times n$ with elements in $\mathcal{X}_1$, where each row is associated with a different pair $(u,\tilde{s}_1)\in\mathcal{U}\times\mathcal{S}$. The
element of the $i$th symbol of the codeword $\mathbf{t}_1(m_1)$ that is associated with the pair $\big(t_{0,i}(m_0,\tilde{s}_{1,i}),\tilde{s}_{1,i}\big)$ is denoted by $t_{1,i}\big(m_1,t_{0,i}(m_0,\tilde{s}_{1,i}),\tilde{s}_{1,i}\big)$.


\par Finally, generate the codebook associated with the private message set $\mathcal{M}_2$ in a manner analogous to codebook generation for $\mathcal{M}_1$, but here, the symbols of each codeword are elements in $\mathcal{X}_2^{|\mathcal{U}|\cdot|\mathcal{S}|^2}$. Namely, the distribution of each codeword $\mathbf{t}_2(m_2)$, $m_2\in\mathcal{M}_2$, is \begin{equation}
\mathbb{P}\big[\mathbf{T}_2=\mathbf{t}_2\big]=\prod_{i=1}^nP_{T_2}\big(t_{2,i}\big),
\end{equation}
where $t_{2,i}\in\mathcal{X}_2^{\mathcal{|U|}\cdot\mathcal{|S|}^2}$, for $i\in\{1,2,\ldots,n\}$, and
\begin{align*}
P_{T_2}&(t_2)=\mathbb{P}\big[T_2=t_2\big]\\
&=\mspace{-3mu}\prod_{(u,\tilde{s}_1,\tilde{s}_2)\in\mathcal{U}\times\mathcal{S}^2}\mspace{-20mu}P_{X_2|U=u,\tilde{S}_1=\tilde{s}_1,\tilde{S}_2=\tilde{s}_2}(x_{2,(u,\tilde{s}_1,\tilde{s}_2)}|u,\tilde{s}_1,\tilde{s}_2),\numberthis
\end{align*}
where $t_2=(x_{2,(u,\tilde{s}_1,\tilde{s}_2)})_{(u,\tilde{s}_1,\tilde{s}_2)\in\mathcal{U}\times\mathcal{S}^2}$. Again, each codeword can be viewed as a matrix of dimension $\big(\mathcal{|U|}\cdot\mathcal{|S|}^2\big)\times n$ with elements in $\mathcal{X}_2$, where each row is associated with a different triplet $(u,\tilde{s}_1,\tilde{s}_2)\in\mathcal{U}\times\mathcal{S}^2$. We denote by $t_{2,i}\big(m_2,t_{0,i}(m_0,\tilde{s}_{1,i}),\tilde{s}_{1,i},\tilde{s}_{2,i}\big)$ the element of the $i$th symbol of the codeword $\mathbf{t}_2(m_2)$ that is associated with the triplet $\big(t_{0,i}(m_0,\tilde{s}_{1,i}),\tilde{s}_{1,i},\tilde{s}_{2,i}\big)$. All codebooks are revealed to both encoders and to the decoder.


\subsubsection{Encoding}
\par To transmit the message triplet $(m_0,m_1,m_2)\in\mathcal{M}_0\times\mathcal{M}_1\times\mathcal{M}_2$, the encoders adhere to the following scheme. First, note that the delayed CSI becomes available at \emph{both} encoders only after the first $d_1$ channel uses. Therefore, the initial $d_1$ symbols transmitted by Encoder $j$, $j\in\{1,2\}$, are arbitrarily chosen from $\mathcal{X}_j$ (note that the choice of these symbols has no effect on the achievable rates since $d_1$ is fixed and finite, see the discussion in the sequel). The manner in which the encoders choose the symbols transmitted at times $i\in\{d_1+1,\ldots,n\}$ is described next.


\par\emph{Encoder 1:} At each time $i\in\{d_1+1,\ldots,n\}$, Encoder 1 has access to $\tilde{s}_{1,i}$. This delayed CSI is first used to choose an element from the codeword associated with the common message $m_0$. Namely, Encoder 1 starts by choosing $u_i(m_0,\tilde{s}_{1,i}) \triangleq t_{0,i}(m_0,\tilde{s}_{1,i})$.
Having $u_i(m_0,\tilde{s}_{1,i})$, Encoder 1 refers to the codeword $\mathbf{t}_1(m_1)$ and chooses $x_{1,i}\big(m_1,u_i(m_0,\tilde{s}_{1,i}),\tilde{s}_{1,i}\big)\triangleq t_{1,i}\big(m_1,u_i(m_0,\tilde{s}_{1,i}),\tilde{s}_{1,i}\big)$.
Encoder 1 then sends $x_{1,i}\big(m_1,u_i(m_0,\tilde{s}_{1,i}),\tilde{s}_{1,i}\big)$ to the channel.


\par\emph{Encoder 2:} Encoder 2 selects its channel input in manner analogous to that used by Encoder 1. First, recall that at each time $i\in\{d_1+1,\ldots,n\}$, Encoder 2 has access to both $(\tilde{s}_{1,i},\tilde{s}_{2,i})$. Thus, using $\tilde{s}_{1,i}$, the encoder first chooses $u_i(m_0,\tilde{s}_{1,i})=t_{0,i}(m_0,\tilde{s}_{1,i})$. The $i$th channel input from Encoder 2 is then chosen to be $x_{2,i}\big(m_2,u_i(m_0,\tilde{s}_{1,i}),\tilde{s}_{1,i},\tilde{s}_{2,i}\big)\triangleq t_{2,i}\big(m_2,u_i(m_0,\tilde{s}_{1,i}),\tilde{s}_{1,i},\tilde{s}_{2,i}\big)$.

\par An illustration of the codewords' structure for some pair $(m_0,m_1)\in\mathcal{M}_0\times\mathcal{M}_1$ and the corresponding transmitted symbols is shown in Fig. \ref{Code_Construction}. The structure of the codewords associated with the message set $\mathcal{M}_2$ is analogous, and is therefore omitted from the figure. We also note that the above construction of the codebooks and transmitted sequences can be regarded as a manifestation of the functional representation lemma \cite[Appendix B]{ElGammalKim10LectureNotes}.


\subsubsection{Decoding Process}

\par Upon receiving the whole channel output sequence $\mathbf{y}$ and the sequence of channel states $\mathbf{s}$ (assumed fully known at the receiver), a joint typicality decision rule is employed to decode the transmitted sequences. Note that since the delayed CSI available to each of the encoders is a deterministic function of the state sequence, the decoder can immediately reconstruct the sequences $\tilde{\mathbf{s}}_1$ and $\tilde{\mathbf{s}}_2$ from the latter.

\par The decoder searches for a triplet $(\hat{m}_0,\hat{m}_1,\hat{m}_2)\in\mathcal{M}_0\times\mathcal{M}_1\times\mathcal{M}_2$ such that (\ref{Typ_test}) at the bottom of the page is satisfied, where $\mathcal{T}_\epsilon^{(n)}(U,X_1,X_2,S,\tilde{S}_1,\tilde{S}_2,Y)$ denotes the jointly $\epsilon$-strongly typical set as defined in \cite[Chapter 2]{ElGammalKim10LectureNotes}, and
\begin{align*}
\mathbf{u}(\hat{m}_0,\tilde{\mathbf{s}}_1)\mspace{-3mu}\triangleq\mspace{-3mu}\big\{u_i(\hat{m}&_0,\tilde{s}_{1,i})\big\}_{i=1}^n\mspace{-3mu}=\mspace{-3mu} \big\{t_{0,i}(\hat{m}_0,\tilde{s}_{1,i})\big\}_{i=1}^n\\ \mathbf{x}_1\big(\hat{m}_1,\mathbf{u}(\hat{m}_0,\tilde{\mathbf{s}}_1),\tilde{\mathbf{s}}_1\big)\mspace{-3mu}&\triangleq\mspace{-3mu}\Big\{x_{1,i}\big(\hat{m}_1,u_i(\hat{m}_0,\tilde{s}_{1,i}),\tilde{s}_{1,i}\big)\Big\}_{i=1}^n\\&\mspace{-3mu}=\mspace{-3mu}\Big\{t_{1,i}\big(\hat{m}_1,u_i(\hat{m}_0,\tilde{s}_{1,i}),\tilde{s}_{1,i}\big)\Big\}_{i=1}^n \end{align*}
\begin{equation*}
\begin{aligned}
\mathbf{x}_2\big(\mspace{-2mu}\hat{m}_2,\mspace{-2mu}\mathbf{u}(\hat{m}_0,\tilde{\mathbf{s}}_1),\tilde{\mathbf{s}}_1,\tilde{\mathbf{s}}_2\mspace{-2mu}\big)\mspace{-4mu}\triangleq\mspace{-4mu}\Big\{\mspace{-2mu}x_{2,i}\mspace{-2mu}\big(\mspace{-2mu}\hat{m}_2,\mspace{-2mu}u_i(\mspace{-2mu}\hat{m}_0,\mspace{-2mu}\tilde{s}_{1,i}\mspace{-2mu}),\mspace{-2mu}\tilde{s}_{1,i},\mspace{-2mu}\tilde{s}_{2,i}\big)\mspace{-3mu}\Big\}_{i=1}^n\\\mspace{-3mu}\hspace{-5mm}=\mspace{-3mu}\Big\{t_{2,i}\big(\hat{m}_2,u_i(\hat{m}_0,\tilde{s}_{1,i}),\tilde{s}_{1,i},\tilde{s}_{2,i}\mspace{-2mu}\big)\Big\}_{i=1}^n. \end{aligned}
\end{equation*}

\noindent If such a unique triplet is found, it is declared as the decoded message triplet; otherwise, the decoder outputs a random message triplet.
\par We note here that although Encoder 1 (respectively, Encoder 2) arbitrarily chooses the first $d_1$ symbols of each codeword, these choices have a vanishing effect on the typicality test (\ref{Typ_test}). This is since the underlying assumption of the channel model is that both $d_1$ and $d_2$ are fixed, while the total block length $n$ can grow without bound. Therefore, to simplify the analysis that follows, we henceforth ignore the fact that the first symbols of each transmitted sequence do not follow the prescribed input distribution (\ref{input_joint_dist}).

\par By error probability analysis (see Appendix \ref{analysis}), we get that for the probability of error to vanish as $n\to\infty$, the rate constraints in (\ref{T1region}) must be satisfied. We have thus shown that if $(R_0,R_1,R_2)$ is inside the rate region specified in Theorem \ref{MAC Common Capacity}, then there exists a sequence of $(n,2^{nR_{0}},2^{nR_{1}},2^{nR_{2}},d_1,d_2)$ codes such that $P_{e}^{(n)}\rightarrow0$ as $n\rightarrow\infty$. This completes the proof of the achievability part.
\end{IEEEproof}

\begin{remark}
The cardinality bound on the auxiliary random variable $U$ is straightforwardly established using the convex cover method (see \cite[Appendix C]{ElGammalKim10LectureNotes} for details) and is therefore omitted.
\end{remark}

\begin{remark}\label{infinite_delay_common_remark}
The capacity region of the corresponding channel in which no CSI is available to Encoder 1 can be obtained from the capacity region in Theorem \ref{MAC Common Capacity} by omitting $\tilde{S}_1$ from all mutual information expressions in (\ref{T1region}) and from the joint distribution over which the union is taken.
Similarity, when considering the case in which the CSI at both encoders is absent, the capacity region can also be obtained from the result in Theorem \ref{MAC Common Capacity} by omitting $\tilde{S}_1$ and $\tilde{S}_2$ from the corresponding expressions.
Now, based on the underlying assumptions on the properties of the Markov state process, when $d_1$ is increased without bound, the delayed channel state $S_{i-d_1}$ becomes independent of the pair $(S_i,S_{i-d_2})$, which, in turn, implies that $\tilde{S}_1$ and $(S,\tilde{S}_2)$ are independent (see (\ref{s1s2sdefinition})). Thus, it can be shown that when $d_1$ is increased without bound, the capacity region in Theorem \ref{MAC Common Capacity} approaches the corresponding region of the case in which no CSI is available to Encoder 1. Using similar arguments, one can show that when $d_2$ is also increased without bound, the capacity region reduces to the corresponding capacity region with no encoder CSI.
\end{remark}

\begin{remark}
Based on practical considerations, a rate-splitting and multiplexing coding scheme as used, e.g., in \cite{Basher_Permuter}, can be considered for the current setting. We note, however, that the single codebooks approach employed here exhibits a simpler construction, and also lends itself more easily to error probability analysis.
\end{remark}


\section{\textsc{The Capacity Region of the FSM-MAC with Partially Cooperative Encoders and Delayed Transmitter CSI}}\label{Conf_Result}

In this section we state the capacity region of the FSM-MAC with partially cooperative encoders and delayed transmitter CSI followed by its proof.

\setcounter{equation}{36}
\begin{theorem}\label{MAC Conference Capacity}
The capacity region of FSM-MAC with partially cooperative encoders, cooperation link capacities $C_{12}$ and $C_{21}$, CSI at the decoder and asymmetrically delayed CSI at the encoders with delays $d_1$ and $d_2$, such that $d_1\ge d_2$, is the union of all sets of rate pairs $(R_1,R_2)\in\mathbb{R}^2_+$ satisfying:
\begin{subequations}
\begin{align}
R_1 &\leq I(X_1;Y|X_2,U,S,\tilde{S}_1,\tilde{S}_2)+C_{12}\\
R_2 &\leq I(X_2;Y|X_1,U,S,\tilde{S}_1,\tilde{S}_2)+C_{21}\\
R_1+R_2 &\leq I(X_1,X_2;Y|U,S,\tilde{S}_1,\tilde{S}_2)+C_{12}+C_{21}\\
R_1+R_2 &\leq I(X_1,X_2;Y|S,\tilde{S}_1,\tilde{S}_2),
\end{align}\label{T2region}
\end{subequations}

\vspace{-4mm}

\noindent where the union is over all distribution $P_{U|\tilde{S}_1}P_{X_1|\tilde{S}_1,U}P_{X_{2}|\tilde{S}_1,\tilde{S}_2,U}$.
The joint distribution of $(S,\tilde{S}_1,\tilde{S}_2)$ is specified in (\ref{s1s2sdefinition}) and $|\mathcal{U}|\leq|\mathcal{X}_1|\cdot|\mathcal{X}_2|\cdot|\mathcal{S}|^3+2$. Furthermore, the capacity region is convex.
\end{theorem}

\begin{IEEEproof}


\subsection{Converse} \label{MAC Conference Converse}
Given an  achievable rate $(R_{1},R_{2})$, we need to show that there exists a joint distribution of the form
$P_{S,\tilde{S}_1,\tilde{S}_2}P_{U|\tilde{S}_1}P_{X_{1}|\tilde{S}_1,U}P_{X_{2}|\tilde{S}_1,\tilde{S}_2,U}P_{Y|X_1,X_2,S}$ such that the inequalities in (\ref{T2region}) are satisfied. Since $(R_1,R_2)$ is an achievable rate-pair, there exists an $(n,l,2^{nR_1},2^{nR_2},d_1,d_2)$ code  with an arbitrarily small error probability $P_{e}^{(n)}$. By Fano's inequality (and with some abuse of notation),
\begin{equation}
H(M_1,M_2|Y^n,S^n)\leq n(R_1+R_2)P_{e}^{(n)}+H(P_{e}^{(n)})\triangleq n\epsilon_n,
\end{equation}
where $\epsilon_n\rightarrow0$ as $P_{e}^{(n)}\rightarrow 0$. It hence follows that
\begin{align}
H(M_1|Y^n,S^n)\leq H(M_1,M_2|Y^n,S^n)\leq n\epsilon_n\label{fano_conf1} \\
H(M_2|Y^n,S^n)\leq H(M_1,M_2|Y^n,S^n)\leq n\epsilon_n.
\end{align}
As in the proof of Theorem \ref{MAC Common Capacity}, we focus on the upper bound on $R_1$ and note that the upper bounds on all other rates can be straightforwardly obtained in an analogous manner. For $R_1$ we have the following:
  \begin{align*}
  nR_{1}&=\mspace{-5mu} H(M_1)\stackrel{(a)}\leq\mspace{-5mu} I(M_1;Y^n,S^{n})+n \epsilon_n\\
        &\hspace{-7.5mm}\stackrel{(b)}=\mspace{-5mu}I(M_1;Y^n|S^{n})+n \epsilon_n\\
        &\hspace{-7.5mm}\stackrel{(c)}\leq\mspace{-5mu} H(M_1|S^n,M_2)\mspace{-2mu}-\mspace{-2mu}H(M_1|V_1^\ell,V_2^\ell,Y^n,S^n,M_2)+n\epsilon_n\\
        &\hspace{-7.5mm}\stackrel{(d)}=\mspace{-5mu} I(M_1;V_1^\ell,V_2^\ell|S^n,M_2)\mspace{-2mu}+\mspace{-2mu}I(M_1;Y^n|V_1^\ell,V_2^\ell,S^n,M_2)\mspace{-2mu}+\mspace{-2mu}n \epsilon_n\\
        &\hspace{-7.5mm}\stackrel{(e)}=\mspace{-5mu} H(V_1^\ell|S^n,M_2)\mspace{-1mu}+\mspace{-4mu}\sum_{i=1}^n{I(M_1;Y_i|V_1^\ell,V_2^\ell,S^n,M_2,Y^{i-1})}\mspace{-2mu}+\mspace{-2mu}n \epsilon_n\\
        &\hspace{-7.5mm}\stackrel{(f)}\leq H(V_1^\ell) +\sum_{i=1}^n\Big[H(Y_i|V_1^\ell,V_2^\ell,S^n,X_2^n,M_2,Y^{i-1})\\
        &\mspace{35mu}-H(Y_i|V_1^\ell,V_2^\ell,S^n,X_1^n,X_2^n,M_1,M_2,Y^{i-1})\Big]\mspace{-2mu}+\mspace{-2mu}n \epsilon_n\\
        &\hspace{-7.5mm}\stackrel{(g)}\leq \sum_{j=1}^\ell{H(V_{1,j})}\\
        &\hspace{-5.1mm}+\sum_{i=1}^n\Big[H(Y_i|X_{2,i},S_i,S_{i-d_1},S_{i-d_2},V_1^\ell,V_2^\ell,S^{i-d_1-1})\\
        &\hspace{-5.1mm}-H(Y_i|X_{1,i},X_{2,i},S_i,S_{i-d_1},S_{i-d_2},V_1^\ell,V_2^\ell,S^{i-d_1-1})\Big]\mspace{-2mu}+\mspace{-2mu}n \epsilon_n \\
        &\hspace{-6.5mm}\leq \sum_{j=1}^\ell{\log|V_{1,j}|}\\
        &\hspace{-5.1mm}+\mspace{-2mu}\sum_{i=1}^n{\mspace{-2mu}I(\mspace{-2mu}X_{1,i};Y_i|X_{2,i},S_i,S_{i-d_1},S_{i-d_2}\mspace{-2mu},V_1^\ell,\mspace{-2mu}V_2^\ell,S^{i-d_1-1})}\mspace{-2mu}+\mspace{-2mu}n \epsilon_n \\
        &\hspace{-7.5mm}\stackrel{(h)}\leq nC_{12}\mspace{-1.5mu}+\mspace{-1.5mu}\sum_{i=1}^n{I(X_{1,i};Y_i|X_{2,i},S_i,S_{i-d_1},S_{i-d_2},U_i)}\mspace{-1.5mu}+\mspace{-1.5mu}n \epsilon_n\numberthis
  \end{align*}
where:\\
  (a) follows from (\ref{fano_conf1});\\
  (b) follows because $M_1$ and $S^{n}$ are independent;\\
  (c) follows because $M_1$ and $M_2$ are independent given $S^{n}$ (first term) and since conditioning reduces entropy (second term);\\
  (d) follows by adding and subtracting the term $H(M_1|V_1^\ell, V_2^\ell, S^n, M_2)$;\\
  (e) follows because $V_1^\ell$ and $V_2^\ell$ are fully determined by $(S^n,M_1,M_2)$ while $V_2^\ell$ is a deterministic function of $(M_2,V_1^\ell)$ (first term), and from the the mutual information chain rule (second term);\\
  (f) follows since conditioning reduces entropy (first term), and because $X_{1}^{n}$ is a deterministic function of $(M_1,V_1^\ell,S^n)$ while $X_{2}^{n}$ is a deterministic function of $(M_2,V_2^\ell,S^n)$ (second and third terms);\\
  (g) follows since conditioning reduces entropy (first and second terms) and because when conditioned on $(X_{1,i},X_{2,i})$ and $S_i$, the channel output at time $i$ is independent of $(\mspace{-2mu}V_1^\ell,V_2^\ell,M_1\mspace{-2mu},\mspace{-2mu}M_2,S^{i-1}\mspace{-3mu},S_{i+1}^n,X_1^{i-1}\mspace{-2mu},X_{1,i+1}^n,X_2^{i-1}\mspace{-2mu},X_{2,i+1}^n,Y^{i-1}\mspace{-2mu})$;\\
  (h) follows from (\ref{finite_cap_def}) and by defining $U_i=(V_1^\ell,V_2^\ell,S^{i-d_1-1})$.

Note that the auxiliary random variable $U$ at time $i$ was defined as $U_i\triangleq (V^\ell_1,V^\ell_2,S^{i-d_1-1})$. Accordingly, it represents the information shared during the conference (i.e., the parts of the private messages available to both encoders) and the common knowledge of the states.
This is completely analogous to the role of $U_i$ in the common message setting (cf.\ Theorem \ref{MAC Common Capacity} and Section \ref{U_def_common}).

\par Applying similar arguments to $R_2$ and $R_1+R_2$, one can conclude that any achievable rate-pair $(R_1,R_2)$ must satisfy the following inequalities:
\begin{gather}
R_{1}\leq \frac{1}{n}\sum_{i=1}^{n}{I(X_{1,i};Y_{i}|X_{2,i},S_{i},S_{i-d_2},S_{i-d_1},U_i)}+C_{12}+\epsilon_n  \label{r1}\\
R_{2}\leq \frac{1}{n}\sum_{i=1}^{n}{I(X_{2,i};Y_{i}|X_{1,i},S_{i},S_{i-d_2},S_{i-d_1},U_i)}+C_{21}+\epsilon_n  \label{r2}\\
\ \nonumber\\ \mspace{-50mu}R_1+R_{2}\leq \frac{1}{n}\sum_{i=1}^{n}{I(X_{1,i},X_{2,i};Y_{i}|S_{i},S_{i-d_2},S_{i-d_1},U_i)}\nonumber\\\mspace{310mu}+C_{12}+C_{21}+\epsilon_n  \label{r1+r2a}\\
R_1+R_{2}\leq \frac{1}{n}\sum_{i=1}^{n}{I(X_{1,i},X_{2,i};Y_{i}|S_{i},S_{i-d_2},S_{i-d_1})}+\epsilon_n.  \label{r1+r2b}
\end{gather}

\par The expressions on the right-hand side of the inequalities in (\ref{r1})-(\ref{r1+r2b}) represent empirical averages of mutual information (taken over the code symbols). These inequalities can be alternatively represented by introducing a new time-sharing random variable $Q$, uniformly distributed over $\{1,\dots,n\}$, as in Subsection \ref{U_def_common}. Starting again with the upper bound on $R_1$, this yields
\begin{align}
R_{1}&\leq \frac{1}{n}\sum_{i=1}^{n}I(Y_{Q};X_{1,Q}|X_{2,Q},S_{Q},S_{Q-d_2},S_{Q-d_1},U_Q,Q=i)\nonumber\\&\mspace{392mu}+C_{12}+ \epsilon_n \nonumber\\
     &= I(Y_{Q};X_{1,Q}|X_{2,Q},S_{Q},S_{Q-d_2},S_{Q-d_1},U_Q,Q)+C_{12}+ \epsilon_n \label{Uconverse}
\end{align}
Applying the same procedure to the rest of the upper bounds, while denoting $X_{1}\triangleq X_{1,Q},\ X_{2} \triangleq X_{2,Q},\ Y \triangleq Y_{Q},\ S\triangleq S_{Q},\ \tilde{S}_1\triangleq S_{Q-d_1},\ \tilde{S}_2\triangleq S_{Q-d_2}$ and  $U\triangleq (U_Q,Q)$, we get
\begin{subequations}
\begin{align}
   R_{1}&\leq I(X_{1};Y|X_{2},U,S,\tilde{S}_1,\tilde{S}_2)+C_{12}+ \epsilon_n \\
   R_{2} &\leq I(X_{2};Y|X_{1},U,S,\tilde{S}_1,\tilde{S}_2)+C_{21} + \epsilon_n\\
   R_{1}+R_{2}&\leq I(X_1,X_2;Y|U,S,\tilde{S}_1,\tilde{S}_2)+C_{12}+C_{21}+ \epsilon_n\\
   R_{1}+R_{2}&\leq I(X_1,X_2;Y|S,\tilde{S}_1,\tilde{S}_2)+ \epsilon_n,\label{r0r1r2_conf}
\end{align}
\end{subequations}
where the justification for (\ref{r0r1r2_conf}) follows similar steps to those presented in (\ref{r0r1r2_q_justification}). Moreover, note that the fact that the obtained region is convex follows from the same arguments given in Section \ref{MAC Common Converse}.
\par Completion of the proof of the converse relies on showing that the following Markov relations hold:
\begin{subequations}
\begin{gather}
U-\tilde{S}_1-(S,\tilde{S}_2)\label{conf_markov1}\\
X_1-(\tilde{S}_1,U)-(S,\tilde{S}_2)\label{conf_markov2}\\
X_2-(\tilde{S}_1,\tilde{S}_2,U)-(X_1,S),\label{conf_markov3}
\end{gather}\label{conf_markov}
\end{subequations}

\vspace{-4mm}
\noindent which is accomplished by applying the same line of arguments employed in Appendix \ref{common_markov_proof}, while replacing $M_0$ with $(V_1^\ell,V_2^\ell)$.


\subsection{Achievability}  \label{MAC Conference Achievability}
To prove the achievability of the capacity region, we need to show that for a
fixed distribution of the form $P_{U|\tilde{S}_1}P_{X_{1}|\tilde{S}_1,U}P_{X_{2}|\tilde{S}_1,\tilde{S}_2,U}$ and for $(R_1,R_2)$ that satisfy the inequalities in (\ref{T2region}), there exists a sequence of $(n,\ell,2^{nR_{1}},2^{nR_{2}},d_1,d_2)$ codes for which $P_{e}^{(n)}\rightarrow0$ as $n\rightarrow\infty$.\\
The idea behind this proof is to convert the conferencing problem into a setting that corresponds to the FSM-MAC with a common message considered in Section \ref{Common_Result}, and then rely on Theorem \ref{MAC Common Capacity} to show that the capacity region with conferencing is, indeed, achievable. This is accomplished by sharing as much of the original private messages $(m_1,m_2)$ as possible through the conferencing links to construct a common message. The parts of the original messages not shared by the encoders constitute the private messages in the new setting. Next, the coding scheme introduced in Section \ref{MAC Common Achievability} for the FSM-MAC with a common message can be employed.
\par We start by defining:
\begin{subequations}
\begin{align}
\tilde{R}_1&=\min\{R_1,C_{12}\}\\
\tilde{R}_2&=\min\{R_2,C_{21}\},
\end{align}
\end{subequations}
and rewriting the inequalities in (\ref{T2region}) as
\begin{subequations}
\begin{align}
&(R_{1}-\tilde{R}_1) \leq I(X_{1};Y|X_{2},U,S,\tilde{S}_1,\tilde{S}_2)\\
&(R_{2}-\tilde{R}_2) \leq I(X_{2};Y|X_{1},U,S,\tilde{S}_1,\tilde{S}_2)\\
&(R_{1}-\tilde{R}_1)+(R_{2}-\tilde{R}_2) \leq I(X_1,X_2;Y|U,S,\tilde{S}_1,\tilde{S}_2)\\
&(\tilde{R}_1\mspace{-2mu}+\mspace{-2mu}\tilde{R}_2)\mspace{-3mu}+\mspace{-3mu}(R_{1}\mspace{-2mu}-\mspace{-2mu}\tilde{R}_1)\mspace{-3mu}+\mspace{-3mu}(R_{2}\mspace{-2mu}-\mspace{-2mu}\tilde{R}_2) \mspace{-2mu}\leq\mspace{-2mu} I(X_1,X_2;Y|S,\tilde{S}_1,\tilde{S}_2).
\end{align}\label{newassumption}
\end{subequations}

In view of this representation, we construct a coding scheme by splitting the sets $\mathcal{M}_j=\{1,2,\ldots,2^{nR_j}\}$, for $j\in\{1,2\}$, into $2^{n\tilde{R}_j}$ cells, each containing $2^{n(R_j-\tilde{R}_j)}$ messages, and introducing the functions
\begin{subequations}
\begin{eqnarray}
c_1 &:& \mathcal{M}_1\rightarrow \{1,2,\ldots,2^{n\tilde{R}_1}\}\\
c_2 &:& \mathcal{M}_2\rightarrow \{1,2,\ldots,2^{n\tilde{R}_2}\}\\
e_1 &:& \mathcal{M}_1\rightarrow \{1,2,\ldots,2^{n(R_1-\tilde{R}_1)}\}\\
e_2 &:& \mathcal{M}_2\rightarrow \{1,2,\ldots,2^{n(R_2-\tilde{R}_2)}\}.
\end{eqnarray}
\end{subequations}
Here, for every message $m_j$, where $j\in\{1,2\}$, $c_j$ returns its cell number, $c_j(m_j)$, while $e_j$ returns its index number, $e_j(m_j)$, within the cell $c_j(m_j)$. For the sake of simplicity, we assume here that $2^{n\tilde{R}_1}$, $2^{n\tilde{R}_2}$, $2^{n(R_1-\tilde{R}_1)}$ and $2^{n(R_2-\tilde{R}_2)}$ are integers, although the same approach can be formalized for real numbers as well. Also note that the partitioning above is deterministic.

\par Now, for every message pair $(m_1,m_2)\in\mathcal{M}_1\times\mathcal{M}_2$ we define the triplet $(m_0',m_1',m_2')$ where
\begin{subequations}
\begin{align}
m_1' &\triangleq e_1(m_1)\\
m_2' &\triangleq e_2(m_2)\\
m_0' &\triangleq \big(c_1(m_1),c_2(m_2)\big).
\end{align}
\end{subequations}
Note that the above definitions dictate that $m_1'\in\{1,2,\ldots,2^{n(R_1-\tilde{R}_1)}\}$, $m_2'\in\{1,2,\ldots,2^{n(R_2-\tilde{R}_2)}\}$ and $m_0'\in\{1,2,\ldots,2^{n\tilde{R}_1}\}\times\{1,2,\ldots,2^{n\tilde{R}_2}\}$. Since by definition $\tilde{R}_1\leq C_{12}$ and $\tilde{R}_2\leq C_{21}$, it is possible for Encoder $1$ to transmit $c_1(m_1)$ to Encoder $2$ and for Encoder $2$ to transmit $c_2(m_2)$ to Encoder $1$ via the respective conferencing links. Therefore, following the conferencing stage, both encoders know $m_0'=\big(c_1(m_1),c_2(m_2)\big)$. $m_1'$ and $m_2'$ are viewed as the new private messages.

\begin{figure*}[!b]
\hrulefill
\normalsize
\setcounter{newcounter}{\value{equation}}
\setcounter{equation}{54}
\begin{subequations}
\begin{align}
R_{1} &\leq \sum_{\tilde{s}_1}\pi(\tilde{s}_1)\sum_{\tilde{s}_2}K^{d_1-d_2}(\tilde{s}_2,\tilde{s}_1)\sum_{s}K^{d_2}(s,\tilde{s}_2)\sum_{i=1}^N\log\Big(1+|g_{1,i}(s)|^2\gamma_{1,i}(\tilde{s}_1)\Big)+C_{12},\label{rate_bound1}\\
R_{2} &\leq \sum_{\tilde{s}_1}\pi(\tilde{s}_1)\sum_{\tilde{s}_2}K^{d_1-d_2}(\tilde{s}_2,\tilde{s}_1)\sum_{s}K^{d_2}(s,\tilde{s}_2)\sum_{i=1}^N\log\Big(1+|g_{2,i}(s)|^2\gamma_{2,i}(\tilde{s}_1,\tilde{s}_2)\Big)+C_{21},\label{rate_bound2}\\
R_{1}+R_{2} &\leq \sum_{\tilde{s}_1}\pi(\tilde{s}_1)\sum_{\tilde{s}_2}K^{d_1-d_2}(\tilde{s}_2,\tilde{s}_1)\sum_{s}K^{d_2}(s,\tilde{s}_2)\sum_{i=1}^N\log\Big(1+|g_{1,i}(s)|^2\gamma_{1,i}(\tilde{s}_1)+|g_{2,i}(s)|^2\gamma_{2,i}(\tilde{s}_1,\tilde{s}_2)\Big)\nonumber\\&\ \ \ \ \ \ \ \ \ \ \ \ \ \ \ \ \ \ \ \ \ \ \ \ \ \ \ \ \ \ \ \ \ \ \ \ \ \ \ \ \ \ \ \ \ \ \ \ \ \ \ \ \ \ \ \ \ \ \ \ \ \ \ \ \ \ \ \ \ \ \ \ \ \ \ \ \ \ \ \ \ \ \ \ \ \ \ \ \ \ \ \ \ \ \ \ \ \ \ +C_{12}+C_{21},\label{rate_bound3}\\
R_{1}+R_{2} &\leq \sum_{\tilde{s}_1}\pi(\tilde{s}_1)\sum_{\tilde{s}_2}K^{d_1-d_2}(\tilde{s}_2,\tilde{s}_1)\sum_{s}K^{d_2}(s,\tilde{s}_2)\sum_{i=1}^N\log\Big(1+|g_{1,i}(s)|^2P_{1,i}(\tilde{s}_1)+|g_{2,i}(s)|^2P_{2,i}(\tilde{s}_2,\tilde{s}_1)\nonumber\\&\ \ \ \ \ \ \ \ \ \ \ \ \ \ \ \ \ \ \ \ \ \ \ \ \ \ \ \ \ \ \ \  +2g_{1,i}(s)g_{2,i}^*(s)\sqrt{(P_{1,i}\big(\tilde{s}_1)-\gamma_{1,i}(\tilde{s}_1)\big)\big(P_{2,i}(\tilde{s}_1,\tilde{s}_2)-\gamma_{2,i}(\tilde{s}_1,\tilde{s}_2)\big)}\Big),\label{rate_bound4}
\end{align}\label{conf_Gauss_region}
\end{subequations}
\setcounter{equation}{52}
\end{figure*}
\par The above setting can hence be viewed as a FSM-MAC with a common message. The messages to be transmitted are given by the triplet $(m_0',m_1',m_2')$, where $m_0'\in\{1,2,\ldots,2^{n\tilde{R}_1}\}\times\{1,2,\ldots,2^{n\tilde{R}_2}\}$, $m_1'\in\{1,2,\ldots,2^{n(R_1-\tilde{R}_1)}\}$ and $m_1'\in\{1,2,\ldots,2^{n(R_1-\tilde{R}_1)}\}$, while (\ref{newassumption}) holds by assumption. By Theorem \ref{MAC Common Capacity}, it now immediately follows that the new message triplet $(m_0',m_1',m_2')$ can be transmitted to the decoder with an arbitrarily small probability of error. The decoder can, therefore, reliably reconstruct the message pair $(m_1,m_2)$ and the rate-region (\ref{T2region}) is therefore achievable.
\end{IEEEproof}

\begin{remark}\label{infinite_delay_cooperation_remark}
Using arguments similar to those presented in Remark \ref{infinite_delay_common_remark}, we obtain that when either (or both) of the delays is increased without bound, the capacity region in Theorem \ref{MAC Conference Capacity} approaches the corresponding region where the CSI at the appropriate encoder(s) is absent.
\end{remark}


\section{\textsc{The Vector Gaussian FSM-MAC with Diagonal Channel Transfer Matrices, Conferencing and Delayed CSI}}\label{Gauss_Diagonal_Model}

In this section we consider the vector Gaussian FSM-MAC with diagonal channel transfer matrices, partially cooperative encoders and delayed CSI. For every time instance $t\in\{1,\dots,n\}$, the channel model under consideration is:
\begin{equation}
\mathbf{Y}_t=\mathrm{G}_1(s_t)\mathbf{X}_{1,t}+\mathrm{G}_2(s_t)\mathbf{X}_{2,t}+\mathbf{Z}_t,\label{Gauss_model}
\end{equation}
where $\big\{\mathrm{G}_1(s)\big\}_{s\in\mathcal{S}}$ and $\big\{\mathrm{G}_2(s)\big\}_{s\in\mathcal{S}}$ are $N \times N$ diagonal matrices, which are deterministic functions of the channel state $S=s$.
We denote the diagonal entries of these matrices by $g_{1,i}(s)$ and $g_{2,i}(s)$, respectively, for $i\in\{1,\ldots,N\}$ and $s\in\mathcal{S}$. Moreover, we assume that $\big\{\mathrm{G}_1(s)\big\}_{s\in\mathcal{S}},\big\{\mathrm{G}_2(s)\big\}_{s\in\mathcal{S}}\subset\mathbb{C}^{N\times N}$. For every $t\in\{1,2,\ldots,n\}$,
$\mathbf{X}_{1,t},\mathbf{X}_{2,t}\in\mathbb{C}^N$ and $\mathbf{Y}_t\in\mathbb{C}^N$ are the channel input vectors and the channel output vector, respectively. $\{\mathbf{Z}_t\}_{t=1}^n$ is a proper complex zero mean additive white Gaussian noise (AWGN) process, independent of $\mathbf{X}_{1,t}$ and $\mathbf{X}_{2,t}$ for every $t\in\{1,2,\ldots,n\}$. Thus, each noise sample is distributed according to $\mathbf{Z}_t\sim \mathcal{CN}(0,\mathrm{I})$, where $\mathrm{I}$ is the identity matrix of dimensions $N\times N$. The input vector signals are assumed to satisfy the average power constraints
\begin{align}
\mathrm{tr}\big(\Sigma_{X_1X_1}\big)\leq\bar{P}_1\ \ ; \ \ \mathrm{tr}\big(\Sigma_{X_2X_2}\big)\leq\bar{P}_2,\label{power_const}
\end{align}
where we use the standard notation $\Sigma_{XY}=\mathbb{E}\big[\mathbf{X}\mathbf{Y}^\dag\big]$, and $\mathrm{A}^\dag$ denotes the conjugate transpose of the matrix $\mathrm{A}$.

\par The motivation for examining the channel model in (\ref{Gauss_model}) stems from the fact that it can be used to represent an OFDM-based communication system, employing single receive and transmit antennas. OFDM is an efficient technique used to mitigate frequency selective fading, which is typical in modern wideband communication systems (see, e.g., \cite{LTE,WIMAX}).
The underlying idea behind OFDM is to split the channel's bandwidth into $N$ separate sub-channels through which orthogonal signals are transmitted. By doing so, not only is the impact of intersymbol interference (ISI) dramatically reduced, but the transfer functions of each of the sub-channels boil down to multiplicative scalar gains. These gains are modeled by the diagonal entries of the channel matrices defined above. In this section we derive the maximization problem that specifies the capacity region for the vector Gaussian channel under consideration and convert it into a convex problem. The solution of this convex maximization problem, which can be easily obtained using  a numerical tool such as CVX \cite{cvx}, also yields the optimal power allocation strategy among the sub-channels, which is another essential factor in an OFDM-based transmission.


\setcounter{equation}{57}
\begin{figure*}[!b]
\hrulefill
\normalsize
\begin{subequations}
\begin{align}
R_{1} &\leq \sum_{\tilde{s}_1}\pi(\tilde{s}_1)\sum_{\tilde{s}_2}K^{d_1-d_2}(\tilde{s}_2,\tilde{s}_1)\sum_{s}K^{d_2}(s,\tilde{s}_2)\sum_{i=1}^N\log\Big(1+|g_{1,i}(s)|^2\gamma_{1,i}(\tilde{s}_1)\Big),\label{crate_bound1}\\
R_{2} &\leq \sum_{\tilde{s}_1}\pi(\tilde{s}_1)\sum_{\tilde{s}_2}K^{d_1-d_2}(\tilde{s}_2,\tilde{s}_1)\sum_{s}K^{d_2}(s,\tilde{s}_2)\sum_{i=1}^N\log\Big(1+|g_{2,i}(s)|^2\gamma_{2,i}(\tilde{s}_1,\tilde{s}_2)\Big),\label{crate_bound2}\\
R_{1}+R_{2} &\leq \sum_{\tilde{s}_1}\pi(\tilde{s}_1)\sum_{\tilde{s}_2}K^{d_1-d_2}(\tilde{s}_2,\tilde{s}_1)\sum_{s}K^{d_2}(s,\tilde{s}_2)\sum_{i=1}^N\log\Big(1+|g_{1,i}(s)|^2\gamma_{1,i}(\tilde{s}_1)+|g_{2,i}(s)|^2\gamma_{2,i}(\tilde{s}_1,\tilde{s}_2)\Big),\label{crate_bound3}\\
R_0+R_{1}+R_{2} &\leq \sum_{\tilde{s}_1}\pi(\tilde{s}_1)\sum_{\tilde{s}_2}K^{d_1-d_2}(\tilde{s}_2,\tilde{s}_1)\sum_{s}K^{d_2}(s,\tilde{s}_2)\sum_{i=1}^N\log\Big(1+|g_{1,i}(s)|^2P_{1,i}(\tilde{s}_1)+|g_{2,i}(s)|^2P_{2,i}(\tilde{s}_2,\tilde{s}_1)\nonumber\\&\ \ \ \ \ \ \ \ \ \ \ \ \ \ \ \ \ \ \ \ \ \ \ \ \ \ \ \ \ \ \ \  +2g_{1,i}(s)g_{2,i}^*(s)\sqrt{(P_{1,i}\big(\tilde{s}_1)-\gamma_{1,i}(\tilde{s}_1)\big)\big(P_{2,i}(\tilde{s}_1,\tilde{s}_2)-\gamma_{2,i}(\tilde{s}_1,\tilde{s}_2)\big)}\Big),\label{crate_bound4}
\end{align}\label{common_Gauss_region}
\end{subequations}
\end{figure*}

\subsection{Capacity Region}

\begin{theorem}\label{conf_cap}
The capacity region of the power-constrained vector Gaussian FSM-MAC with diagonal channel transfer matrices, partially cooperative encoders, cooperation link capacities $C_{12}$ and $C_{21}$, delayed CSI and average power constraints
$(\Bar{P}_1,\Bar{P}_2)$ is the union of all sets of rate pairs $(R_1,R_2)\in\mathbb{R}^2_+$ satisfying (\ref{conf_Gauss_region}) at the bottom of the page, where the union is over all $\big\{\gamma_{1,i}(\tilde{s}_1)\big\}_{i\in\{1,\ldots,N\},\tilde{s}_1\in\mathcal{S}}$, $\big\{\gamma_{2,i}(\tilde{s}_1,\tilde{s}_2)\big\}_{i\in\{1,\ldots,N\},(\tilde{s}_1,\tilde{s}_2)\in\mathcal{S}}$,   $\big\{P_{1,i}(\tilde{s}_1)\big\}_{i\in\{1,\ldots,N\},\tilde{s}_1\in\mathcal{S}}$, $\big\{P_{2,i}(\tilde{s}_1,\tilde{s}_2)\big\}_{i\in\{1,\ldots,N\},(\tilde{s}_1,\tilde{s}_2)\in\mathcal{S}}\mspace{-3mu}\subset\mspace{-3mu}\mathbb{R}$ that satisfy the constraints:
\setcounter{equation}{55}
\begin{subequations}
\begin{gather}
\sum_{\tilde{s}_1}\pi(\tilde{s}_1)\sum_{i=1}^NP_{1,i}(\tilde{s}_1)\leq \Bar{P}_1\label{const1}\\
\sum_{\tilde{s}_1}\pi(\tilde{s}_1)\sum_{\tilde{s}_2}K^{d_1-d_2}(\tilde{s}_2,\tilde{s}_1)\sum_{i=1}^NP_{2,i}(\tilde{s}_1,\tilde{s}_2)\leq\Bar{P}_2\label{const2}\\
0 \leq \gamma_{1,i}(\tilde{s}_1) \leq P_{1,i}(\tilde{s}_1),\ \forall\ i\in\{1,\ldots,N\},\ \tilde{s}_1\in\mathcal{S}\\
0 \leq \gamma_{2,i}(\tilde{s}_1,\tilde{s}_2) \leq P_{2,i}(\tilde{s}_1,\tilde{s}_2),\ \forall\mspace{1mu}i\in\{1,\ldots,N\},\mspace{1mu} (\tilde{s}_1,\tilde{s}_2)\mspace{-2mu}\in\mspace{-2mu}\mathcal{S}^2.\label{const4}
\end{gather}\label{conf_Gauss_const}
\end{subequations}
\end{theorem}

The corresponding capacity region for the analogous setting with a common message can be obtained from Theorem \ref{conf_cap} by taking:
\begin{subequations}
\begin{align}
\tilde{R}_0&=C_{12}+C_{21}\\
\tilde{R}_1&=\max\{0,R_1-C_{12}\}\\
\tilde{R}_2&=\max\{0,R_2-C_{21}\},
\end{align}
\end{subequations}
where $\tilde{R}_0$ denotes the common message rate, and $\tilde{R}_1$ and $\tilde{R}_2$ denote the rates of the private messages (according to the common message channel definition in Section \ref{CommonDef}). The result is summarized in the following Corollary.

\begin{corollary}\label{common_cap}
The capacity region of the power-constrained vector Gaussian FSM-MAC with diagonal channel transfer matrices, a common message, delayed CSI and average power constraints $(\Bar{P}_1,\Bar{P}_2)$ is the union of all sets of rate triplets $(R_0,R_1,R_2)\in\mathbb{R}^3_+$ satisfying (\ref{common_Gauss_region}) at the bottom of the page, where the union is over the domain satisfying the constraints (\ref{conf_Gauss_const}).
\end{corollary}

\par Note that the capacity regions in Theorem \ref{conf_cap} and Corollary \ref{common_cap} are both given in the form of a convex optimization problem, which can be solved efficiently using numerical tools. In the following proof we first derive a slightly different, yet equivalent, region for the Gaussian conferencing model. This equivalent capacity region involves a nonconvex optimization problem that we then convert into a convex problem by an appropriate change of optimization variables.
\setcounter{equation}{58}

\begin{IEEEproof}  A straightforward extension of the result stated in Theorem \ref{MAC Conference Capacity} yields the capacity region of the \textit{general} vector FSM-MAC with partially cooperative encoders, delayed CSI and power constraints as in (\ref{power_const}). The region is given by the closure of the set of rate pairs $(R_1,R_2)\in\mathbb{R}^2_+$ that satisfy (cf. (\ref{T2region}))
\begin{subequations}
\begin{align}
R_1&\leq I(\mathbf{X}_1;\mathbf{Y}|\mathbf{X}_2,\mathbf{U},S,\tilde{S}_1,\tilde{S}_2)+C_{12}\\
R_2&\leq I(\mathbf{X}_2;\mathbf{Y}|\mathbf{X}_1,\mathbf{U},S,\tilde{S}_1,\tilde{S}_2)+C_{21}\\
R_1+R_2&\leq I(\mathbf{X}_1,\mathbf{X}_2;\mathbf{Y}|\mathbf{U},S,\tilde{S}_1,\tilde{S}_2)+C_{12}+C_{21}\\
R_1+R_2&\leq I(\mathbf{X}_1,\mathbf{X}_2;\mathbf{Y}|S,\tilde{S}_1,\tilde{S}_2),
\end{align}\label{general_region}
\end{subequations}
for some joint distribution of the form
\begin{equation}
P_{S\tilde{S}_1\tilde{S}_2}P_{\mathbf{U}|\tilde{S}_1}P_{\mathbf{X}_{1}|\tilde{S}_1,\mathbf{U}}P_{\mathbf{X}_{2}|\tilde{S}_1,\tilde{S}_2,\mathbf{U}}\label{PDF},
\end{equation}
where $\mathbf{U}$ is an auxiliary random vector with bounded cardinality. The convexity of the capacity region in (\ref{general_region}) follows from arguments of a nature similar to those presented in Section \ref{MAC Common Converse}, namely, by relying on a time-sharing random variable. Note that the structure of the conditional PDF in (\ref{PDF}) implies the Markov relations:
\begin{subequations}
\begin{gather}
\mathbf{U}-\tilde{S}_1-(S,\tilde{S}_2)\label{umarkov1}\\
\mathbf{X}_1-(\mathbf{U},\tilde{S}_1)-(S,\tilde{S}_2)\label{umarkov2}\\
\mathbf{X}_2-(\mathbf{U},\tilde{S}_1,\tilde{S}_2)-(S,\mathbf{X}_1).\label{umarkov3}
\end{gather}\label{umarkov}
\end{subequations}
The proof of Theorem \ref{conf_cap} consists of two main parts. First, we provide an outer bound for the general capacity region in (\ref{general_region}). Then, by choosing a jointly proper complex Gaussian distribution for $(\mathbf{X}_1,\mathbf{U},\mathbf{X}_2)$, we show that the upper bound is indeed achievable and thus characterizes the actual capacity region.

\par The outer bound for the capacity region is obtained by substituting the random vectors $(\mathbf{X}_1,\mathbf{U},\mathbf{X}_2)$ in (\ref{general_region}) with appropriately chosen jointly proper complex Gaussian random vectors $(\mathbf{X}_1^G,\mathbf{V}^G,\mathbf{X}_2^G)$, which satisfy a certain Markovian relation. We conclude that the chosen random vectors $(\mathbf{X}_1^G,\mathbf{V}^G,\mathbf{X}_2^G)$ indeed admit the desired Markov relation using the following lemma \cite[Section 2, Theorem 1]{Markov_Gauss}.

\begin{lemma}\label{lemma_mimo}
Let $(\mathbf{A},\mathbf{B},\mathbf{C})$ be jointly proper complex Gaussian random vectors. Then $(\mathbf{A},\mathbf{B},\mathbf{C})$ form a Markov chain $\mathbf{A}-\mathbf{B}-\mathbf{C}$ if and only if their covariance matrices satisfy:
\begin{equation}
\Sigma_{AC}=\Sigma_{AB}\Sigma_{BB}^{-1}\Sigma_{BC}.\label{cov_markov_cond}
\end{equation}
\end{lemma}
\par As before, we restrict the detailed derivation to the upper bound on $R_1$, while noting that all other bounds in (\ref{general_region}) can be straightforwardly treated in an analogous manner. To this end, we rewrite the bound on $R_1$ as (cf. (\ref{s1s2sdefinition}))
\begin{align*}
R_1&\leq \sum_{\tilde{s}_1\in\mathcal{S}}\pi(\tilde{s}_1)\sum_{\tilde{s}_2\in\mathcal{S}}K^{d_1-d_2}(\tilde{s}_2,\tilde{s}_1)\\
&\times\sum_{s\in\mathcal{S}}K^{d_2}(s,\tilde{s}_2)I(\mathbf{X}_1;\mathbf{Y}|\mathbf{X}_2,\mathbf{U},s,\tilde{s}_1,\tilde{s}_2)+C_{12}\numberthis\label{UB1}
\end{align*}
and proceed with upper bounding each of the mutual information terms in the sum. Consider:
\begin{align}
I(&\mathbf{X}_{1};\mathbf{Y}|\mathbf{X}_{2},\mathbf{U},s,\tilde{s}_1,\tilde{s}_2)\mspace{-3mu}\stackrel{(a)}{=} \mspace{-3mu} h(\mathrm{G}_1(s)\mathbf{X}_1+\mathbf{Z}|\mathbf{U},s,\tilde{s}_1)\mspace{-3mu}-\mspace{-3mu}h(\mathbf{Z})\nonumber\\
&\stackrel{(b)}{\leq} h(\mathrm{G}_1(s)\mathbf{X}_1+\mathbf{Z}|\mathbf{V},s,\tilde{s}_1)-h(\mathbf{Z})\nonumber\\
&\stackrel{(c)}{\leq} h(\mathrm{G}_1(s)\mathbf{X}^G_1+\mathbf{Z}|\mathbf{V}^G,s,\tilde{s}_1)-h(\mathbf{Z})\label{mimo_r1_ub_basic}\\
&\leq \sum_{i=1}^N\Big\{ h(g_{1,i}(s)X^G_{1,i}+Z_i,V^G_i|s,\tilde{s}_1)-h(V^G_i|\tilde{s}_1)\Big\}-h(\mathbf{Z})\nonumber\\
&= \sum_{i=1}^N\Biggl\{\log\Biggl((\pi e)^2\Biggl[1+|g_{1,i}(s)|^2\Biggl(\mathbb{E}\big[|X_{1,i}^G|^2\big|\tilde{s}_1\big]\nonumber\\&\mspace{130mu}-\frac{\Big|\mathbb{E}\big[X_{1,i}^G(V_i^G)^*\big|\tilde{s}_1\big]\Big|^2}{\mathbb{E}\big[|V_i^G|^2\big|\tilde{s}_1\big]}\Biggr)\Biggr]\cdot\mathbb{E}\big[|V_i^G|^2\big|\tilde{s}_1\big]\Biggr)\nonumber\\&\mspace{170mu}-\log\Big((\pi e)\mathbb{E}\big[|V_i^G|^2\big|\tilde{s}_1\big]\Big)-\log\big(\pi e\big)\Biggr\}\nonumber\\
&\stackrel{(d)}= \mspace{-3mu}\sum_{i=1}^N\log\left(\mspace{-3mu}1\mspace{-3mu}+\mspace{-3mu}|g_{1,i}(s)|^2P_{1,i}(\tilde{s}_1)\left[\mspace{-3mu}1\mspace{-3mu}-\mspace{-3mu}\frac{\Big|\mathbb{E}\big[X_{1,i}^G(V_i^G)^*\big|\tilde{s}_1\big]\Big|^2}{P_{1,i}(\tilde{s}_1)\mathbb{E}\big[|V_i^G|^2\big|\tilde{s}_1\big]}\mspace{-3mu}\right]\right)\nonumber\\
&\stackrel{(e)}{=}\sum_{i=1}^N\log\Big(1+|g_{1,i}(s)|^2\beta_{1,i}(\tilde{s}_1)P_{1,i}(\tilde{s}_1)\Big)\label{mimo_R1_ub}
\end{align}
where:\\
(a) follows from (\ref{Gauss_model}) and the Markov relations (\ref{umarkov});\\
(b) follows by substituting the random vector $\mathbf{U}$, for any given $\tilde{S}_1=\tilde{s}_1$, with a new random vector: $\mathbf{V}(\tilde{s}_1)\triangleq\mathbb{E}\big[\mathbf{X}_1\big|\mathbf{U},\tilde{s}_1\big]$. Note that this is the optimal estimator in the minimum mean square error (MMSE) sense of $\mathbf{X}_1$ given $\mathbf{U}$, for each specified delayed CSI $\tilde{S}_1=\tilde{s}_1$. By substituting $\mathbf{U}$ (for some $\tilde{S}_1=\tilde{s}_1$) with $\mathbf{V}(\tilde{s}_1)$ we increase the first entropy term in view of the fact that $\mathbf{V}(s_1)$ is a deterministic function of the pair $(\mathbf{U},\tilde{s}_1)$, while $h(\mathbf{Z})$ is not affected by the substitution. Moreover, one can easily confirm that $(\mathbf{X}_1,\mathbf{V},\mathbf{X}_2)$ satisfy the covariance condition (\ref{cov_markov_cond}), i.e., the relation
\begin{equation}
\Sigma_{X_1X_2}(\tilde{s}_1,\tilde{s}_2)=\Sigma_{X_1V}(\tilde{s}_1)\Sigma_{VV}^{-1}(\tilde{s}_1)\Sigma_{VX_2}(\tilde{s}_1,\tilde{s}_2)\label{cov_cond}
\end{equation}
holds for every $(\tilde{s}_1,\tilde{s}_2)\in\mathcal{S}^2$. Note that the dependance of the covariance matrices on the states is induced by the Markov relations (\ref{umarkov});\\
(c) follows from the maximum differential entropy lemma \cite[Section 2.2]{ElGammalKim10LectureNotes} and by introducing the triplet $(\mathbf{X}_1^G,\mathbf{V}^G,\mathbf{X}_2^G)$ of zero-mean jointly proper complex Gaussian random vectors with the same auto- and cross- covariance matrices as those of $(\mathbf{X}_1,\mathbf{V},\mathbf{X}_2)$. Replacing $(\mathbf{X}_1,\mathbf{V},\mathbf{X}_2)$ with $(\mathbf{X}_1^G,\mathbf{V}^G,\mathbf{X}_2^G)$ thus increases the first entropy term. Moreover, by Lemma \ref{lemma_mimo}, we conclude that the Gaussian triplet $(\mathbf{X}_1^G,\mathbf{V}^G,\mathbf{X}_2^G)$, for any given $(S,\tilde{S}_1,\tilde{S}_2)=(s,\tilde{s}_1,\tilde{s}_2)$, is Markov, i.e., the relation $\mathbf{X}_1^G(\tilde{s}_1)-\mathbf{V}^G(\tilde{s}_1)-\mathbf{X}_2^G(\tilde{s}_1,\tilde{s}_2)$ holds.\\
(d) follows from defining $P_{1,i}(\tilde{s}_1)\triangleq \mathbb{E}\big[|X_{1,i}|^2\big|\tilde{s}_1\big]$ and $ P_{2,i}(\tilde{s}_1,\tilde{s}_2)\triangleq \mathbb{E}\big[|X_{2,i}|^2\big|\tilde{s}_1,\tilde{s}_2\big]$ (note that these are in fact the $i$-th diagonal entries of the covariance matrices $\Sigma_{X_1^GX_1^G}(\tilde{s}_1)$ and $\Sigma_{X_2^GX_2^G}(\tilde{s}_1,\tilde{s}_2)$, respectively. For this reason, the constraints in (\ref{const1})-(\ref{const2}) follow immediately from (\ref{power_const}) by applying the law of total expectation);\\
(e) follows from defining
\begin{align*}
\bar{\beta}_{1,i}(\tilde{s}_1)&=\left|\frac{\mathbb{E}\big[V^G_i(X_{1,i}^G)^*\big|\tilde{s}_1\big]}{\sqrt{\mathbb{E}\big[|X_{1,i}^G|^2\big|\tilde{s}_1\big]\mathbb{E}\big[|V^G_i|^2\big|\tilde{s}_1\big]}}\right|^2\\&=\frac{\Big|E\big[V^G_i(X_{1,i}^G)^*\big|\tilde{s}_1\big]\Big|^2}{P_{1,i}(\tilde{s}_1)\mathbb{E}\big[|V_i^G|^2\big|\tilde{s}_1\big]},\numberthis\label{beta_vec1}
\end{align*}
where we use the notation $\bar{\alpha}=1-\alpha$, $\alpha\in \mathbb{R}$. We also introduce the definition
\begin{align*}
\bar{\beta}_{2,i}(\tilde{s}_1,\tilde{s}_2)&=\left|\frac{\mathbb{E}\big[V^G_i(X_{2,i}^G)^*\big|\tilde{s}_1,\tilde{s}_2\big]}{\sqrt{\mathbb{E}\big[|X_{2,i}^G|^2\big|\tilde{s}_1,\tilde{s}_2\big]\mathbb{E}\big[|V^G_i|^2\big|\tilde{s}_1\big]}}\right|^2\\
&=\frac{\Big|E\big[V^G_i(X_{2,i}^G)^*\big|\tilde{s}_1,\tilde{s}_2\big]\Big|^2}{P_{2,i}(\tilde{s}_1,\tilde{s}_2)\mathbb{E}\big[|V_i^G|^2\big|\tilde{s}_1\big]},\numberthis\label{beta_vec2}
\end{align*}
that will be used to represent the additional rate constraints in (\ref{conf_Gauss_region}). Note that $\bar{\beta}_{1,i}(\tilde{s}_1)$ (respectively, $\bar{\beta}_{2,i}(\tilde{s}_1,\tilde{s}_2)$) is defined to be the squared correlation coefficient between $X_{1,i}^G$ (respectively, $X_{2,i}^G$) and $V^G_i$ for a given delayed CSI
$\tilde{S}_1=\tilde{s}_1$ (respectively, delayed CSI pair $(\tilde{S}_1,\tilde{S}_2)=(\tilde{s}_1,\tilde{s}_2)$). Accordingly, we have that $\beta_{1,i}(\tilde{s}_1),\beta_{2,i}(\tilde{s}_1,\tilde{s}_2)\in[0,1]$ for every $i\in\{1,\ldots,N\}$.
The upper bounds on $R_2$, and both upper bounds on the sum-rate $R_1+R_2$, are similarly constructed.

\setcounter{figure}{4}
\begin{figure*}[ht]
\begin{center}
\begin{psfrags}
    \psfrag{A}[][][0.7]{$\ \ \ \ \ \ \ \ C=0$}
    \psfrag{B}[][][0.7]{$\ \ \ \ \ \ \ \ \ C=0.1$}
    \psfrag{C}[][][0.7]{$\ \ \ \ \ \ \ \ \ C=0.3$}
    \psfrag{D}[][][0.7]{$\ \ \ \ \ \ \ \ \ C=0.5$}
    \psfrag{E}[][][0.7]{$\ \ \ \ \ \ \ \ \ C=0.9$}
    \psfrag{F}[][][0.7]{$\ \ \ \ \ \ \ \ \ \ \ C_{12}=0$}
    \psfrag{G}[][][0.7]{$\ \ \ \ \ \ \ \ \ \ \ \ \ C_{12}=0.1$}
    \psfrag{H}[][][0.7]{$\ \ \ \ \ \ \ \ \ \ \ \ \ C_{12}=0.3$}
    \psfrag{I}[][][0.7]{$\ \ \ \ \ \ \ \ \ \ \ \ \ C_{12}=0.5$}
    \psfrag{J}[][][0.7]{$\ \ \ \ \ \ \ \ \ \ \ \ \mspace{-12mu}C_{12}\mspace{-5mu}=\mspace{-3mu}0.9$}
    \psfrag{K}[][][0.7]{$\ \ \ \ \ \ \ \ \ \ \ C_{12}=0$}
    \psfrag{L}[][][0.7]{$\ \ \ \ \ \ \ \ \ \ \ \ \ C_{12}=0.1$}
    \psfrag{M}[][][0.7]{$\ \ \ \ \ \ \ \ \ \ \ \ \ C_{12}=0.3$}
    \psfrag{N}[][][0.7]{$\ \ \ \ \ \ \ \ \ \ \ \ \ C_{12}=0.5$}
    \psfrag{O}[][][0.7]{$\ \ \ \ \ \ \ \ \ \ \ \ \ C_{12}=0.9$}
    \psfrag{T}[][][0.7]{Symmetrical Capacities $C_{12}=C_{21}$ and Delay $d=2$}
    \psfrag{P}[][][0.7]{Infinity Capacity $C_{12}<C_{21}=\infty$ and Delay $d=2$}
    \psfrag{S}[][][0.7]{Single Capacity $C_{12}\geq C_{21}=0$ and Delay $d=2$}
    \psfrag{X}[][][0.8]{$R_1$ [bits/symbol]}
    \psfrag{Y}[][][0.8]{$R_2$ [bits/symbol]}
\subfloat[]{\includegraphics[width=6cm]{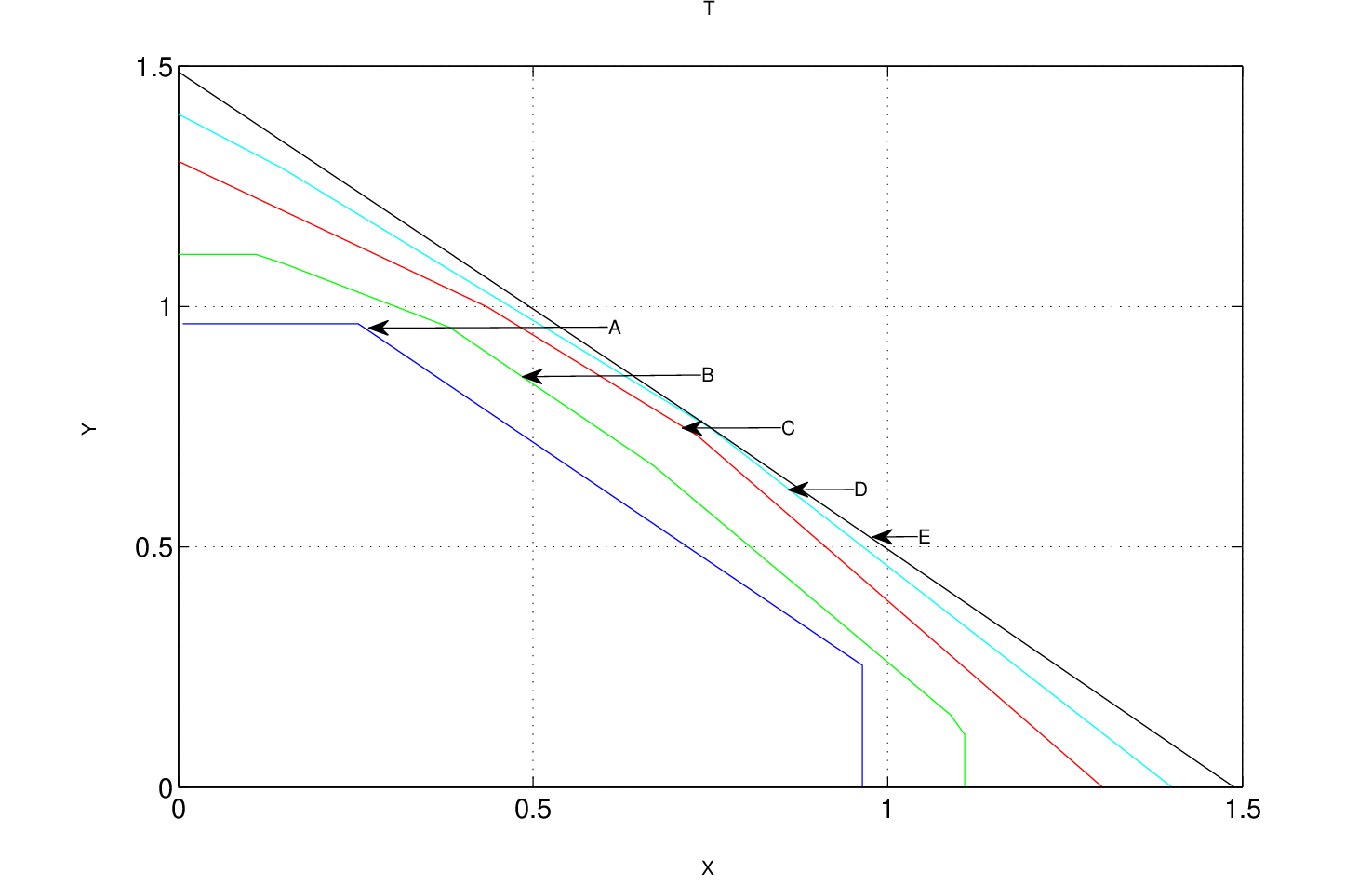}}
\subfloat[]{\includegraphics[width=6cm]{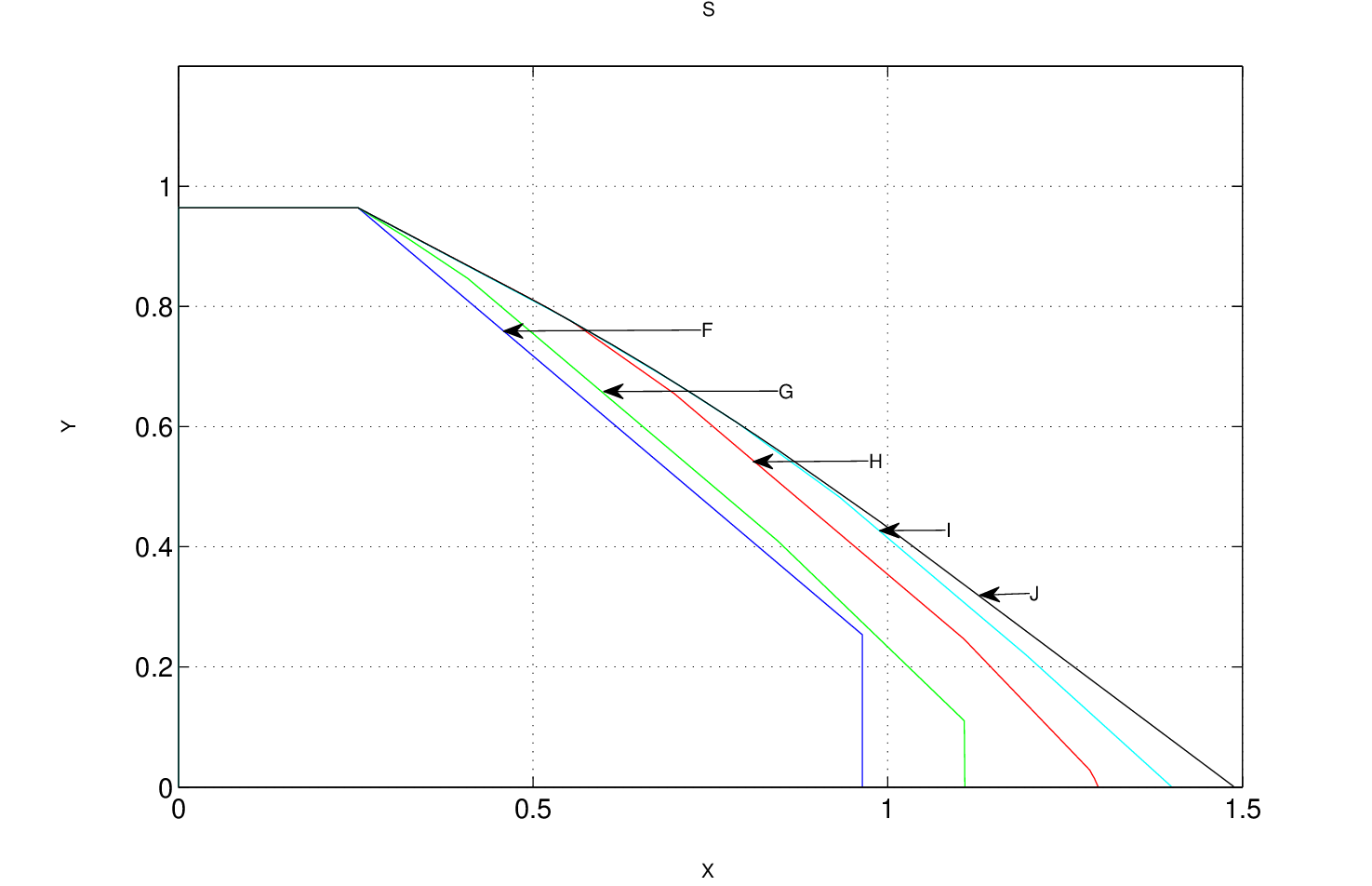}}
\subfloat[]{\includegraphics[width=6cm]{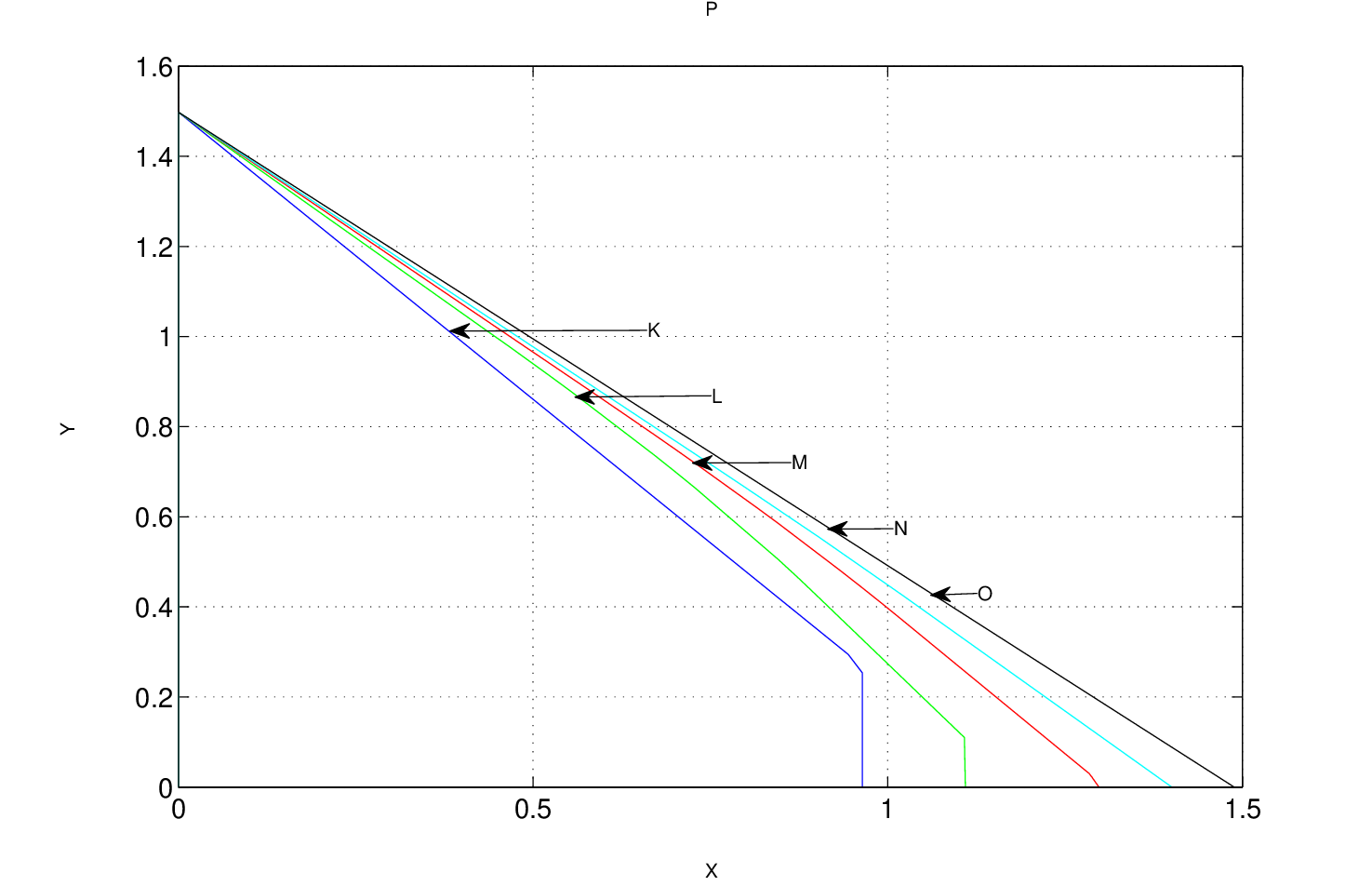}}
\caption{Capacity region for the two-state AWGN-MAC: (a) symmetrical, $C_{12}=C_{21}\triangleq C$; (b) single cooperation link, $C_{12}\geq C_{21}=0$; (c) infinite cooperation link, $C_{12}<C_{21}=\infty$.} \label{symmetric_cap_region}
\end{psfrags}
\end{center}
\end{figure*}
\setcounter{figure}{3}

\par Next, we show that the upper bounds are also achievable. We take  $(\mathbf{X}_1,\mathbf{U},\mathbf{X}_2)$ to be zero-mean jointly proper complex Gaussian random vectors that admit the Markov relations (\ref{umarkov}) and for which the auto- and cross- covariance matrices $\Sigma_{X_1X_1}(\tilde{s}_1)$, $\Sigma_{X_2X_2}(\tilde{s}_1,\tilde{s}_2)$, $\Sigma_{UU}(\tilde{s}_1)$, $\Sigma_{X_1U}(\tilde{s}_1)$ and $\Sigma_{X_2U}(\tilde{s}_1,\tilde{s}_2)$ are diagonal for every $(\tilde{s}_1,\tilde{s}_2)\in\mathcal{S}^2$. Specifically, we take
\begin{subequations}
\begin{align}
\Sigma_{X_1X_1}(\tilde{s}_1)&=\mbox{diag}\Big(\big\{P_{1,i}(\tilde{s}_1)\big\}_{i=1}^N\Big)\\
\Sigma_{X_2X_2}(\tilde{s}_1,\tilde{s}_2)&=\mbox{diag}\Big(\big\{P_{2,i}(\tilde{s}_1,\tilde{s}_2)\big\}_{i=1}^N\Big),
\end{align}
\end{subequations}
and denote the diagonal entries of the covariance matrices $\Sigma_{UU}(\tilde{s}_1),\ \Sigma_{X_1U}(\tilde{s}_1)$ and $\Sigma_{X_2U}(\tilde{s}_1,\tilde{s}_2)$ by $\sigma_{U_i}^2(\tilde{s}_1)$, $\mathbb{E}\big[X_{1,i}U_i^*\big|\tilde{s}_1\big]$ and $\mathbb{E}\big[X_{2,i}U_i^*\big|\tilde{s}_1,\tilde{s}_2\big]$, respectively, where $i\in\{1,2,\ldots,n\}$.
Moreover, $(\mathbf{X}_1,\mathbf{U},\mathbf{X}_2)$ are chosen to have the same entry-wise correlations as $(\mathbf{X}_1^G,\mathbf{V}^G,\mathbf{X}_2^G)$, that is
\begin{subequations}
\begin{align}
\frac{\Big|E\big[U_iX_{1,i}^*\big|\tilde{s}_1\big]\Big|^2}{P_{1,i}(\tilde{s}_1)\sigma_{U_i}^2(\tilde{s}_1)}&=\bar{\beta}_{1,i}(\tilde{s}_1)\\
\frac{\Big|E\big[U_iX_{2,i}^*\big|\tilde{s}_1,\tilde{s}_2\big]\Big|^2}{P_{2,i}(\tilde{s}_1,\tilde{s}_2)\sigma_{U_i}^2(\tilde{s}_1)}&=\bar{\beta}_{2,i}(\tilde{s}_1,\tilde{s}_2).
\end{align}
\end{subequations}
It can now be shown that this choice for the random vectors $(\mathbf{X}_1,\mathbf{U},\mathbf{X}_2)$ achieves the upper bounds (\ref{conf_Gauss_region}). For conciseness, we present only the calculation for $R_1$ and note that the proof for the remaining rate bounds is similar. As in (\ref{mimo_r1_ub_basic}), using the channel model and the Markov relations (\ref{umarkov}), we have that:
\begin{align}
I(\mathbf{X}_{1};&\mathbf{Y}|\mathbf{X}_{2},\mathbf{U},s,\tilde{s}_1,\tilde{s}_2)\mspace{-3mu}=\mspace{-3mu}h(\mathrm{G}_1(s)\mathbf{X}_1\mspace{-6mu}+\mspace{-4mu}\mathbf{Z}|\mathbf{U},s,\tilde{s}_1\mspace{-2mu})\mspace{-3mu}-\mspace{-3mu}h(\mathbf{Z})\nonumber\\
&=h(\mathrm{G}_1(s)\mathbf{X}_1+\mathbf{Z},\mathbf{U}|s,\tilde{s}_1)-h(\mathbf{U}|\tilde{s}_1)-h(\mathbf{Z})\label{vec_R1_ach}.
\end{align}
Clearly
\begin{align}
h(\mathbf{Z})&=\log\Big((\pi e)^N\Big)\label{vec_hZ}\\
h(\mathbf{U}|\tilde{s}_1)&=\log\Big((\pi e)^N\prod_{i=1}^N{\sigma_{U_i}^2(\tilde{s}_1)}\Big)\label{vec_hU}.
\end{align}
Therefore, it is left to obtain an explicit expression for
\begin{equation}
h(\mathrm{G}_1(s)\mathbf{X}_1+\mathbf{Z},\mathbf{U}|s,\tilde{s}_1)=\log\Big((\pi e)^{2N}\left|\tilde{\Sigma}(s,\tilde{s}_1)\right|\Big)\label{vec_entR1},
\end{equation}
where $\tilde{\Sigma}(s,\tilde{s}_1)$ is a block matrix of the structure
\begin{equation}
\tilde{\Sigma}(\mspace{-2mu}s,\mspace{-2mu}\tilde{s}_1\mspace{-2mu})\mspace{-4mu}=\mspace{-4mu}\left(\mspace{-7mu} \begin{array}{cc}
\mathrm{I}+\mathrm{G}_1(s)\Sigma_{X_1X_1}(\tilde{s}_1)\mathrm{G}_1^\dag(s) & \mathrm{G}_1(s)\Sigma_{X_1U}(\tilde{s}_1)\\
\Sigma_{X_1U}^\dag(\tilde{s}_1)\mathrm{G}_1^\dag(s) & \Sigma_{UU}(\tilde{s}_1)
\end{array}\mspace{-7mu}\right).
\end{equation}

\noindent After some algebra it can be shown that:

\begin{align}
&\left|\tilde{\Sigma}(s,\tilde{s}_1)\right|=\left(\prod_{i=1}^N\left[|g_{1,i}(s)|^2P_{1,i}(\tilde{s}_1)+1\right]\right)\nonumber\\&\times\left(\prod_{i=1}^N\left[\sigma_{U_i}^2(\tilde{s}_1)-\frac{|g_{1,i}(s)|^2P_{1,i}(\tilde{s}_1)\bar{\beta}_{1,i}(\tilde{s}_1)\sigma_{U_i}^2(\tilde{s}_1)}{|g_{1,i}(s)|^2P_{1,i}(\tilde{s}_1)+1}\right]\right)\nonumber\\
&=\left(\prod_{i=1}^N{\sigma_{U_i}^2(\tilde{s}_1)}\right)\cdot\left(\prod_{i=1}^N\left[|g_{1,i}(s)|^2\beta_{1,i}(\tilde{s}_1)P_{1,i}(\tilde{s}_1)+1\right]\right).\label{vec_detR1}
\end{align}

\par Substituting (\ref{vec_detR1}) along with (\ref{vec_hZ}), (\ref{vec_hU}) and (\ref{vec_entR1}) into (\ref{vec_R1_ach}) and summing the mutual information terms over all state triplets $(S,\tilde{S}_1,\tilde{S}_2)=(s,\tilde{s}_1,\tilde{s}_2)$ in (\ref{UB1}), we achieve the upper bound for $R_1$ conforming with (\ref{mimo_R1_ub}). In a similar manner, all other upper bounds can be shown to be achievable. This characterizes the maximization problem defining the capacity region for the diagonal vector Gaussian FSM-MAC with partially cooperative encoders and delayed CSI. Note that through this proof we have shown the optimality of the proper complex Gaussian multivariate input distribution for this model.
\par We note that the problem of maximizing the achievable rate region obtained using the above steps is not convex since the argument of the logarithms involve products of the optimization variables (e.g., the product $P_{1,i}(\tilde{s}_1)\bar{\beta}_{1,i}(\tilde{s}_1)$ in (\ref{vec_detR1})). Thus, to convert it into a convex maximization problem we further substitute
\begin{subequations}
\begin{align}
\gamma_{1,i}(\tilde{s}_1)&=\beta_{1,i}(\tilde{s}_1)P_{1,i}(\tilde{s}_1),\ \forall \tilde{s}_1\in\mathcal{S}\\
\gamma_{2,i}(\tilde{s}_1,\tilde{s}_2)&=\beta_{2,i}(\tilde{s}_1,\tilde{s}_2)P_{2,i}(\tilde{s}_1,\tilde{s}_2),\ \forall (\tilde{s}_1,\tilde{s}_2)\in\mathcal{S}^2,
\end{align}\label{gamma}
\end{subequations}

\vspace{-3mm}
\noindent for every $i\in\{1,\ldots,N\}$. This substitution yields the rate bounds in (\ref{conf_Gauss_region}) and concludes the proof.
\end{IEEEproof}


\subsection{Two-State Scalar AWGN Channel Example}\label{Gauss_Example}

\par
\begin{figure}[!h]
\begin{center}
\begin{psfrags}
    \psfrag{g}[][][0.7]{ $g$ }  \psfrag{b}[][][0.7]{$b$}
    \psfrag{B}[][][0.7]{ $B$ }  \psfrag{G}[][][0.7]{$G$}
    \psfrag{a}[][][0.7]{\ \ \ \  $1-g$ }  \psfrag{d}[][][0.7]{$1-b$}
\includegraphics[scale=0.45]{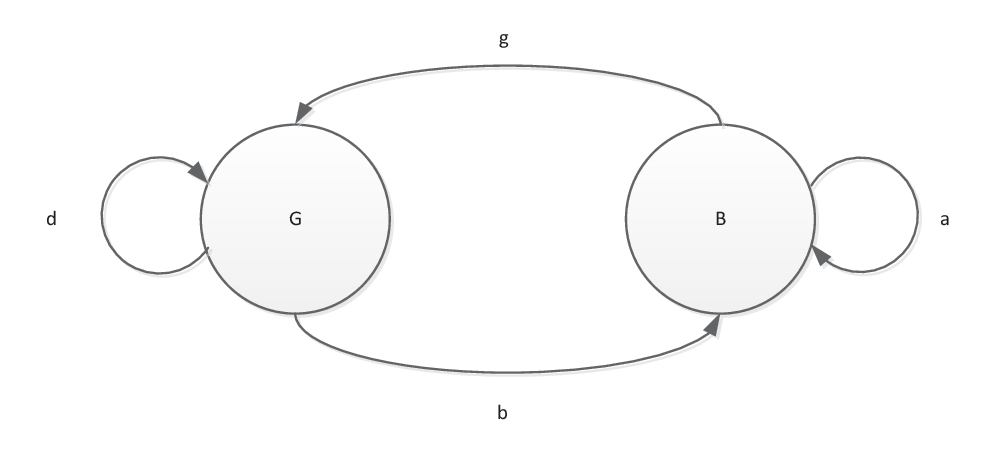}\vspace{-3 mm}\ \ \ \ \ \ \ \
\caption{Two-state AWGN channel.} \label{Two-state AGN channel}
\end{psfrags}
\end{center}
\end{figure}

 To gain some intuition on the capacity region of the MAC with partially cooperative encoders and delayed CSI, we now consider the scalar Gaussian channel with only two possible states. The scalar channel corresponds to taking $N=1$ in the diagonal vector channel definition in (\ref{Gauss_model}). We denote the two possible channel states by $G$ and $B$ (where $G$ stands for `Good' and $B$ for `Bad'), thus, $\mathcal{S}=\{G,B\}$. The two states differ in their associated channel gains. When $S=G$, the gains are $g_1(s=G)=g_2(s=G)\triangleq g_G$, whereas when $S=B$ the gains are $g_1(s=B)=g_2(s=B)\triangleq g_B$. We assume without loss of generality that $g_G>g_B$. The Markov model of the state process is illustrated in Fig. \ref{Two-state AGN channel}.

\setcounter{figure}{5}
\noindent The state process is specified by the the transition probability matrix:
\begin{equation}
K=\left(\begin{array}{cc} P(G|G) & P(B|G)\\ P(G|B) & P(B|B)\end{array}\right)=\left(\begin{array}{cc} 1-b & b\\ g & 1-g\end{array}\right),
\end{equation}
which induces the following stationary distribution:
\begin{equation}
\pi=\left(\begin{array}{cc} \pi(G) & \pi(B)\end{array}\right)=\left(\begin{array}{cc} \frac{g}{g+b} & \frac{b}{g+b} \end{array}\right).
\end{equation}

\begin{figure*}[ht]
\begin{center}
\begin{psfrags}
    \psfrag{T}[][][0.7]{A-Symmetrical Delays $0=d_2\leq d_1=2$}
    \psfrag{L}[][][0.7]{Symmetrical Delays $d_1=d_2=2$}
    \psfrag{C}[][][0.7]{Infinite Delay $2=d_2<d_1=\infty$}
    \psfrag{X}[][][0.7]{$C_{12}=C_{21}$ [bits/symbol]}
    \psfrag{Y}[][][0.7]{$R_1+R_2$ [bits/symbol]}
\subfloat[]{\includegraphics[width=6cm]{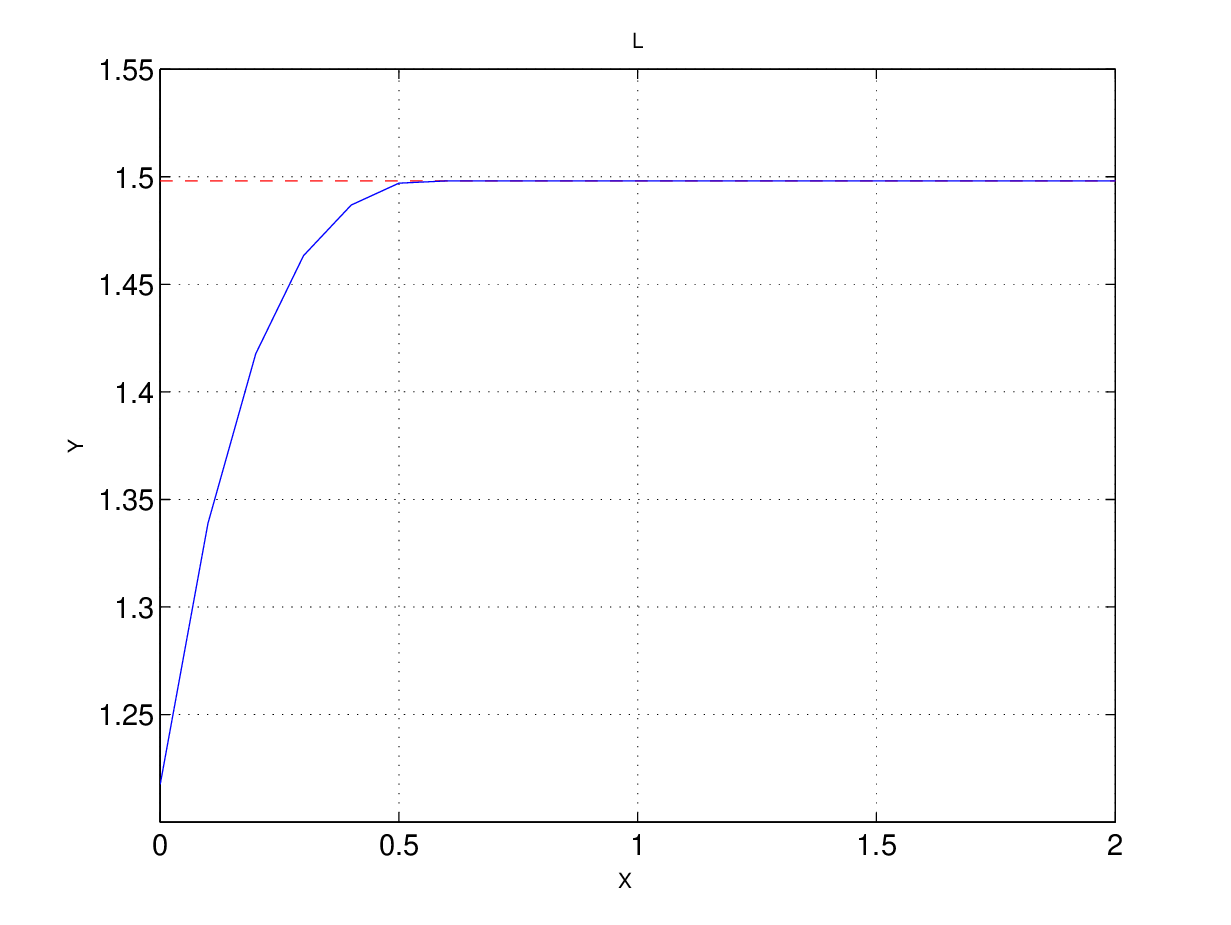}}
\subfloat[]{\includegraphics[width=6cm]{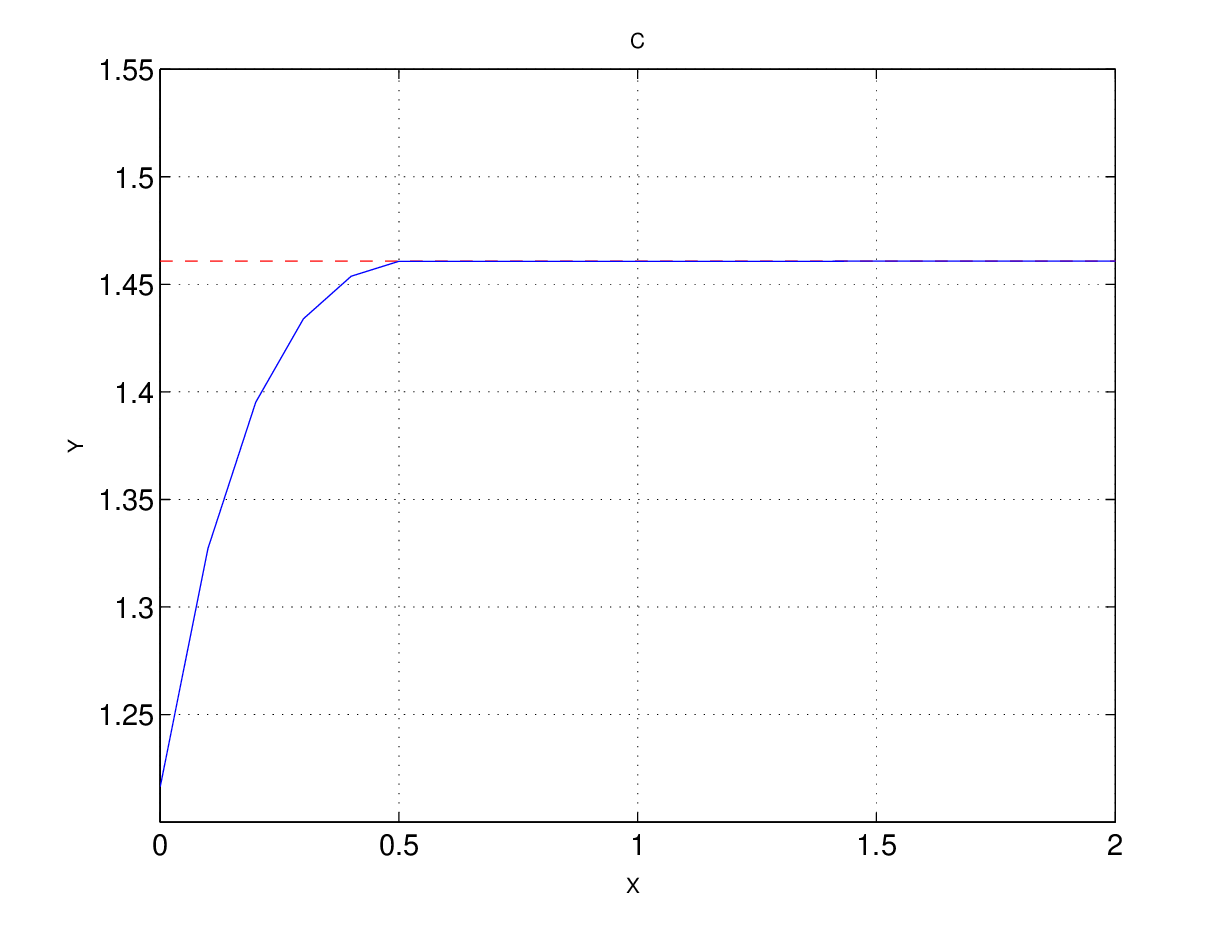}}
\subfloat[]{\includegraphics[width=6cm]{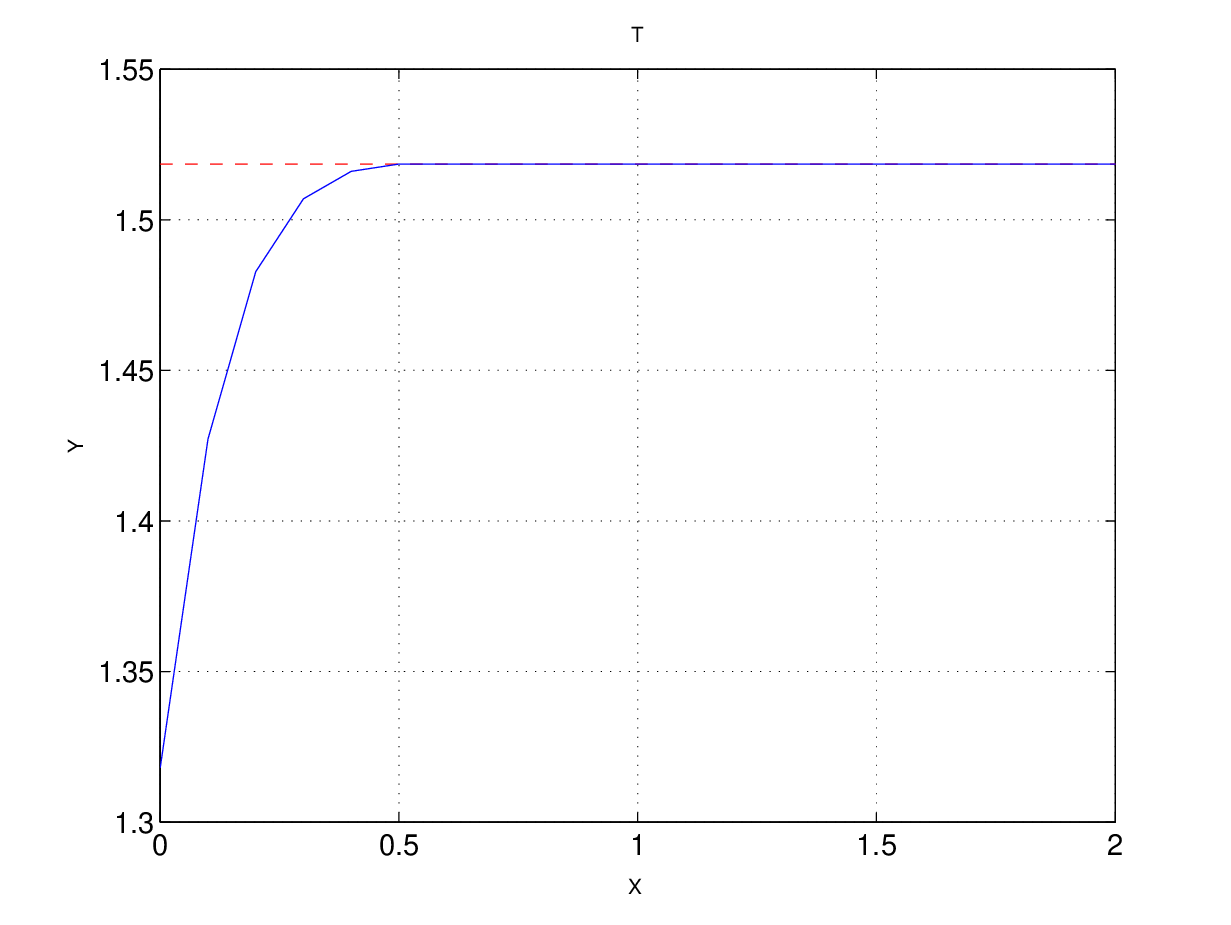}}
\caption{The sum-rate versus the cooperation link capacities $C_{12}=C_{21}$ for three different cases of delayed CSI: (a) symmetrical delays, $d_2=d_2=2$; (b) infinite delay, $2=d_2<d_1=\infty$; (c) asymmetrical delays, $0=d_2\leq d_1=2$. The dashed line corresponds to the case where $C_{12}=C_{21}=\infty$.}\label{sumratevscap}
\end{psfrags}
\end{center}
\end{figure*}

We start by examining the impact of the cooperation link capacities, $C_{12}$ and $C_{21}$, on the capacity regions in the particular case of symmetric CSI delays, i.e., $d_1=d_2\triangleq d$. Note that since $d_1=d_2$, it immediately follows that $\tilde{S}_1=\tilde{S}_2\triangleq \tilde{S}$. The capacity region is presented in Fig. \ref{symmetric_cap_region} for three different cases: (a) symmetrical capacities, represented by, $C_{12}=C_{21}\triangleq C$, (b) single cooperation link, represented by, $C_{12}\geq C_{21}=0$ and (c) one infinite cooperation link, represented by, $C_{12}<C_{21}=\infty$. The capacity regions were calculated by numerically solving the optimization problem induced by Theorem \ref{conf_cap} for the above three cases using CVX \cite{cvx}. Throughout this example we assume $\Bar{P}_1=\Bar{P}_2=10$, $g_B=0.01$, $g_G=1$, $g=b=0.1$ and $d=2$  (results of similar nature were observed for $g_B=0.2$ and $g_B=0.3$).

\par Note that in Fig. \ref{symmetric_cap_region}(a), which presents the region for the symmetrical case, as $C$ grows without bound, the capacity region increases and eventually adopts a triangular shape. This outcome is because the first three constraints on the rates $(R_1,R_2)$, as given by (\ref{rate_bound1})-(\ref{rate_bound3}), also grow without bound, and thus, the binding constraint is the sum-rate constraint of (\ref{rate_bound4}).
For the case of a single cooperation link shown in Fig. \ref{symmetric_cap_region}(b), the upper bound on $R_2$ remains fixed as $C_{12}$ grows, since the constraint in (\ref{rate_bound2}) does not change with $C_{12}$ and stays fixed at approximately $0.9642$. Finally, for the case of infinite cooperation link capacity $C_{21}=\infty$, as shown in Fig. \ref{symmetric_cap_region}(c), we have that the constraint on $R_2$ in (\ref{rate_bound2}) and the first constraint on the sum-rate in (\ref{rate_bound3}) are both redundant. Hence, the only meaningful constraint on $R_2$ is (\ref{rate_bound4}), which does not involve $C_{21}$ (or $C_{12}$).

\par Next, we demonstrate that the capacity region of this setting grows as the cooperation link capacities grow, regardless of the specific assumptions on the relation between the delays of the CSI available at the encoders. To do so, we present the maximum sum-rate versus the cooperation link capacities for three different possible relations between the delays: (a) $d_1=d_2=2$, (b) $2=d_2<d_1=\infty$ and (c) $2=d_1\geq d_2=0$. For all three cases we assume $C_{12}=C_{21}$ and use the same values of the channel gains as before. The curves are shown in Fig. \ref{sumratevscap}(a)-(c).
\par As expected, The sum-rate of case (c) (which exhibits the best CSI properties of the three) reaches the highest value as the capacities grow, whereas the sum-rate for case (b) (which exhibits the worst CSI properties) reaches the lowest value. Moreover, we note the correspondence between Fig. \ref{sumratevscap}(a) and Fig. \ref{symmetric_cap_region}(a) (both corresponding to the case of symmetrical delays and equal cooperation link capacities). Evidence of this correspondence is the fact that when $C_{12}=C_{21}$ grow, the sum-rate, in both figures, approaches its maximal value, which is approximately $1.5$ bits per symbol.

\par
\begin{figure*}[ht]
\begin{center}
\begin{psfrags}
    \psfrag{A}[][][0.8]{Correlation vs. SNR for $C_{12}=C_{21}=0$}
    \psfrag{B}[][][0.8]{Correlation vs. SNR for $C_{12}=C_{21}=0.1$}
    \psfrag{C}[][][0.8]{Correlation vs. SNR for $C_{12}=C_{21}=0.2$}
    \psfrag{D}[][][0.8]{Correlation vs. SNR for $C_{12}=C_{21}=0.3$}
    \psfrag{E}[][][0.8]{Correlation vs. SNR for $C_{12}=C_{21}=0.4$}
    \psfrag{F}[][][0.8]{Correlation vs. SNR for $C_{12}=C_{21}=0.5$}
    \psfrag{X}[][][0.8]{SNR [dB]}
    \psfrag{Y}[][][0.8]{Correlation}
    \psfrag{G}[][][0.8]{\ \ \ \ \ \ \ \ \ \ \ \ \ \ \ \ \ \ \ \ \ \ \ \ \ \ \ \ \ \ \ \ \ \ $\mathrm{SNR}^{\mathrm{Crit}}=10\log_{10}\Big(\frac{2^{2(C_{12}+C_{21})}-1}{4}\Big)[\mathrm{dB}]$}
    \psfrag{H}[][][0.8]{\ \ \ \ \ \ \ \ $\rho^*_\infty=\sqrt{1-\frac{2}{2^{2(C_{12}+C_{21})}+1}}$}
\centerline{\includegraphics[scale = .67]{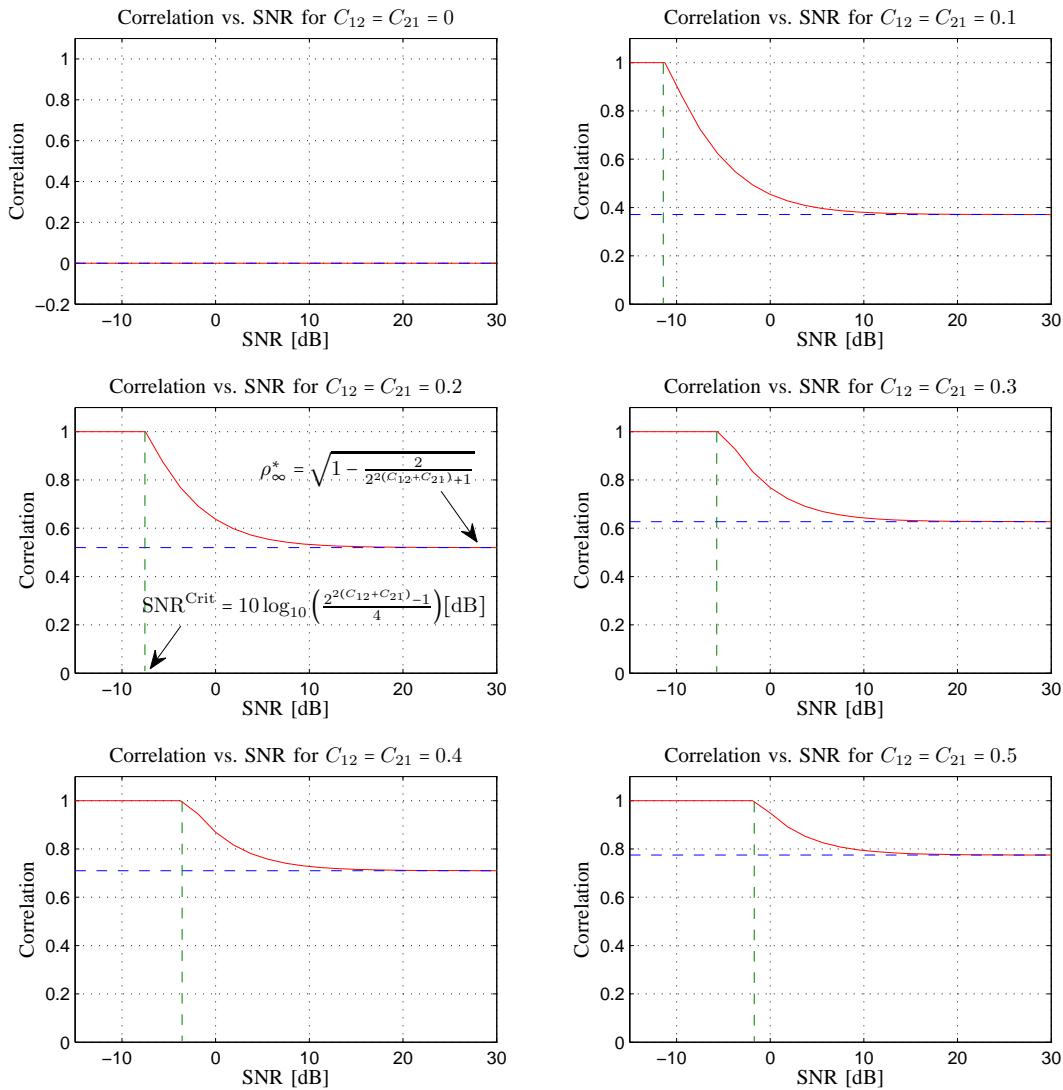}}
\caption{Correlation as a function of SNR for different values of the capacities $C_{12}=C_{21}$.} \label{betavsSNR}
\end{psfrags}
\end{center}
\end{figure*}

 Another interesting aspect of the Gaussian channel example is the impact of the signal-to-noise ratio (SNR) on the correlations between the auxiliary random variable, $U$, and the random variables $X_1$ and $X_2$. These correlations are associated with the level of cooperation used in the scheme. We assume that the transmit powers satisfy $P_1=P_2\triangleq P$ and that $g_1=g_2=1$, so that the SNR, in fact, equals $P$, and restrict the analysis to the case where $|\mathcal{S}|=1$, i.e., a single and constant channel state \cite{Wigger_gaussian_cop}. We use throughout the same notations and expressions for the rate bounds as in \cite{Wigger_gaussian_cop}. Note that for the case where $|\mathcal{S}|=1$, the maximization problem in (\ref{conf_Gauss_region}) turns out to be concave even without the transformation (\ref{gamma}); thus no transformation is needed. The remaining optimization variables are $\beta_1$ and $\beta_2$, which are defined through (cf., (\ref{beta_vec1})-(\ref{beta_vec2}))
\begin{subequations}
\begin{align}
&\sqrt{1-\beta_1}=\left|\frac{\mathbb{E}[UX_1]}{\sqrt{\mathbb{E}[X_1^2]\mathbb{E}[U^2]}}\right|\triangleq \rho_1\\
&\sqrt{1-\beta_2}=\left|\frac{\mathbb{E}[UX_2]}{\sqrt{\mathbb{E}[X_2^2]\mathbb{E}[U^2]}}\right|\triangleq \rho_2.
\end{align}\label{rho_def}
\end{subequations}
We consider the case of symmetrical cooperation link capacities, i.e., $C_{12}=C_{21}$. By the symmetry of the maximization problem in $(\beta_1,\beta_2)$, optimality is achieved when $\beta_1=\beta_2$. For this reason we use the notation $\beta_1=\beta_2\triangleq \beta$ and plot a single curve representing both correlations (which are calculated directly from $\beta$ according to (\ref{rho_def})). The numerical results are shown in Fig. \ref{betavsSNR}. The dashed blue and green lines designate the asymptotic value of the correlation and the critical SNR at which the correlation drops from unity, respectively. Results are shown for six different values of $C_{12}=C_{21}$.


Although the effect of the SNR on the correlations could not be calculated analytically, we use asymptotic evaluations to gain some additional insight. Namely, we demonstrate that the optimal correlation admits
\begin{equation}
\rho^*=\begin{cases}
\ \ \ \ \ \ \ \ \ \ \ \ 1\ \ \ \ \ \ \ \ \ \ \ \ \ ,& \mathrm{SNR}\leq \mathrm{SNR}^\mathrm{Crit}\\
\sqrt{1-\frac{2}{2^{2(C_{12}+C_{21})}+1}}\ \ \ \ ,& \mathrm{SNR}\rightarrow\infty
\end{cases},\label{opt_corr}
\end{equation}
where $\mathrm{SNR}^\mathrm{Crit}=10\log_{10}\Big(\frac{2^{2(C_{12}+C_{21})}-1}{4}\Big)[\mathrm{dB}]$.
\par We start by justifying the observation that the correlation approaches $1$ for small SNR values. For some positive value of $C_{12}=C_{21}$ and for $P_1,P_2\ll 1$, consider (cf. (\ref{rate_bound3})-(\ref{rate_bound4})):
\ \\
\ \\
\begin{align}
R_{1}+R_{2} & \leq\mspace{-2mu} \min\mspace{-2mu} \left\{\begin{array}{ll} \mspace{-11mu}\frac{1}{2}\log\Big(1+\beta(g_1^2P_1+g_2^2P_2)\Big)+C_{12}+C_{21},\\  \mspace{-11mu}\frac{1}{2}\log\Big(1+g_1^2P_1+g_2^2P_2+2g_1g_2\sqrt{\bar{\beta}^2P_1P_2})\end{array}\mspace{-11.6mu}\right\}\nonumber\\
&{=} \frac{1}{2}\log\Big(1+g_1^2P_1+g_2^2P_2+2g_1g_2\sqrt{\bar{\beta}^2P_1P_2}\Big)\label{asymp_small}
\end{align}
Now note that the last term in (\ref{asymp_small}) is maximized for $\beta^*=0$, which, in turn, implies that the correlation is equal to unity. As shown in Fig. \ref{betavsSNR}, for smaller values of SNR the correlation is indeed higher, indicating that the scheme compensates for the low SNR via cooperation.


\par The asymptotic evaluation for low SNRs is valid up to some critical SNR value at which the correlation drops from its maximal value of unity. We define this critical value of SNR as
\begin{equation}
\mathrm{SNR}^{\mathrm{Crit}}=\mbox{sup}\ \big\{P\ \big|\  \rho^*(P)=1\big\}.\label{SNR^crit}
\end{equation}
To calculate $\mathrm{SNR}^\mathrm{Crit}$ we restrict the analysis to the segment of SNRs at which the correlation is maximal (or equivalently, $\beta^*=0$) and consider (\ref{asymp_small}) taken for $P_1=P_2=P$ and $g_1=g_2=1$. As shown in (\ref{asymp_small}), when $\beta^*=0$ and $P\to 0$, the second logarithm achieves the minimum between the two terms. Fixing $\beta^*=0$ and increasing $P$ increases the second logarithm in (\ref{asymp_small}) while the first term remains unchanged and equals $C_{12}+C_{21}$. As long as
\begin{equation}
\frac{1}{2}\log\Big(1+2P+2\bar{\beta}P\Big)\Big|_{\beta=0}<C_{12}+C_{21}\label{crit_ineq},
\end{equation}
the optimum in achieved for $\beta^*=0$. However, when (\ref{crit_ineq}) is no longer valid, the optimal value of $\beta$ must vary from 0. Thus, calculating $\mathrm{SNR}^{\mathrm{Crit}}$ reduces to solving the following equation:
\begin{equation}
\frac{1}{2}\log\Big(1+2P+2\bar{\beta}P\Big)\Big|_{\beta=0}=C_{12}+C_{21}\label{crit_eq},
\end{equation}
yielding,
\begin{equation}
\mathrm{SNR}^{\mathrm{Crit}}=\frac{2^{2(C_{12}+C_{21})}-1}{4}\label{SNRcrit}.
\end{equation}
\par The value of $\mathrm{SNR}^{\mathrm{Crit}}\ [\mathrm{dB}]$ is represented by the perpendicular dashed green line in the plots shown in Fig. \ref{betavsSNR} and is observed to agree with the numerical results. Note that as the capacities $C_{12}=C_{21}$ grow, so does the value of $\mathrm{SNR}^{\mathrm{Crit}}$, and hence, the transition between the low- and high-SNR regimes occurs at higher SNR values.


\par As the SNR grows, the correlation asymptotically approaches some value in the interval $(0,1)$; this value is denoted by $\rho^*_\infty$. To find this asymptotic correlation, we present the following analysis for the high-SNR regime (assuming $P_1,P_2\gg1$). We start by excluding $\beta^*=0$ as a possible solution for this case (a fact which will be used subsequently). Fixing $C_{12}=C_{21}$ and substituting $\beta=0$ into the sum-rate bounds on $R_1+R_2$ yields (cf. (\ref{rate_bound3})-(\ref{rate_bound4})):
\begin{align}
R_1+R_2  &\leq \min \left\{\begin{array}{ll}
C_{12}+C_{21},\\ \frac{1}{2}\log\Big(1+g_1^2P_1+g_2^2P_2+2g_1g_2\sqrt{P_1P_2}\Big)\end{array}\right\}\nonumber\\
&\stackrel{(a)}{=} C_{12}+C_{21}
\end{align}
where (a) follows from the fact that $P_1,P_2\gg1$. We thus get that for $\mathrm{SNR}\to\infty$, by taking $\beta=0$, the sum-rate is bounded by the sum of the cooperation link capacities. However, since $C_{12}+C_{21}$ is a constant that does not depend on the powers $P_1$ and $P_2$, we conclude that $\beta^*$ cannot be equal to zero.
\par Next, assuming $\beta^*>0$, we calculate $\beta^*$ by using some approximations that are easily justified at a high SNR. First, note that the first and second logarithms in (\ref{asymp_small}) are monotonically increasing and decreasing, respectively, in $\beta$. This implies that the optimum is achieved at the value of $\beta$ at which the functions intersect, that is
\begin{align*}
\frac{1}{2}\log\Big(1&+\beta(g_1^2P_1+g_2^2P_2)\Big)+C_{12}+C_{21}\\&=\frac{1}{2}\log\Big(1+g_1^2P_1+g_2^2P_2+2g_1g_2\sqrt{\bar{\beta}^2P_1P_2}\Big)\numberthis\label{eq_asym_SNR}.
\end{align*}
Using the fact that for high SNR we have:
\begin{align*}
\frac{1}{2}\log\Big(1+\beta(g_1^2P_1+&g_2^2P_2)\Big)+C_{12}+C_{21}\\&\approx\frac{1}{2}\log\Big(\beta(g_1^2P_1+g_2^2P_2)\Big)+C_{12}+C_{21}\\
&\mspace{-163mu}\frac{1}{2}\log\Big(1+g_1^2P_1+g_2^2P_2+2g_1g_2\sqrt{\bar{\beta}^2P_1P_2}\Big)\\&\approx\frac{1}{2}\log\Big(g_1^2P_1\mspace{-3mu}+\mspace{-3mu}g_2^2P_2\mspace{-3mu}+\mspace{-3mu}2g_1g_2\sqrt{\bar{\beta}^2P_1P_2}\Big),
\end{align*}
the equation in (\ref{eq_asym_SNR}) reduces to:
\begin{equation*}
\beta(g_1^2P_1+g_1^2P_2)2^{2(C_{12}+C_{21})}=g_1^2P_1+g_2^2P_2+2g_1g_2\sqrt{\bar{\beta}^2P_1P_2}\ .
\end{equation*}
To further simplify the analysis we again assume a unit channel gain, that is, $g_1=g_2=1$. After some algebra we obtain that the intersection point is given by
\begin{equation}
\beta^*=\frac{(\sqrt{P_1}+\sqrt{P_2})^2}{2^{2(C_{12}+C_{21})}(P_1+P_2)+2\sqrt{P_1P_2}}\ ,\label{beta_comp}
\end{equation}
which by taking $P_1=P_2=P$, reduces to
\begin{equation}
\beta^*=\frac{2}{2^{2(C_{12}+C_{21})}+1}\ .
\end{equation}
Therefore, the optimal correlation at infinite SNR is given by
\begin{equation}
\rho^*_\infty=\sqrt{1-\beta^*}=\sqrt{1-\frac{2}{2^{2(C_{12}+C_{21})}+1}}\ .\label{rho_opt}
\end{equation}
The value of $\rho^*_\infty$, for each value of the cooperation link capacities $C_{12}$ and $C_{21}$, is represented by the horizontal dashed blue line in the plots shown in Fig. \ref{betavsSNR}. Note that the numerical calculations indeed meet the asymptotic results for large values of SNR.

\begin{figure*}[!b]
\hrulefill
\normalsize
\setcounter{newcounter}{\value{equation}}
\setcounter{equation}{94}
\begin{equation}
\mathcal{E}(i,j,k,\mathbf{s})=\bigg\{\Big(\mathbf{U}(i,\tilde{\mathbf{s}}_1),\mathbf{X}_1\big(j,\mathbf{U}(i,\tilde{\mathbf{s}}_1),\tilde{\mathbf{s}}_1\big),\mathbf{X}_2\big(k,\mathbf{U}(i,\tilde{\mathbf{s}}_1),\tilde{\mathbf{s}}_1,\tilde{\mathbf{s}}_2\big),\mathbf{s},\tilde{\mathbf{s}}_1,\tilde{\mathbf{s}}_2,\mathbf{Y}\Big)\in\mathcal{T}_\epsilon^{(n)}(U,X_1,X_2,S,\tilde{S}_1,\tilde{S}_2,Y)\bigg\}\label{error_event}
\end{equation}

\end{figure*}

\par To conclude, we interpret the numerical and analytical results in terms of the optimal transmission strategies of the users for each SNR regime. Recall that the symbols of the codewords transmitted by the users are modeled by the random variables $X_1$ and $X_2$. The fact that for low SNR the correlation is at its maximal value of unity implies that both users tend to transmit the same codewords, which, in turn, indicates that they transmit the same message. However, the only common information the users share is the common message that they have created using the conference. Therefore, we conclude that when the channel quality is low, the best strategy for the users is to transmit the common message exclusively and to abandon their private messages (i.e., the parts of their original messages that they have not managed to share). As the SNR grows beyond $\mathrm{SNR}^{\mathrm{Crit}}$, the correlation between the code symbols decreases to some positive value $\rho^*\in(0,1)$, asymptotically approaching (\ref{rho_opt}). This decrease in correlation is the result, when a higher quality channel is experienced, of each user transmitting not only the common (correlated) message, but also the private (uncorrelated) message.

\par One can also get some additional insight by examining the behavior of the correlation coefficient from the rate perspective. As long as the sum-rate falls below the sum of the cooperation link capacities, i.e., $R_1+R_2\leq C_{12}+C_{21}$, the transmission consists only of the correlated common message; namely, the users are fully cooperative. However, once the sum-rate crosses this threshold value, the transmitted codewords incorporate both the common and private messages, leading to a decrease of the optimal correlation coefficient.


\section{\textsc{Summary and Concluding Remarks}}\label{summary}
In this paper we considered the FSM-MAC with partially cooperative encoders and delayed CSI, and derived its capacity region. The achievability proof used another result of this paper, namely, the capacity region of the FSM-MAC with a common message and delayed CSI. The latter result was obtained by providing a coding scheme that relies on strategy letters. Nonetheless, using the fact that the decoder has access to full CSI, it was also shown that optimal codes can be constructed directly over the input alphabet. Thus, a single codebook was constructed, a fact that formed the basis for simultaneous joint decoding. This approach not only successfully avoids the unnecessary complexity of a coding scheme based on rate-splitting and multiplexing (in contrast to previous works involving delayed CSI \cite{Viswanathan99,Basher_Permuter}), but it also circumvents the need to rely on the corner points of the capacity region, which can render the analysis cumbersome and inefficient when the number of corner points is large.

\par The general conferencing result was then applied to the special case of the Gaussian vector MAC with diagonal channel transfer matrices, which models OFDM-based communication systems. The corresponding capacity region was given in the form of a convex optimization problem and the optimality of Gaussian Markovian inputs was established. This result serves as a generalization of \cite{Wigger_gaussian_cop} to the vector state-dependant case. Focusing on a two-state Gaussian FSM-MAC example, the crucial role of cooperation for low SNR values was demonstrated.

\par We finally note that an extension of the results to a more general state-dependant MAC with partially cooperative encoders and CSI at both transmitters and at the receiver (as, e.g., in \cite{Das_Narayan_MAC2002}) is currently being investigated. Extensions of the results for the Gaussian vector FSM-MAC to general MIMO settings (see, e.g., \cite{ThreeUserMAC_Conf}) and to the ISI channel are also being considered.


\appendices

\setcounter{equation}{91}

\section{\textsc{Proof of the Markov Relation in (\ref{common_markov})}}\label{common_markov_proof}
We prove the Markov relation (\ref{common_markov}) using the following claims. The Markov property in (\ref{common_markov1}) follows from the fact that $(M_0,S^{q-d_1-1})-S_{q-d_1}-S_{q-d_2}-S_{q}$, $\forall q\in\{1,\ldots,n\}$ and $d_1>d_2$, and thus, due to the stationary property of the state process, also ($M_0,S_{1}^{Q-d_1-1},Q) - S_{Q-d_1} -S_{Q-d_2} - S_{Q}$.\\
To show (\ref{common_markov2}) consider the following relations
\begin{align}
P&(x_{1,q}|s_q,s_{q-d_1},s_{q-d_2},u_q,q)\nonumber\\
 &= P(x_{1,q}|s_q,s_{q-d_1},s_{q-d_2},m_0,s^{q-d_1-1},q)\nonumber\\
 &= \sum_{m_1\in\mathcal{M}_1}{P(m_1,x_{1,q}|s_q,s_{q-d_1},s_{q-d_2},m_0,s^{q-d_1-1},q)}\nonumber\\
 &=\sum_{m_1\in\mathcal{M}_1}P(m_1|s_q,s_{q-d_1},s_{q-d_2},m_0,s^{q-d_1-1},q)\nonumber\\
 &\mspace{100mu}\times P(x_{1,q}|s_q,s_{q-d_1},s_{q-d_2},m_0,m_1,s^{q-d_1-1},q)\nonumber\\                                     &\stackrel{(a)}=\sum_{m_1\in\mathcal{M}_1}P(m_1|s_{q-d_1},m_0,s^{q-d_1-1},q)\nonumber\\                                  &\mspace{180mu}\times P(x_{1,q}|s_{q-d_1},m_0,m_1,s^{q-d_1-1},q)\nonumber\\
 &= \sum_{m_1\in\mathcal{M}_1}{P(m_1,x_{1,q}|s_{q-d_1},m_0,s^{q-d_1-1},q)}\nonumber\\
 &= P(x_{1,q}|s_{q-d_1},m_0,s^{q-d_1-1},q),
\end{align}
where (a) follows from the facts that $M_1$ is independent of $S^n$ given $M_0$ and $X_{1,q}$ is a deterministic function of $(M_0,M_1,S_{q-d_1},S^{q-d_1-1})$.
Now, since this is true for all $q\in\{1,\ldots,n\}$ and because the auxiliary random variable is defined as $U=(M_0,S^{Q-d_1-1},Q)$, we conclude that
\begin{equation}
P(x_1|s,\tilde{s}_1,\tilde{s}_2,u)=P(x_1|\tilde{s}_1,u).
\end{equation}
Finally, to show (\ref{common_markov3}) we use the following relations
\begin{align}
P&(x_{2,q}|x_{1,q},s_q,s_{q-d_1},s_{q-d_2},u_q,q)\nonumber\\&
=P(x_{2,q}|x_{1,q},s_q,s_{q-d_1},s_{q-d_2},m_0,s^{q-d_1-1},q)\nonumber\\
&=\sum_{m_2\in\mathcal{M}_2}P(m_2,x_{2,q}|x_{1,q},s_q,s_{q-d_1},s_{q-d_2},m_0,s^{q-d_1-1},q)\nonumber\\
&=\sum_{m_2\in\mathcal{M}_2}P(m_2|x_{1,q},s_q,s_{q-d_1},s_{q-d_2},m_0,s^{q-d_1-1},q)\nonumber\\
&\mspace{70mu}\times P(x_{2,q}|x_{1,q},s_q,s_{q-d_1},s_{q-d_2},m_0,m_2,s^{q-d_1-1},q)\nonumber\\
&\stackrel{(a)}=\sum_{m_1\in\mathcal{M}_1}P(m_2|s_{q-d_1},s_{q-d_2},m_0,s^{q-d_1-1},q)\nonumber\\
&\mspace{142mu}\times P(x_{2,q}|s_{q-d_1},s_{q-d_2},m_0,m_2,s^{q-d_1-1},q)\nonumber\\
&= \sum_{m_1\in\mathcal{M}_1}{P(m_2,x_{2,q}|s_{q-d_1},s_{q-d_2},m_0,s^{q-d_1-1},q)}\nonumber\\
&= P(x_{2,q}|s_{q-d_1},s_{q-d_2},m_0,s^{q-d_1-1},q),
\end{align}
where (a) follows from the facts that $M_2$ is independent of $(X_{1,q},S^n)$ given $M_0$ and $X_{2,i}$ is independent of ($X_{1,q},S_q)$ given $(M_0,M_2,S_{q-d_1},S_{q-d_2},S^{q-d_1-1})$.
Again, the above holds for every $q\in\{1,\ldots,n\}$, and by the definition of the random variable $U$, we conclude that
\begin{equation*}
P(x_2|x_1,s,\tilde{s}_1,\tilde{s}_2,u)=P(x_2|\tilde{s}_1,\tilde{s}_2,u).
\end{equation*}


\section{\textsc{Error Probability Analysis for the Achievability Proof of Theorem \ref{MAC Common Capacity}}}\label{analysis}

We need to show that for the coding scheme presented in Section \ref{MAC Common Achievability} and for a rate triplet $(R_0,R_1,R_2)$ as given in Theorem \ref{MAC Common Capacity}, $P_e^{(n)}\rightarrow 0$ as $n\rightarrow\infty$. Define the event in (\ref{error_event}) at the bottom of the page for any $\mathbf{s}\in\mathcal{S}^n$ (recall that a fixed state sequence $\mathbf{s}$ induces a fixed pair of delayed state sequences $(\tilde{\mathbf{s}}_1,\tilde{\mathbf{s}}_2)$). Denote the transmitted messages by $(m_0,m_1,m_2)$. Using (\ref{error_event}), the probability of error, when averaged over the ensemble of codebooks, can be written as in (\ref{Perror}) at the bottom of the next page. By the union bound, (\ref{Perror}) is further upper bounded by (\ref{Perror_bound}). We proceed with the following steps:

\begin{figure*}[!b]
\setcounter{equation}{95}
\hrulefill
\begin{align*}
P_e^{(n)}&=\mathbb{P}\biggl[\mathcal{E}^C(m_0,m_1,m_2,\mathbf{s})\medcup\Big\{\bigcup_{\tilde{m}_0\neq m_0}\mathcal{E}(\tilde{m}_0,m_1,m_2,\mathbf{s})\Big\}\medcup\Big\{\bigcup_{(\tilde{m}_0,\tilde{m}_1)\neq (m_0,m_1)}\mathcal{E}(\tilde{m}_0,\tilde{m}_1,m_2,\mathbf{s})\Big\}\\
&\mspace{40mu}\medcup\Big\{\bigcup_{(\tilde{m}_0,\tilde{m}_2)\neq (m_0,m_2)}\mathcal{E}(\tilde{m}_0,m_1,\tilde{m}_2,\mathbf{s})\Big\}\medcup\Big\{\bigcup_{(\tilde{m}_0,\tilde{m}_1,\tilde{m}_2)\neq (m_0,m_1,m_2)}\mathcal{E}(\tilde{m}_0,\tilde{m}_1,\tilde{m}_2,\mathbf{s})\Big\}\\
&\mspace{40mu}\medcup\Big\{\bigcup_{\tilde{m}_1\neq m_1}\mathcal{E}(m_0,\tilde{m}_1,m_2,\mathbf{s})\Big\}\medcup\Big\{\bigcup_{\tilde{m}_2\neq m_2}\mathcal{E}(m_0,m_1,\tilde{m}_2,\mathbf{s})\Big\}\medcup\Big\{\bigcup_{(\tilde{m}_1,\tilde{m}_2)\neq (m_1,m_2)}\mathcal{E}(m_0,m_1,\tilde{m}_2,\mathbf{s})\Big\}\biggr]\numberthis\label{Perror}\\
&\leq\underbrace{\mathbb{P}\Big[\mathcal{E}^C(m_0,m_1,m_2,\mathbf{s})\Big]}_{P_e^{[1]}}+\underbrace{\sum_{\tilde{m}_0\neq m_0}\mathbb{P}\Big[\mathcal{E}(\tilde{m}_0,m_1,m_2,\mathbf{s})\Big]}_{P_e^{[2]}}+\underbrace{\sum_{(\tilde{m}_0,\tilde{m}_1)\neq (m_0,m_1)}\mathbb{P}\Big[\mathcal{E}(\tilde{m}_0,\tilde{m}_1,m_2,\mathbf{s})\Big]}_{P_e^{[3]}}\\
&\underbrace{+\sum_{(\tilde{m}_0,\tilde{m}_2)\neq (m_0,m_2)}\mathbb{P}\Big[\mathcal{E}(\tilde{m}_0,m_1,\tilde{m}_2,\mathbf{s})\Big]}_{P_e^{[4]}}+\underbrace{\sum_{(\tilde{m}_0,\tilde{m}_1,\tilde{m}_2)\neq (m_0,m_1,m_2)}\mathbb{P}\Big[\mathcal{E}(\tilde{m}_0,\tilde{m}_1,\tilde{m}_2,\mathbf{s})\Big]}_{P_e^{[5]}}\\
&+\underbrace{\sum_{\tilde{m}_1\neq m_1}\mathbb{P}\Big[\mathcal{E}(m_0,\tilde{m}_1,m_2,\mathbf{s})\Big]}_{P_e^{[6]}}+\underbrace{\sum_{\tilde{m}_2\neq m_2}\mathbb{P}\Big[\mathcal{E}(m_0,m_1,\tilde{m}_2,\mathbf{s})\Big]}_{P_e^{[7]}}+\underbrace{\sum_{(\tilde{m}_1,\tilde{m}_2)\neq (m_1,m_2)}\mathbb{P}\Big[\mathcal{E}(m_0,m_1,\tilde{m}_2,\mathbf{s})\Big]}_{P_e^{[8]}}\numberthis\label{Perror_bound}
\end{align*}
\end{figure*}

\setcounter{equation}{97}
\begin{enumerate}
\item

    $P_e^{[1]}\rightarrow0$ as $n\rightarrow\infty$ by the law of large numbers.\\

\item
To upper bound $P_e^{[5]}$ consider the following:
\begin{align}
P_e^{[5]}&\stackrel{(a)}\leq\sum_{(\tilde{m}_0,\tilde{m}_1,\tilde{m}_2)\neq (m_0,m_1,m_2)}\mspace{-40mu}2^{-n\big(I(U,X_1,X_2;Y|S,\tilde{S}_1,\tilde{S}_2)-\delta_\epsilon\big)}\nonumber\\
&\leq2^{n(R_0+R_1+R_2)}2^{-n\big(I(U,X_1,X_2;Y|S,\tilde{S}_1,\tilde{S}_2)-\delta_\epsilon\big)}\nonumber\\
&=2^{n\big(R_0+R_1+R_2-I(U,X_1,X_2;Y|S,\tilde{S}_1,\tilde{S}_2)+\delta_\epsilon\big)}\label{UB_r0r1r2}
\end{align}

where step (a) is proven in App. \ref{indepenence}, and $\delta_\epsilon\to 0$ as $n\to\infty$. Hence, for the probability $P_e^{[5]}$ to vanish as $n\to\infty$, the following must hold:
\begin{equation}
     R_0+R_1+R_2 < I(U,X_1,X_2;Y|S,\tilde{S}_1,\tilde{S}_2).\label{r0r1r2_bound}
\end{equation}
The mutual information term in (\ref{r0r1r2_bound}) can be rewritten as,
\begin{align}
&I(U,X_1,X_2;Y|S,\tilde{S}_1,\tilde{S}_2)\nonumber\\&\stackrel{(a)}=I(X_1,X_2;Y|S,\tilde{S}_1,\tilde{S}_2)+I(U;Y|X_1,X_2,S,\tilde{S}_1,\tilde{S}_2)\nonumber\\
           &\stackrel{(b)}=I(X_1,X_2;Y|S,\tilde{S}_1,\tilde{S}_2),\label{r0r1r2_bound_final}
\end{align}
where (a) follows from the mutual information chain rule and (b) follows from the fact that $Y$ is independent of $U$ given $(X_1,X_2,S)$, by the underlying channel model (see Section \ref{MAC Common Achievability}).\\

\item
The upper bounds on $P_e^{[2]}$, $P_e^{[3]}$ and $P_e^{[4]}$ are all observed to be redundant, since in all three types of events the codeword $\mathbf{U}$ is assumed incorrect, which immediately implies that the codewords $\mathbf{X}_1$ and $\mathbf{X}_2$ are also incorrect. Hence, requiring the probability of error to vanish as $n\to\infty$ produces the same upper bound as in (\ref{r0r1r2_bound}) but with respect to the partial sum-rates $R_0$, $R_0+R_1$ and $R_0+R_2$. It can therefore be concluded that the upper bound in (\ref{r0r1r2_bound}) is the dominating constraint.\\

\item
To upper bound $P_e^{[6]}$ consider the following steps:
\begin{align}
P_e^{[6]}&\stackrel{(a)}\leq\sum_{\tilde{m}_1\neq m_1}2^{-n\big(I(X_1;Y|X_2,U,S,\tilde{S}_1,\tilde{S}_2)-\delta_\epsilon\big)}\nonumber\\
&\leq2^{nR_1}2^{-n\big(I(X_1;Y|X_2,U,S,\tilde{S}_1,\tilde{S}_2)-\delta_\epsilon\big)}\nonumber\\
&=2^{n\big(R_1-I(X_1;Y|X_2,U,S,\tilde{S}_1,\tilde{S}_2)+\delta_\epsilon\big)}\label{UB_r1}
\end{align}
where the proof of step (a) is provided in App. \ref{indepenence}, and $\delta_\epsilon\to 0$ as $n\to\infty$. It hence follows that $P_e^{[6]}\rightarrow 0$ as $n\rightarrow\infty$ as long as,
\begin{equation}
     R_1 < I(X_1;Y|X_2,U,S,\tilde{S}_1,\tilde{S}_2).\label{r1_bound}
\end{equation}


\item
Using similar arguments it can be shown that to guarantee that $P_e^{[7]}$ and $P_e^{[8]}$ vanish as $n\to\infty$ the following conditions must hold,
\begin{align}
R_2&<I(X_2;Y|X_1,U,S,\tilde{S}_1,\tilde{S}_2)\label{r2_bound}\\
R_1+R_2&<I(X_1,X_2;Y|U,S,\tilde{S}_1,\tilde{S}_2).\label{r1r2_bound}
\end{align}
\end{enumerate}

\par Summarizing the above results, we get that the right-hand side of (\ref{Perror_bound}) goes to zero as the blocklength $n\to\infty$ if the rate bounds in (\ref{T1region}) are satisfied.

\begin{figure*}[!b]
\hrulefill
\normalsize
\setcounter{newcounter}{\value{equation}}
\setcounter{equation}{105}
\begin{align*}
P\big(\mathbf{y}|\mathbf{u}(\tilde{m}_0),\mathbf{x}_1&(\tilde{m}_0,\tilde{m}_1),\mathbf{x}_2(\tilde{m}_0,\tilde{m}_2),\mathbf{s},\tilde{\mathbf{s}}_1,\tilde{\mathbf{s}}_2\big)\\
&=\sum_{\big(\mathbf{x}_1(m_0,m_1),\mathbf{x}_2(m_0,m_2)\big)}P\big(\mathbf{x}_1(m_0,m_1),\mathbf{x}_2(m_0,m_2),\mathbf{y}|\mathbf{u}(\tilde{m}_0),\mathbf{x}_1(\tilde{m}_0,\tilde{m}_1),\mathbf{x}_2(\tilde{m}_0,\tilde{m}_2),\mathbf{s},\tilde{\mathbf{s}}_1,\tilde{\mathbf{s}}_2\big)\\
&=\sum_{\big(\mathbf{x}_1(m_0,m_1),\mathbf{x}_2(m_0,m_2)\big)}P\big(\mathbf{x}_1(m_0,m_1),\mathbf{x}_2(m_0,m_2)|\mathbf{u}(\tilde{m}_0),\mathbf{x}_1(\tilde{m}_0,\tilde{m}_1),\mathbf{x}_2(\tilde{m}_0,\tilde{m}_2),\mathbf{s},\tilde{\mathbf{s}}_1,\tilde{\mathbf{s}}_2\big)\\
&\ \ \ \ \ \ \ \ \ \ \ \ \ \ \ \ \ \ \ \ \ \ \ \ \ \ \ \ \ \times P\big(\mathbf{y}|\mathbf{x}_1(m_0,m_1),\mathbf{x}_2(m_0,m_2),\mathbf{u}(\tilde{m}_0),\mathbf{x}_1(\tilde{m}_0,\tilde{m}_1),\mathbf{x}_2(\tilde{m}_0,\tilde{m}_2),\mathbf{s},\tilde{\mathbf{s}}_1,\tilde{\mathbf{s}}_2\big)\\
&\stackrel{(a)}=\sum_{\big(\mathbf{x}_1(m_0,m_1),\mathbf{x}_2(m_0,m_2)\big)}P\big(\mathbf{x}_1(m_0,m_1),\mathbf{x}_2(m_0,m_2)|\mathbf{s},\tilde{\mathbf{s}}_1,\tilde{\mathbf{s}}_2\big)\\
&\ \ \ \ \ \ \ \ \ \ \ \ \ \ \ \ \ \ \ \ \ \ \ \ \ \ \ \ \ \times P\big(\mathbf{y}|\mathbf{x}_1(m_0,m_1),\mathbf{x}_2(m_0,m_2),\mathbf{u}(\tilde{m}_0),\mathbf{x}_1(\tilde{m}_0,\tilde{m}_1),\mathbf{x}_2(\tilde{m}_0,\tilde{m}_2),\mathbf{s},\tilde{\mathbf{s}}_1,\tilde{\mathbf{s}}_2\big)\\
&\stackrel{(b)}=\sum_{\big(\mathbf{x}_1(m_0,m_1),\mathbf{x}_2(m_0,m_2)\big)}P\big(\mathbf{x}_1(m_0,m_1),\mathbf{x}_2(m_0,m_2)|\mathbf{s},\tilde{\mathbf{s}}_1,\tilde{\mathbf{s}}_2)P(\mathbf{y}|\mathbf{x}_1(m_0,m_1),\mathbf{x}_2(m_0,m_2),\mathbf{s},\tilde{\mathbf{s}}_1,\tilde{\mathbf{s}}_2\big)\\
&=P(\mathbf{y}|\mathbf{s},\tilde{\mathbf{s}}_1,\tilde{\mathbf{s}}_2)\numberthis\label{r0r1r2_y_ind}
\end{align*}
\end{figure*}


\section{\textsc{Proof of (\ref{UB_r0r1r2}) and (\ref{UB_r1}) in Appendix \ref{analysis}}}\label{indepenence}
For simplicity, the codewords associated with a message triplet $(m_0,m_1,m_2)\in\mathcal{M}_0\times\mathcal{M}_1\times\mathcal{M}_2$ are denoted by $\big(\mathbf{u}(m_0),\mathbf{x}_1(m_0,m_1),\mathbf{x}_2(m_0,m_2)\big)$, thus omitting the functional dependence of the codewords on the delayed state sequences $(\tilde{\mathbf{s}}_1,\tilde{\mathbf{s}}_2)$. Moreover, when referring to the conditional $\epsilon$-strongly typical set $\mathcal{T}_\epsilon^{(n)}(U,X_1,X_2,Y|\mathbf{s},\tilde{\mathbf{s}}_1,\tilde{\mathbf{s}}_2)$, we sometimes use the shortened notation $\mathcal{T}$, to save space.
\setcounter{equation}{104}
\subsection{Proof of Step (a) in (\ref{UB_r0r1r2})}\label{r0r1r2_proof}
Consider the equalities,
\begin{align*}
&\mathbb{P}\Big[\big(\mathbf{U}(\tilde{m}_0),\mathbf{X}_1(\tilde{m}_0,\tilde{m}_1),\mathbf{X}_2(\tilde{m}_0,\tilde{m}_2),\mathbf{s},\tilde{\mathbf{s}}_1,\tilde{\mathbf{s}}_2,\mathbf{Y}\big)\\&\mspace{215mu}\in\mathcal{T}_\epsilon^{(n)}(U,X_1,X_2,S,\tilde{S}_1,\tilde{S}_2,Y)\Big]\\
&=\mspace{-170mu}\sum_{\mspace{200mu}\big(\mathbf{u}(\tilde{m}_0),\mathbf{x}_1(\tilde{m}_0,\tilde{m}_1),\mathbf{x}_2(\tilde{m}_0,\tilde{m}_2),\mathbf{y}\big)\in\mathcal{T}}\mspace{-190mu}P\big(\mathbf{u}(\tilde{m}_0),\mathbf{x}_1(\tilde{m}_0,\tilde{m}_1),\mathbf{x}_2(\tilde{m}_0,\tilde{m}_2),\mathbf{y}|\mathbf{s},\tilde{\mathbf{s}}_1,\tilde{\mathbf{s}}_2\big)\\
&\ \\
&=\mspace{-170mu}\sum_{\mspace{200mu}\big(\mathbf{u}(\tilde{m}_0),\mathbf{x}_1(\tilde{m}_0,\tilde{m}_1),\mathbf{x}_2(\tilde{m}_0,\tilde{m}_2),\mathbf{y}\big)\in\mathcal{T}}\mspace{-190mu}P\big(\mathbf{u}(\tilde{m}_0),\mathbf{x}_1(\tilde{m}_0,\tilde{m}_1),\mathbf{x}_2(\tilde{m}_0,\tilde{m}_2)|\mathbf{s},\tilde{\mathbf{s}}_1,\tilde{\mathbf{s}}_2\big)\\
&\mspace{20mu}\times P\big(\mathbf{y}|\mathbf{u}(\tilde{m}_0),\mathbf{x}_1(\tilde{m}_0,\tilde{m}_1),\mathbf{x}_2(\tilde{m}_0,\tilde{m}_2),\mathbf{s},\tilde{\mathbf{s}}_1,\tilde{\mathbf{s}}_2\big).\numberthis\label{r0r1r2_proof_general}
\end{align*}
For the second term in the right-hand side (RHS) of the last equality in (\ref{r0r1r2_proof_general}) we have (\ref{r0r1r2_y_ind}) at the bottom of the page. Step (a) in (\ref{r0r1r2_y_ind}) follows since given $(\tilde{\mathbf{s}}_1,\tilde{\mathbf{s}}_2)$, $\big(\mathbf{x}_1(m_0,m_1),\mathbf{x}_2(m_0,m_2)\big)$ were drawn independently of $\big(\mathbf{x}_1(\tilde{m}_0,\tilde{m}_1),\mathbf{x}_2(\tilde{m}_0,\tilde{m}_2)\big)$, and
(b) follows because the channel output is independent of the incorrect inputs given the correct inputs and states.

\par Substituting (\ref{r0r1r2_y_ind}) into (\ref{r0r1r2_proof_general}) we get
\begin{align*}
&\mathbb{P}\Big[\big(\mathbf{U}(\tilde{m}_0),\mathbf{X}_1(\tilde{m}_0,\tilde{m}_1),\mathbf{X}_2(\tilde{m}_0,\tilde{m}_2),\mathbf{s},\tilde{\mathbf{s}}_1,\tilde{\mathbf{s}}_2,\mathbf{Y}\big)\\&\mspace{230mu}\in\mathcal{T}_\epsilon^{(n)}(U,X_1,X_2,S,\tilde{S}_1,\tilde{S}_2,Y)\Big]\\
&=\mspace{-170mu}\sum_{\mspace{200mu}\big(\mathbf{u}(\tilde{m}_0),\mathbf{x}_1(\tilde{m}_0,\tilde{m}_1),\mathbf{x}_2(\tilde{m}_0,\tilde{m}_2),\mathbf{y}\big)\in\mathcal{T}}\mspace{-180mu}P\big(\mathbf{u}(\tilde{m}_0),\mathbf{x}_1(\tilde{m}_0,\tilde{m}_1),\mathbf{x}_2(\tilde{m}_0,\tilde{m}_2)|\mathbf{s},\tilde{\mathbf{s}}_1,\tilde{\mathbf{s}}_2\big)\\&\mspace{370mu}\times P(\mathbf{y}|\mathbf{s},\tilde{\mathbf{s}}_1,\tilde{\mathbf{s}}_2)\\
&\leq |\mathcal{T}|\cdot2^{-n\big(H(U,X_1,X_2|S,\tilde{S}_1,\tilde{S}_2)-\delta^{(1)}_\epsilon\big)}\cdot2^{-n\big(H(Y|S,\tilde{S}_1,\tilde{S}_2)-\delta^{(2)}_\epsilon\big)}\\
&\leq2^{n\big(H(U,X_1,X_2,Y|S,\tilde{S}_1,\tilde{S}_2)+\delta^{(3)}_\epsilon\big)}\mspace{-5mu}\cdot\mspace{-2mu}2^{-n\big(H(U,X_1,X_2|S,\tilde{S}_1,\tilde{S}_2)-\delta^{(1)}_\epsilon\big)}\\&\mspace{300mu}\times2^{-n\big(H(Y|S,\tilde{S}_1,\tilde{S}_2)-\delta^{(2)}_\epsilon\big)}\\
&=2^{-n\big(I(U,X_1,X_2;Y|S,\tilde{S}_1,\tilde{S}_2)-\delta_\epsilon\big)}
\end{align*}
where $\delta_\epsilon=\sum_{i=1}^3\delta^{(i)}_\epsilon$. This completes the proof of step (a) in (\ref{UB_r0r1r2}).

\subsection{Proof of Step (a) in (\ref{UB_r1})}\label{r1_proof}
\setcounter{equation}{106}
Analogous to the previous subsection we first write:
\begin{align*}
&\mathbb{P}\Big[\big(\mathbf{U}(m_0),\mathbf{X}_1(m_0,\tilde{m}_1),\mathbf{X}_2(m_0,m_2),\mathbf{s},\tilde{\mathbf{s}}_1,\tilde{\mathbf{s}}_2,\mathbf{Y}\big)\\&\mspace{215mu}\in\mathcal{T}_\epsilon^{(n)}(U,X_1,X_2,S,\tilde{S}_1,\tilde{S}_2,Y)\Big]\\
&\ \\
&=\mspace{-170mu}\sum_{\mspace{200mu}\big(\mathbf{u}(m_0),\mathbf{x}_1(m_0,\tilde{m}_1),\mathbf{x}_2(m_0,m_2),\mathbf{y}\big)\in\mathcal{T}}\mspace{-170mu}P\big(\mathbf{u}(m_0),\mathbf{x}_1(m_0,\tilde{m}_1),\mathbf{x}_2(m_0,m_2)|\mathbf{s},\tilde{\mathbf{s}}_1,\tilde{\mathbf{s}}_2\big)\\&\mspace{20mu}\times P\big(\mathbf{y}|\mathbf{u}(m_0),\mathbf{x}_1(m_0,\tilde{m}_1),\mathbf{x}_2(m_0,m_2),\mathbf{s},\tilde{\mathbf{s}}_1,\tilde{\mathbf{s}}_2\big)\numberthis\label{r1_proof_general}.
\end{align*}
By similar arguments to those used to obtain (\ref{r0r1r2_y_ind}), the second term in the RHS of (\ref{r1_proof_general}) can be shown to satisfy
\begin{align*}
P\big(\mathbf{y}|\mathbf{u}(m_0),\mathbf{x}_1&(m_0,\tilde{m}_1),\mathbf{x}_2(m_0,m_2),\mathbf{s},\tilde{\mathbf{s}}_1,\tilde{\mathbf{s}}_2\big)\\
&=P\big(\mathbf{y}|\mathbf{u}(m_0),\mathbf{x}_2(m_0,m_2),\mathbf{s},\tilde{\mathbf{s}}_1,\tilde{\mathbf{s}}_2\big)\numberthis\label{r1_y_ind}.
\end{align*}

\par Substituting (\ref{r1_y_ind}) into (\ref{r1_proof_general}) yields
\begin{align*}
&\mathbb{P}\Big[\big(\mathbf{U}(m_0),\mathbf{X}_1(m_0,\tilde{m}_1),\mathbf{X}_2(m_0,m_2),\mathbf{s},\tilde{\mathbf{s}}_1,\tilde{\mathbf{s}}_2,\mathbf{Y}\big)\\&\mspace{230mu}\in\mathcal{T}_\epsilon^{(n)}(U,X_1,X_2,S,\tilde{S}_1,\tilde{S}_2,Y)\Big]\\
&=\mspace{-170mu}\sum_{\mspace{200mu}\big(\mathbf{u}(m_0),\mathbf{x}_1(m_0,\tilde{m}_1),\mathbf{x}_2(m_0,m_2),\mathbf{y}\big)\in\mathcal{T}}\mspace{-155mu}P\big(\mathbf{u}(m_0),\mathbf{x}_1(m_0,\tilde{m}_1),\mathbf{x}_2(m_0,m_2)|\mathbf{s},\tilde{\mathbf{s}}_1,\tilde{\mathbf{s}}_2\big)\\
&\mspace{194mu}\times P\big(\mathbf{y}|\mathbf{u}(m_0),\mathbf{x}_2(m_0,m_2),\mathbf{s},\tilde{\mathbf{s}}_1,\tilde{\mathbf{s}}_2\big)\\
&\leq |\mathcal{T}|\mspace{-3mu}\cdot\mspace{-5mu}2^{-n\big(\mspace{-3mu}H(U,X_1,X_2|S,\tilde{S}_1,\tilde{S}_2)-\delta^{(1)}_\epsilon\mspace{-3mu}\big)}\mspace{-7mu}\cdot\mspace{-3mu}2^{-n\big(\mspace{-3mu}H(Y|X_2,U,S,\tilde{S}_1,\tilde{S}_2)-\delta^{(2)}_\epsilon\mspace{-3mu}\big)}\\
&\leq2^{n\big(H(U,X_1,X_2,Y|S,\tilde{S}_1,\tilde{S}_2)+\delta^{(3)}_\epsilon\big)}\mspace{-4mu}\cdot\mspace{-2mu}2^{-n\big(H(U,X_1,X_2|S,\tilde{S}_1,\tilde{S}_2)-\delta^{(1)}_\epsilon\big)}\\
&\mspace{259mu}\times2^{-n\big(H(Y|X_2,U,S,\tilde{S}_1,\tilde{S}_2)-\delta^{(2)}_\epsilon\big)}\\
&=2^{-n\big(I(X_1;Y|X_2,U,S,\tilde{S}_1,\tilde{S}_2)-\delta_\epsilon\big)}
\end{align*}
where $\delta_\epsilon=\sum_{i=1}^3\delta^{(i)}_\epsilon$. This establishes step (a) in~(\ref{UB_r1}).


\section*{\textsc{Acknowledgments}}\label{acknowledgments}
The authors would like to thank the associate editor and the anonymous reviewers for their careful reading of the paper and their helpful comments. Especially, we thank the anonymous Reviewer 1 for his suggestion to simplify the achievability proof of Theorem 1, a suggestion which led to the proof presented in the current version of the paper.

\bibliographystyle{IEEEtran}
\bibliography{IEEEabrv,ref}

\begin{IEEEbiographynophoto}{Ziv Goldfeld}
(S'13) received his B.Sc.\@ (summa cum laude) degree in Electrical and Computer Engineering from the Ben-Gurion University, Israel, in 2012. He is currently a student in the direct Ph.D. program for honor
students in Electrical and Computer Engineering at that same institution.

Between 2003 and 2006, he served in the intelligence corps of the Israeli Defense Forces.

Ziv is a recipient of the Dean's List Award, the Basor Fellowship for honor students in the direct Ph.D. program, the Lev-Zion fellowship and the Minerva Short-Term Research Grant (MRG).
\end{IEEEbiographynophoto}


\begin{IEEEbiographynophoto}{Haim H. Permuter}
(M'08-SM'13) received his B.Sc.\@ (summa cum laude) and M.Sc.\@ (summa cum laude) degrees in Electrical and Computer Engineering from the Ben-Gurion University, Israel, in 1997 and 2003, respectively, and the Ph.D. degree in Electrical Engineering from Stanford University, California in 2008.

Between 1997 and 2004, he was an officer at a research and development unit of the Israeli Defense Forces. Since 2009 he is with the department of Electrical and Computer Engineering at Ben-Gurion University where he is currently an associate professor.

Prof. Permuter is a recipient of several awards, among them the Fullbright Fellowship, the Stanford Graduate Fellowship (SGF), Allon Fellowship, and and the U.S.-Israel Binational Science Foundation Bergmann Memorial Award. Haim is currently serving on the editorial boards of the IEEE Transactions on Information Theory
\end{IEEEbiographynophoto}


\begin{IEEEbiographynophoto}{Benjamin  M. Zaidel}
(S’94-M’07) received the B.Sc. and M.Sc. degrees from Tel Aviv University, Israel, in 1990 and 1996, respectively, and the Ph.D. degree from the Technion-Israel Institute of Technology, Haifa, Israel, in 2006, all in electrical engineering.

During 1990-1997, he worked with a communications research group responsible for conducting feasibility studies of communication systems, and in particular cellular systems and other mobile communications networks. During 2001-2007 and 2008-2012, he has been with the Government Research Laboratories in the capacity of a senior research engineer. During the years 2007–2008, he held a Postdoctoral position in the Department of Electronics and Telecommunications, Norwegian University of Science and Technology (NTNU), Trondheim, Norway. He is currently an independent researcher.

His research interests include information-theoretic aspects of multiuser detection techniques, multi-input multi-output channels, cooperative processing in wireless networks, and the application of random matrix theory and statistical physics tools to problems in communications and information theory.
\end{IEEEbiographynophoto}

\end{document}